\documentclass[aps,showpacs,superscriptaddress,nofootinbib,amsmath,amssymb,10pt]{revtex4}
 \usepackage{graphicx}
\usepackage{dcolumn}
\usepackage{bm}
\usepackage{lipsum}
\usepackage{adjustbox}
\usepackage{multirow}
\usepackage{amsmath}
\usepackage{amssymb}
\usepackage{gensymb}
\usepackage{textcomp}
\usepackage{enumerate}

\graphicspath{./figs}
\begin{document}
\title{Perturbative and nonperturbative QCD corrections in polarized nucleon structure functions and spin asymmetries of nucleons}
\author{F. Zaidi}
\author{M. Sajjad Athar\footnote{Corresponding author: sajathar@gmail.com}}
\author{S. K. Singh}
\affiliation{Department of Physics, Aligarh Muslim University, Aligarh - 202002, India}

\begin{abstract}
We have studied the deep inelastic scattering (DIS) of polarized charged leptons from polarized nucleon targets and evaluated 
the polarized nucleon structure functions $g_{1N,2N}(x,Q^2)$ as well as the nucleon asymmetries $A_{1N,2N}(x,Q^2)$ for protons and neutrons.
The higher order perturbative corrections up to the Next-to-Next-to-the-Leading Order (NNLO), using the 
parameterization of Polarized Parton Distribution Functions (PPDFs) given by 
Borsa, Stratmann, Vogelsang, de Florian and Sassot (BDSSV24) in the 3-flavor $\overline{\textrm{MS}}$ scheme, along with the nonperturbative corrections$-$namely the twist-3 corrections
and the Target Mass Corrections (TMC)$-$have been included in the calculations. The numerical results for the polarized nucleon structure functions,
the nucleon asymmetries and the sum rule integrals of the nucleon structure functions$-$corresponding
to the Ellis-Jaffe, Bjorken, and Burkhardt-Cottingham sum rules$-$have been evaluated numerically and are 
found to be in agreement with the experimental results from SLAC, CERN, DESY and JLab. The benchmarking of the PPDFs of BDSSV24 at NNLO 
using the present data on polarized nucleon structure functions and other observables will be useful in studying the nuclear medium effects in the 
scattering of the charged leptons from nuclei at the JLab, EIC, DESY, etc., and the scattering of the (anti)neutrinos from polarized nucleons and nuclei at the proposed neutrino factories.
\end{abstract}
\pacs{13.15.+g,13.60.Hb,21.65.+f,24.10.-i}
\maketitle

\section{Introduction}
The results from the pioneering experiments by the European Muon Collaboration (EMC) at CERN
on the spin structure function of the proton i.e., 
$g_{1p}(x,Q^2)$~\cite{EuropeanMuon:1987isl}, (where $x$ and $Q^2$ being the standard kinematic variables to describe the Deep
Inelastic Scattering (DIS)), when interpreted in light of the predictions of the naive 
Quark Parton Model (QPM)~\cite{Feynman:1969ej, Feynman:1973xc}, along with the results of neutron and hyperon 
$\beta$ decays analyzed assuming the $SU(3)$ symmetry, showed that the simple picture of proton spin in terms of the quark spins is not valid.
It was found that
{\bf (i)} the total spin contribution from quark spins accounts for only a small fraction of the proton's spin,
and {\bf (ii)} the contribution from the strange quark sea is not negligible, but large and negative.
 This surprising result is known in the literature as the ``proton spin crisis''. The resolution of the ``proton spin crisis'' needs very precise experimental
 measurements of the polarized structure function $g_{1p}(x,Q^2)$, not only from the DIS of charged leptons but also 
 from the DIS of (anti)neutrinos from polarized nucleons. These measurements are essential to obtain a full flavor decomposition of nucleon spin and its interpretation 
 based on the theoretical analyses going beyond the QPM  
 by including the contributions from the higher order perturbative and nonperturbative corrections.
 In view of this, extensive efforts have been made to make further measurements of the polarized spin structure functions of the nucleon over a
wide kinematic range of $x$ and $Q^2$, through various experiments on deep inelastic scattering
done with the charged leptons at SLAC~\cite{E142:1993hql, E142:1996thl, E143:1998hbs, E154:1997xfa, E154:1997eyc, E155:1999eug, E155:2000qdr, E155:2002iec}, 
CERN~\cite{SpinMuonSMC:1997voo, SpinMuonSMC:1997mkb, SpinMuon:1998zdf, SpinMuon:1998eqa,
COMPASS:2010wkz, COMPASS:2015mhb, COMPASS:2016jwv}, DESY~\cite{HERMES:1997hjr, HERMES:1998cbu, HERMES:2006jyl, HERMES:2011xgd} and 
JLAB~\cite{CLAS:2003rjt, JeffersonLabHallA:2003joy, JeffersonLabHallA:2004tea, Kramer:2005qe, CLAS:2006ozz, RSS:2006tbm, CLAS:2014qtg, Deur:2014vea, JeffersonLabHallA:2016neg, CLAS:2017qga}.
The measurement of these structure functions off the scattering of (anti)neutrinos from polarized nucleon targets is a subject of considerable interest
at future neutrino factories~\cite{Aschenauer:2013iia, Delahaye:2018yfq, Bogacz:2022xsj, Borsa:2022irn, Riedl:2022pad}.
Theoretically, many attempts have been made to study the polarized spin structure functions of the nucleon
by including various corrections to the results obtained in the QPM for charged lepton
scattering~\cite{Altarelli:1988nr, Anselmino:1994gn, Ball:1995td, Altarelli:1998nb, Lampe:1998eu, 
Hughes:1999wr, Bass:2004xa, Bissey:2005kd, Kuhn:2008sy, Blumlein:2012bf, Bertone:2024taw, Borsa:2024mss},
and in the case of (anti)neutrinos scattering from
nucleons~\cite{Anselmino:1994gn, Anselmino:1996cd, Forte:2001ph, Bass:2004xa, Blumlein:2012bf, Timoshin:2014ina, Riedl:2022pad}.

In the naive QPM calculations, the charged particles are assumed to scatter from the point like individual constituents of nucleon called quark
partons, which are considered to be free, non-interacting and massless spin 1/2 targets. In the asymptotic limit of kinematic variables $Q^2$ and $\nu$, i.e., $Q^2\to \infty,\nu\to\infty$ such
that $\frac{Q^2}{2M\nu}=x$, the nucleon structure functions are found to scale, i.e., they depend only on one variable $x$ and not on $Q^2$, thus satisfying the 
phenomenon called Bjorken scaling~\cite{Bjorken:1968dy}. However, experimentally the measured structure functions $F_{1N, 2N}(x,Q^2)$ in the case of 
scattering from the unpolarized nucleon targets, and $g_{1N, 2N}(x,Q^2)$; ($N=p,n$) in the case of polarized nucleon targets exhibit mild $Q^2$
dependence~\cite{Friedman:1990ur, Whitlow:1990gk, NewMuon:1996fwh, HERMES:2011yno}. This mild $Q^2$ dependence implies that the quark partons
taking part in the scattering process are 
not completely free but are interacting with other quark partons through gluon exchange as
envisaged in the QCD picture of the nucleon. Inclusion of this interaction in the kinematic region of very high but finite non-asymptotic $Q^2$ and $\nu$
introduces some $Q^2$ dependence in the nucleon structure functions. Moreover, the gluons
themselves are self interacting inside the nucleon which also contribute to the nucleon structure functions with a $Q^2$ dependence. 
The corrections to the nucleon structure functions 
due to the interactions of quark partons through the exchange of gluons and the self interaction of gluons inside the nucleon are
calculated using the methods of perturbative QCD treating it in increasing orders of strong coupling $\alpha_s$ of
the quarks and gluons interactions, generally, labeled as the Leading Order (LO), Next-to-Leading Order (NLO), Next-to-Next-to-Leading Order (NNLO)
depending on the higher order terms in the perturbative
QCD. The theoretical formulation for calculating these corrections is based on the 
Dokshitzer-Gribov-Lipatov-Altarelli-Parisi (DGLAP) evolution equations of QCD~\cite{Altarelli:1977zs} as well as 
the methods discussed in Refs.~\cite{Vogt:2008yw, Moch:2014sna, Moch:2015usa, Blumlein:2021enk, Blumlein:2021ryt, Vogt:2004ns}, which 
describes the $Q^2$ evolution of the nucleon structure functions as one 
moves towards lower $Q^2$ from the asymptotic limits of $Q^2\to\infty$. These constitute the higher order perturbative corrections.

In addition, there are nonperturbative corrections such as TMC effect which is related to the finite mass of 
the target nucleon and result in the modification of the scattering kinematics as well as some $Q^2$ dependence, and the corrections 
arising due to the contribution of Higher Twist (HT) terms like twist-3, twist-4, etc., in the 
Operator Product Expansion (OPE) of the hadronic tensor which describes the effect of parton (quark and gluon) correlations beyond the twist-2 
operators in the calculation of nucleon structure functions~\cite{Wilson:1969zs, Brandt:1970kg, Christ:1972ms}. 
These nonperturbative effects can be significant in the kinematic region characterized by high $x$ and low to 
moderate $Q^2$, particularly for $W \leq 2  \; \textrm{GeV}$.  
This region of moderate $Q^2$ and $W$ includes the inclusive production of charged leptons produced along with 
pions through inelastic non-resonant Born processes, which start at $W= 1.08$ GeV, corresponding 
to the threshold for pion production. As  $W$ increases, resonance excitations begin with the excitation of the 
$\Delta$-resonance at  $W= 1.232$ GeV and continue up to $W= 2$ GeV and beyond into the third resonance region producing pions and other mesons.
In the absence of constraints on $Q^2$  
and $W$, the inclusive production of charged leptons in a DIS process, therefore, includes contributions 
from the inelastic region of pion production both from the non-resonant and resonant excitation processes which are described by the hadron dynamics,
in addition to 
contributions from genuine DIS processes described by perturbative QCD.  In order to apply perturbative QCD 
predictions to explain various observables in DIS processes, some kinematic constraints on  $Q^2$  
and $W$  are necessary to define a safe DIS region, for which there is no consensus in the community. However, some 
experimental analyses have used a cut of  $W \ge 2$ GeV for a given $Q^2$ to analyze their data and compare with 
theoretical predictions of perturbative QCD, with and without higher-order perturbative and non-perturbative QCD 
corrections. Recently, Athar and Morfin~\cite{SajjadAthar:2020nvy} discussed in detail the problems associated with defining such 
constraints and suggested a value of  $W \ge 2$ GeV and $Q^2 \ge 1$ GeV$^2$ to define a safe DIS region. Therefore, caution 
is needed when comparing DIS data without a cut on $W$ and $Q^2$ with perturbative QCD predictions.  
In the present work, such kinematic constraints are specifically mentioned when discussing our theoretical 
results and their comparison with experimental data.

Various theoretical groups like the groups of 
Amiri et al.~\cite{Salimi-Amiri:2018had}, Khanpour et al.~\cite{Khanpour:2017cha} 
Blumlein et al.~\cite{Blumlein:2010rn}, Leader et al.~\cite{Leader:2005ci}, Nocera et al.~\cite{Nocera:2014gqa}, 
Ethier et al.~\cite{Ethier:2017zbq} and de Florian et al.~\cite{DeFlorian:2019xxt}, 
etc. have phenomenologically determined
the polarized parton distribution functions using the experimental data on the DIS of charged leptons from the polarized nucleons at NLO. 
Khanpour et al.~\cite{Khanpour:2017cha}, have compared the results for the PPDFs obtained by various authors and find that there 
is significant differences among them. They arise due to the differences in the experimental inputs and the phenomenology used by the various authors.
This induces considerable model dependence in the determination of PPDFs at NLO.
In recent years the determination of the PPDFs using experimental data on various observables in the DIS of the polarized electrons on 
polarized protons, semi inclusive electron DIS i.e. SIDIS and in some cases also the proton-proton scattering has been done at NNLO by some authors like
Mirjalili et al.~\cite{Mirjalili:2022cal}, Borsa et al.~\cite{Borsa:2022irn, Borsa:2024mss}, 
Bertone et al.~\cite{Bertone:2024taw}, etc. For example, Mirjalili et al.~\cite{Mirjalili:2022cal} have evaluated 
the spin structure functions $g_{1N}(x,Q^2)$ and $g_{2N}(x,Q^2)$ by analyzing the updated world data up to NNLO using Jacobi 
polynomials expansion approach. 
They have also included the 
TMC effect following Ref.~\cite{Blumlein:1998nv} and the higher-twist corrections following the BLMP model~\cite{Braun:2011aw} in which nucleon structure 
functions are fitted phenomenologically. 
In most recent determination of PPDFs by the MAPPDF~\cite{Bertone:2024taw} and BDSSV24~\cite{Borsa:2024mss} groups at NLO and NNLO, a comparison 
has been made of the values of the PPDFs obtained at the NLO and NNLO. It has been found that the differences in the values of PPDFs 
evaluated at NLO and NNLO are quite small suggesting stability of the 
higher order perturbative corrections. 
Such a perturbation stability has been also reported in a review of PPDFs at NLO and NNLO done  
by Chiefa~\cite{Chiefa:2024yyr} who also confirms the model dependence of the PPDFs at NLO found earlier by Khanpour et al.~\cite{Khanpour:2017cha}.
However, the model dependence of the PPDFs at NNLO has not been studied. We have used the PPDFs at NLO and NNLO
given by Borsa, Stratmann, Vogelsang, de Florian and Sassot (BDSSV24)~\cite{Borsa:2024mss} 
to study the higher order perturbative corrections by including the nonperturbative corrections due 
to the TMC and HT (twist-3) on the nucleon structure functions and other observables in the DIS of polarized electrons on polarized nucleons.

In this work, we focus on the theoretical calculation of obtaining the polarized structure functions by including the higher order 
perturbative and nonperturbative QCD corrections to the QPM results. Using the unpolarized structure function $F_{1N}(x,Q^2)$, and the polarized 
structure functions $g_{1N, 2N}(x,Q^2)$, the nucleon asymmetries $A_{1N, 2N}(x,Q^2)$; $N=p,n$ have been obtained and compared 
with the experimental results. We have also obtained the integrated spin dependent structure functions to obtain the various sum rule 
integrals corresponding to the Ellis-Jaffe~\cite{Ellis:1973kp}, Bjorken~\cite{Bjorken:1968dy} and Burkhardt-Cottingham~\cite{Burkhardt:1970ti} sum rules and compared them with the experimental results.
In the case of $g_{1N, 2N}(x,Q^2)$,
the higher order perturbative corrections up to NNLO have been incorporated by using the polarized PDFs grids of BDSSV24 parameterization~\cite{Borsa:2024mss}.
The nonperturbative corrections from twist-3 operators and the target mass corrections have been included in a 3-quark flavor $\overline{\textrm{MS}}$
scheme~\cite{Zijlstra:1993sh, Vogelsang:1996im, Blumlein:1996vs, Blumlein:1998nv}. 
The aim of the present work is to perform an analysis of the polarized nucleon structure functions
and other observables in the DIS of polarized electrons from polarized nucleons:

 \begin{enumerate}[(i)]
 \item to study the model dependence of PPDFs at NLO using various parametrizations for this available 
  in literature as well as to study the perturbative stability at NLO and NNLO using the PPDFs from BDSSV24~\cite{Borsa:2024mss}.
 \item to include the nonperturbative corrections due to the TMC and HT (twist-3) effects to describe the experimental data.
 \item  to benchmark a set of PPDFs at NNLO using experimental data on the polarized nucleon structure functions in the DIS 
 of charged leptons, nucleon asymmetries and various sum rule integrals involving these structure functions
 in order to apply them in future to study:
 \begin{enumerate}[(a)]
 \item  the polarized nucleon structure functions in the case of DIS of the (anti)neutrino from polarized nucleon targets as done by 
  us earlier in the case of unpolarized structure functions~\cite{Zaidi:2019asc}.
  \item the nuclear effects in the polarized nucleon structure functions in the case of DIS of charged leptons and (anti)neutrinos 
  from nuclear targets following a field theoretical approach as done in our earlier work~\cite{Zaidi:2019mfd}.
 \end{enumerate}
 \end{enumerate}

In section~\ref{form}, we describe the formulae used to incorporate the higher order perturbative and nonperturbative corrections 
to the results obtained in the QPM.
In section \ref{res}, we present the numerical results for the polarized structure functions $g_{1N, 2N}(x,Q^2)$,
 the nucleon asymmetries $A_{1N,2N}(x,Q^2)$, and the sum rule integrals corresponding to the
Ellis-Jaffe, Bjorken and Burkhardt-Cottingham sum rules
and compare them with the experimental results. We discuss explicitly the role of higher order perturbative and nonperturbative QCD corrections
in the relevant region of kinematic variables $x$ and $Q^2$. The significance of choosing a kinematic cut on the variables $Q^2$ and $W$ (the 
center of mass energy) in defining a true DIS process in the calculation of the nucleon structure functions has also been discussed. 
In section~\ref{summary}, we give conclusion of our work. 

\section{Perturbative and Nonperturbative Corrections to the QPM results}\label{form}
\subsection{Structure Functions in the Quark Parton Model}
The quark-parton-model calculation of nucleon structure functions considers scattering only from the free, non-interacting 
quark partons in the nucleon and neglects
the contribution of scattering from gluons, as well as from quark partons while interacting with gluons and other
partons within the nucleon. This can be considered as 
the leading order calculation in the perturbative QCD description of nucleon structure. In this model, 
$F_{1N}(x,Q^2)$ and $g_{1N,2N}(x,Q^2)$ are found to be independent of $Q^2$.  $F_{1N}(x,Q^2)$ is given by:
\begin{eqnarray}
 F_{1N}(x,Q^2)&=& \frac{F_{2N}(x,Q^2)}{2x}=\frac{1}{2}\; \sum_{i=1}^{n_f} e_i^2\;\Big(q_i(x)+\bar q_i(x)\Big),
 \label{f2a}
\end{eqnarray}
where $n_f$ is the number of quark flavors, $e_i$ is the charge of the corresponding quark $q_i$, the sum over $i$ extends over
all flavors of quarks and antiquarks in the nucleon, $q_i(x)$ and $\bar q_i(x)$ are respectively the quarks 
and antiquarks density distribution. The polarized structure functions $g_{1N, 2N}(x,Q^2)$ are 
given by:
\begin{eqnarray}\label{g1a}
 g_{1N}(x,Q^2)&=& \frac{1}{2}\sum_{i=1}^{n_f} e_i^2(\Delta q_i(x)+\Delta \bar q_i(x)),
\end{eqnarray}
\begin{eqnarray}\label{g2a}
  g_{2N}(x,Q^2)&=& 0,
\end{eqnarray}
with $\Delta q_i(x)$ and $\Delta \bar q_i(x)$ defined as
\begin{eqnarray}
 \Delta q_i(x)&=&q^+(x)-q^-(x),\hspace{3 mm} \Delta \bar q_i(x)=\bar q^+(x)-\bar q^-(x),
\end{eqnarray}
where $q^+(x) (\bar q^+(x))$ and $q^-(x) (\bar q^-(x))$ are the quarks (antiquarks) momentum distribution having spins parallel and antiparallel
to the nucleon spin.
\subsection{Higher order perturbative corrections}\label{qpm_corr}

Beyond the leading order contributions from the scattering of leptons off free, non-interacting quarks, there are contributions
from gluons and the corrections due to the quark partons interacting with gluons and other quark partons in the nucleon. Moreover, gluons
are also self interacting, further modifying their contribution. These effects induce a $Q^2$ dependence in the structure functions and $Q^2$ evolution is performed using the 
DGLAP evolution equations~\cite{Altarelli:1977zs} as well as the methods discussed
in Refs.~\cite{Kodaira:1979ib, Kodaira:1978sh, Bodwin:1989nz, Vogelsang:1990ug, Gorishnii:1985xm, Vogt:2008yw, Moch:2014sna, Moch:2015usa, Blumlein:2021enk, Blumlein:2021ryt, Vogt:2004ns}. 
In general, the nucleon structure functions are calculated by convoluting the parton distribution functions of the unpolarized (polarized) partons with the 
parton coefficient function $C_{f}(x)$ ($\Delta C_{f}(x)$) and summing over all the parton contributions $f(=q,g)$ $(\Delta f(=\Delta q, \Delta g))$ 
in the nucleon. In the case of 
unpolarized structure functions they are discussed in detail in Ref.~\cite{Zaidi:2019mfd}. 
$g_{1N} (x,Q^2)$ is written as~\cite{Zijlstra:1993sh, Vogelsang:1996im}:
\begin{eqnarray}\label{f2_conv}
 g_{1N} (x,Q^2) &=& \sum_{f=q,g} \Delta C_{f}^{(n)}(x,Q^2) \otimes \Delta f(x,Q^2)\;\;,
\end{eqnarray}
where $n=0,1,2,...$ is the order in the perturbative expansion of QCD corresponding to LO, NLO, NNLO, etc., and $\otimes$ represents the Mellin 
transform defined as:
\begin{eqnarray}
 \Delta C_{f,i}(x,Q^2)\otimes \Delta f(x,Q^2) &=&\int_x^1 \Delta C_{f,i}(y,Q^2)\; \Delta f\left(\frac{x}{y},Q^2 \right) {dy \over y}\nonumber
\end{eqnarray}
The above expression (Eq.~\ref{f2_conv}) for $g_{1N} (x,Q^2)$ may be decomposed in terms of singlet ($S$), non-singlet ($NS$) and gluonic ($g$) 
terms as~\cite{Vogelsang:1996im, Zijlstra:1993sh}:
\begin{eqnarray}
 g_{1N} (x,Q^2) &=& \frac{1}{n_f}\sum_{j=1}^{n_f} e_j^2 \underbrace{\Big(\Delta C_{q,S}^{(n)}(x,Q^2) \otimes  \Delta \Sigma(x,Q^2)}_{singlet}+ 
 \underbrace{\Delta C_{g}^{(n)}(x,Q^2) \otimes  \Delta G(x,Q^2)\Big)}_{gluonic}\nonumber\\
 &+& \underbrace{\Delta C_{q,NS}^{(n)}(x,Q^2) \otimes \Delta^{NS}(x,Q^2)}_{nonsinglet},\nonumber
\end{eqnarray}
where $\Delta\Sigma(x,Q^2)=\sum_{i=1}^{n_f}\;\Big(\Delta q_i(x,Q^2)+\Delta \bar q_i(x,Q^2)\Big)$ is the singlet ($S$) 
quark PPDFs, $\Delta G$ is the polarized 
gluonic distribution function, 
$\Delta_{NS}(x,Q^2)=\sum_{i=1}^{n_f} \big(e_i^2-\frac{1}{n_f}\sum_{j=1}^{n_f}\;e_j^2\big)\;\Big(\Delta q_i(x,Q^2)+\Delta \bar q_i(x,Q^2)\Big)$ is 
the non-singlet ($NS$) quark PPDFs, $\Delta C_{q,S}^{(n)}$ and $\Delta C_{q,NS}^{(n)}$ are the singlet and non-singlet 
coefficient functions for the polarized quarks/antiquarks, respectively, and $\Delta C_g^{(n)}$ is the coefficient function
for the gluons with $n=1,2,3...$ for N$^{(n)}$LO~\cite{Vogelsang:1996im, Zijlstra:1993sh, Forte:2001ph}. Introducing a renormalization scale 
$\mu$ in the parton coefficient functions and densities, the above expression at $N^{(n)}LO$; ($n=1,2,...$) is written as~\cite{Vogelsang:1996im, Zijlstra:1993sh}: 
\begin{eqnarray}\label{g1b}
 g_{1N}(x,Q^2)&=& \frac{1}{2}\int_x^1\;\frac{dy}{y}\left[\frac{1}{n_f}\sum_{j=1}^{n_f} e_j^2\left\{\Delta\Sigma\big(\frac{x}{y},\mu^2\big)\;\Delta C_{q,S}^{(n)}\big(y,\frac{Q^2}{\mu^2}\big) + 
  \Delta G\big (\frac{x}{y},\mu^2\big)\;\Delta C_g^{(n)}\big(y,\frac{Q^2}{\mu^2}\big)\right\}\right.\nonumber\\
  &+&\left.\Delta^{NS}\big(\frac{x}{y},\mu^2\big)\;\Delta C_{q,NS}^{(n)}\big(y,\frac{Q^2}{\mu^2}\big) \right].
\end{eqnarray}

The singlet coefficient function $\Delta C_{q,S}^{(n)}$ for the quarks is further decomposed in terms of the non-singlet and the pure singlet ($PS$) contributions as:
\begin{equation}
 \Delta C_{q,S}^{(n)}(x,Q^2)=\Delta C_{q,NS}^{(n)}(x,Q^2)+\Delta C_{q,PS}^{(n)}(x,Q^2).
\end{equation}
At NLO, the pure singlet coefficient function $\Delta C_{q,PS}^{(1)}=0$~\cite{Zijlstra:1993sh}, and 
therefore, $\Delta  C_{q,S}^{(1)}=\Delta C_{q, NS}^{(1)}$, and Eq.~\ref{g1b} may be written as:
\begin{eqnarray}
g_{1N}(x,Q^2)&=& \frac{1}{2}\int_x^1\;\frac{dy}{y}\left[\left\{\frac{1}{n_f}\sum_{j=1}^{n_f} e_j^2 \Delta\Sigma\big(\frac{x}{y},\mu^2\big) + \Delta_{NS}\big(\frac{x}{y},\mu^2\big)\right\}\;
\Delta C_{q,NS}^{(1)}\big(y,\frac{Q^2}{\mu^2}\big)\right.\nonumber\\
   &+&\left.\frac{1}{n_f}\sum_{j=1}^{n_f} e_j^2 \Delta G\big(\frac{x}{y},\mu^2\big)\;\Delta C_g^{(1)}\big(y,\frac{Q^2}{\mu^2}\big)\right] 
\end{eqnarray}
Following Refs.~\cite{Vogelsang:1996im, Zijlstra:1993sh, Forte:2001ph}, the expressions for the non-singlet and gluonic coefficient functions which have been used
in the present calculations are obtained as:
\begin{eqnarray}
 \Delta C_{q,NS}^{(1)}(y,Q^2)&=& \frac{\alpha_s(Q^2)}{4 \pi} \;C_F\;\left[ln\big(\frac{Q^2}{\mu^2} \big) \Big\{ \frac{4}{(1-y)_+}-2 (1+y) +3 \delta(1-y)\Big\}+4\frac{ln(1-y)}{(1-y)_+}- \frac{3}{(1-y)_+}\right.\nonumber\\
 &&\left.-2(1+y) ln(1-y)-2\frac{(1+y^2)}{(1-y)}\;ln(y)+4+2 y+\delta(1-y)\Big(-\frac{4\pi^2}{6}-9 \Big) \right],\\
  \Delta C_{g}^{(1)}(y,Q^2)&=& \frac{\alpha_s(Q^2)}{4 \pi} \;T_F\;n_f\left[ 4(2 y-1) ln\big(\frac{Q^2}{\mu^2} \big) + 4 (2y-1)\big(ln(1-y)-ln(y) \big)+4(3-4 y) \right],
\end{eqnarray}
with $C_F=3$, $T_f=\frac{1}{2}$ and $\alpha_s(Q^2)$ as the strong coupling constant.
In the limit of $\mu^2=Q^2$, the above coefficient functions reduce to the coefficient functions given in Ref.~\cite{Vogelsang:1996im}. 
Following the works of Vogelsang et al.~\cite{Vogelsang:1996im} and Zijlstra et al.~\cite{Zijlstra:1993sh}, we
have performed the numerical evaluation of $g_{1N}(x,Q^2)$ up to NNLO. The simplified expressions for the coefficient functions
at NNLO are lengthy enough~\cite{Zijlstra:1993sh}, 
therefore, we have not given them here explicitly.

The other polarized structure function $g_{2N}(x,Q^2)$ may be written as:
\begin{equation}\label{g2ht}
 g_{2N}(x,Q^2)=g_{2N}^{WW}(x,Q^2) + \bar g_{2N}(x,Q^2),
\end{equation}
where $g_{2N}^{WW}(x,Q^2)$ corresponds to the leading twist (twist-2) term which is given by Wandzura-Wilczek (WW) relation~\cite{Wandzura:1977qf}:
 \begin{eqnarray}\label{wwrel}
 g_{2N}^{WW}(x,Q^2)&=& - g_{1N}(x,Q^2) + \int_{x}^{1}  g_{1N}(y,Q^2) \frac{dy}{y}
\end{eqnarray}
and $\bar g_{2N}(x,Q^2)$ is the contribution from the higher twist terms. 
In the present work, we have evaluated the higher twist correction for $g_{2N}(x,Q^2)$ up to the twist-3 term only
and has been discussed in brief in section~\ref{sec_ht}.

Following a similar analogy, the unpolarized nucleon structure functions $F_{iN}(x,Q^2)~(i=2,L)$ are also expressed in terms of the
convolution of parton coefficient function ($C_{i,f}^{(n)}(x);~(f=q,g)$)
with the parton density distribution ($f(x)$) inside the nucleon. The details of the evolution of the unpolarized structure functions 
and the nonperturbative QCD corrections have been discussed in Refs.~\cite{Athar:2020kqn}.

\subsection{Nonperturbative Corrections}\label{sec_ht}
After performing the perturbative evolution of parton densities up to NNLO, we have incorporated the nonperturbative target mass
correction effect as well as the twist-3 effect following Refs.~\cite{Hekhorn:2024tqm, Blumlein:1998nv, Blumlein:1996vs}.
\subsubsection{Twist-3 corrections}
We have used twist-3 corrected structure functions which are given by~\cite{Blumlein:1996vs, Blumlein:1998nv}:
\begin{eqnarray}
  g_{1N}^{\tau=3}(x,Q^2)&=& \frac{\gamma^2}{(1+\gamma^2)^{3/2}}\;g_{1N}(\xi,Q^2)-\frac{3\;\gamma^2}{(1+\gamma^2)^2}\;\int_\xi^1\;\frac{dv}{v}\;g_{1N}(v,Q^2)\nonumber\\
  &+&\frac{\gamma^2\;(2-\gamma^2)}{(1+\gamma^2)^{5/2}}\;\int_\xi^1\;\frac{dv}{v}\;log\big(\frac{v}{\xi} \big)\;g_{1N}(v,Q^2)
\end{eqnarray}
and 
\begin{eqnarray}
  g_{2N}^{\tau=3}(x,Q^2)&=& \frac{g_{2N}(\xi,Q^2)}{(1+\gamma^2)^{3/2}}\;-\frac{(1-2\;\gamma^2)}{(1+\gamma^2)^2}\;\int_\xi^1\;\frac{dv}{v}\;g_{2N}(v,Q^2)\nonumber\\
  &-&\frac{3\gamma^2}{(1+\gamma^2)^{5/2}}\;\int_\xi^1\;\frac{dv}{v}\;log\big(\frac{v}{\xi} \big)\;g_{2N}(v,Q^2),
\end{eqnarray}
with $\gamma^2=\frac{4 M^2 x^2}{Q^2}$ and $\xi=\frac{2 x}{1+\sqrt{1+\gamma^2}}$.

\subsubsection{Target Mass Corrections}
The target mass corrected polarized structure functions are given by~\cite{Hekhorn:2024tqm, Blumlein:1998nv}: 
\begin{eqnarray}
 g_{1N}^{TMC}(x,Q^2)&=&\frac{1}{(1+\gamma^2)^{3/2}}\;\frac{x}{\xi}\;g_{1N}(\xi,Q^2) + \frac{\gamma^2}{(1+\gamma^2)^2}\;\int_\xi^1\;\frac{dv}{v}\;\left\{\frac{x+\xi}{\xi} \right.\nonumber\\
 &+&\left.\frac{\gamma^2-2}{2 \sqrt{1+\gamma^2}}\;log\big(\frac{v}{\xi} \big)\right\}\;g_{1N}(v,Q^2),\\
  g_{2N}^{TMC}(x,Q^2)&=&-\frac{1}{(1+\gamma^2)^{3/2}}\;\frac{x}{\xi}\;g_{2N}(\xi,Q^2) + \frac{x}{\xi}\; \frac{(1-\frac{\xi}{x}\;\gamma^2)}{(1+\gamma^2)^2}\;\int_\xi^1\;\frac{dv}{v}\;g_{2N}(v,Q^2)\;\nonumber\\
 &+&\frac{3}{2}\;\frac{\gamma^2}{(1+\gamma^2)^{5/2}}\int_\xi^1\;\frac{dv}{v}\;log\big(\frac{v}{\xi} \big)\;g_{2N}(v,Q^2),
\end{eqnarray}

\section{Results and discussion}\label{res}
With the perturbative and nonperturbative QCD corrections as discussed in section~\ref{form}, the polarized nucleon structure 
functions ($g_{iN}(x,Q^2); ~(i=1-2)$) are given by
\begin{eqnarray}\label{g1full}
 g_{iN}(x,Q^2)&=&\underbrace{\underbrace{g_{iN}^{(n)}(x,Q^2)}_{pQCD(n=0,1,2,...)}+g_{iN}^{TMC}(x,Q^2)}_{leading~twist~ (\tau=2)} + \underbrace{g_{iN}^{\tau=3}(x,Q^2)}_{higher~twist}.
\end{eqnarray}
We have obtained the numerical results for the polarized nucleon structure functions
$g_{1N}(x,Q^2)$, $g_{2N}(x,Q^2)$ using Eq.~\ref{g1full}, and for the nucleon asymmetries $A_{1N}(x,Q^2)$ and $A_{2N}(x,Q^2)$  
using Eqs.~\ref{asymg} and \ref{asymg2}. We have also studied
various sum rule integrals defined for $g_{1p,1n}(x,Q^2)$ and $g_{2p}(x,Q^2)$ given by
Ellis-Jaffe~\cite{Ellis:1973kp}, Bjorken~\cite{Bjorken:1968dy}, 
and Burkhardt-Cottingham~\cite{Burkhardt:1970ti} in section~\ref{res_sr}. The numerical results for $F_{1N}(x,Q^2)$ 
needed to evaluate the nucleon asymmetries $A_{1N,2N}(x,Q^2)$ are taken from our earlier work~\cite{Zaidi:2019asc}.
In the present work, the unpolarized 
structure functions are evaluated up to NNLO with TMC and HT effects using the PDF parameterization of Martin, 
Motylinski, Harland-Lang, Thorne (MMHT) 2014 for the nucleons~\cite{Harland-Lang:2014zoa}.

We have performed the numerical calculations by taken into account the higher order perturbative corrections up to NNLO 
with and without a cut on the center of mass energy $W=\sqrt{(q+p)^2}$ (where $p$ is the four momentum
of the polarized nucleon target). All the results are obtained for the kinematic region of $Q^2\ge 1$ GeV$^2$
and incorporating the nonperturbative target mass correction effect. Furthermore, the nonperturbative HT (twist-3) corrections are also included
(wherever mentioned).
The numerical results obtained at NNLO with the TMC and twist-3 effects are labeled as 
``full model'' in the text.
These results are presented and discussed below. The present results have been compared with 
the experimental results from EMC, E142, E143, E154, E155, HERMES, SMC, CLAS, EG1b, CEBAF, E99-117, E06-014,  
COMPASS, etc.~\cite{EuropeanMuon:1987isl, E142:1993hql, E142:1996thl, E143:1998hbs, E154:1997xfa, E154:1997eyc, E155:1999eug, E155:2000qdr, 
E155:2002iec, SpinMuonSMC:1997voo, SpinMuon:1998zdf, SpinMuon:1998eqa, SpinMuonSMC:1997mkb, HERMES:1997hjr, HERMES:1998cbu, HERMES:2006jyl, 
HERMES:2011xgd, CLAS:2003rjt, JeffersonLabHallA:2003joy, JeffersonLabHallA:2004tea, Kramer:2005qe, CLAS:2006ozz, RSS:2006tbm, CLAS:2014qtg, Deur:2014vea, JeffersonLabHallA:2016neg, CLAS:2017qga,
COMPASS:2010wkz, COMPASS:2015mhb, COMPASS:2016jwv}.
Some of the experimental results have been obtained with a kinematic cut of 2 GeV on $W$ 
and therefore, we have also obtained the numerical results with a cut of $W\ge 2$ GeV.

\subsection{Polarized nucleon structure functions}
\subsubsection{  $g_{1p}(x,Q^2)$}\label{sec_g1p}

\begin{figure}[h]
\includegraphics[height=6 cm, width=16 cm]{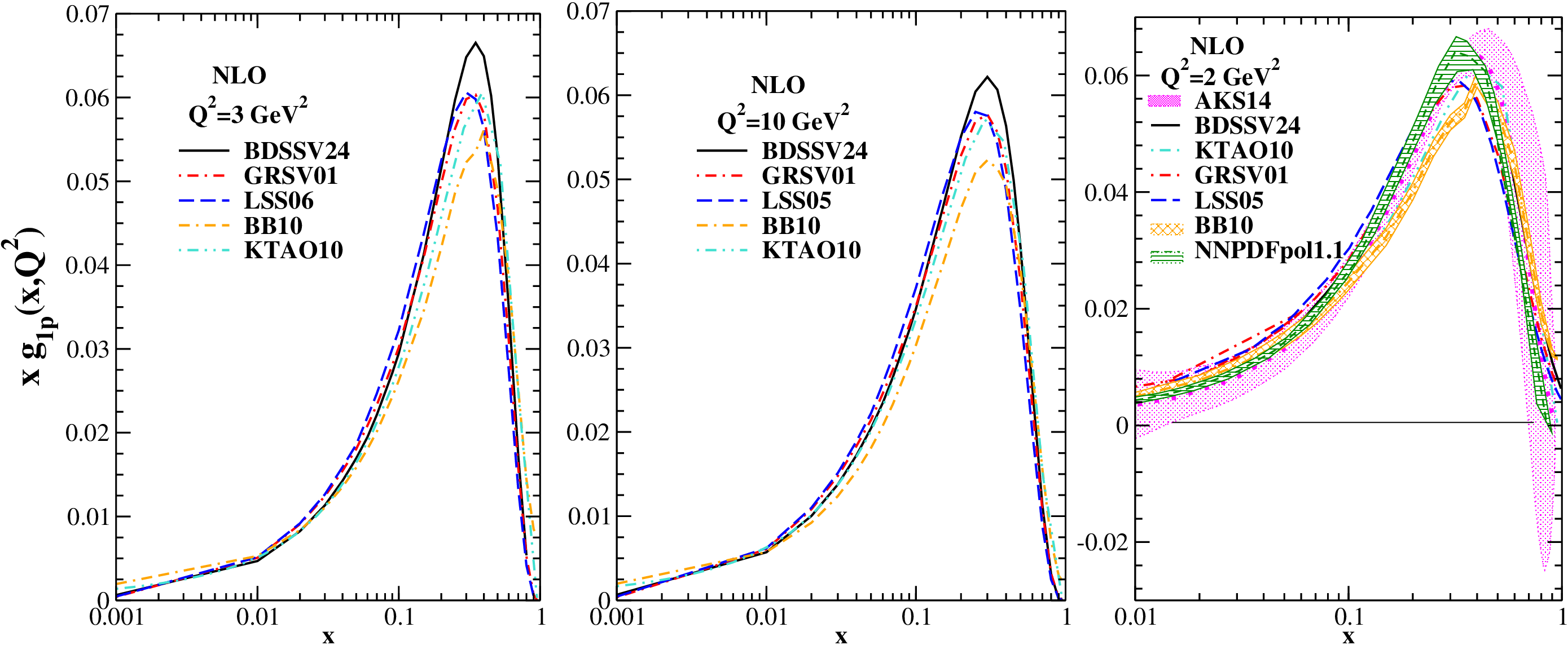}
\caption{ $xg_{1p}(x,Q^2)$ vs $x$ using different PPDFs 
parameterizations viz. Gluck et al.~\cite{Gluck:2000dy} (dash-dotted line), Leader et al.~\cite{Leader:2005ci} (dashed line), 
Blumlein et al.~\cite{Blumlein:2010rn} (double dash-dotted line),
Khorramian et al.~\cite{Khorramian:2010qa} (dash double dotted line) and Borsa et al.~\cite{Borsa:2024mss} (solid line)
at NLO for $Q^2=3$ GeV$^2$ (left panel) and 
$Q^2=10$ GeV$^2$ (middle panel) without applying any kinematical cut on $W$. In the right panel, we have shown the results for $xg_{1p}(x,Q^2)$ 
at $Q^2=2$ GeV$^2$ using different PPDFs parameterizations along with the results of Blumlein et al.~\cite{Blumlein:2010rn} (cross pattern band),
Hekhorn et al.~\cite{Hekhorn:2024tqm} (short dashed line: central value; horizontal line pattern band) and Arbabifar et al.~\cite{Arbabifar:2023hok} (dotted line: central value; dotted pattern band)
with uncertainty bands.}
  \label{figp1}
\end{figure}

 In the literature, different parameterizations are available for the polarized parton distribution functions at NLO, for example, by 
Gluck et al.~\cite{Gluck:1995yr, Gluck:2000dy},
Leader et al.~\cite{Leader:2001kh, Leader:2005ci, Leader:2006xc, Leader:2014uua}, Blumlein et al.~\cite{Blumlein:2002qeu, Blumlein:2010rn}, 
De Florian et al.~\cite{deFlorian:2000bm, deFlorian:2005mw, deFlorian:2008mr, deFlorian:2014yva}, Borsa et al.~\cite{Borsa:2024mss, Borsa:2022irn},
Hirai et al.~\cite{Hirai:2003pm, Hirai:2008aj}, etc. 
To observe the dependence of various PPDFs parameterization, on the evaluation of
polarized structure functions, in Fig.~\ref{figp1}, we have presented the results for $xg_{1p}(x,Q^2)$ vs $x$ at 
$Q^2=3$ GeV$^2$ (left panel) and $Q^2=10$ GeV$^2$ (middle panel) using the central values of
different PPDFs parameterizations at NLO$-$namely,
GRSV01 by Gluck et al.~\cite{Gluck:2000dy} (dash dotted line), LSS05 by Leader et al.~\cite{Leader:2005ci} (dashed line), 
KTAO10 by Khorramian et al.~\cite{Khorramian:2010qa} (double dotted-dash line), 
BB10 by Blumlein et al.~\cite{Blumlein:2010rn} (double dash-dotted line) and BDSSV24 by Borsa et al.~\cite{Borsa:2024mss} (solid line).
We find that there is a significant variation in $xg_{1p}(x,Q^2)$ especially in the region of low and intermediate $x$, for example, 
at $Q^2=3$ GeV$^2$ the difference in the numerical results obtained using BB10~\cite{Blumlein:2010rn} and BDSSV24~\cite{Borsa:2024mss}
parameterizations is approximately 11-20\% in the peak region 
of $x (0.2\le x\le 0.4)$ while it becomes 12-16\% at $Q^2=10$ GeV$^2$. It is noticeable that the difference 
in the numerical results due to the model dependence of PPDFs decreases as $Q^2$ increases though it remains significant even at high $Q^2$. 
These PPDFs parameterizations have uncertainty bands and 
there is some overlap among the different PPDFs parameterizations in certain kinematic regions of $x$ and $Q^2$
when errors are taken into account along with the central values, for example, shown in
Fig.(1) of Chiefa~\cite{Chiefa:2024yyr} and in Fig.(4.5) of Martinez et al.~\cite{Cruz-Martinez:2025ahf}. 
This is consistent with earlier conclusions about the PPDFs parameterizations~\cite{Khanpour:2017cha, Salimi-Amiri:2018had}. Furthermore, in
Ref.~\cite{Arbabifar:2023hok} authors have used 
various PPDFs parameterizations like AKS14~\cite{Arbabifar:2023hok}, DSSV09~\cite{deFlorian:2009vb}, BB10~\cite{Blumlein:2010rn}, etc., 
to obtain the numerical results for $xg_{1p}(x,Q^2)$, and have noted that although the results with the central values of these PPDFs show differences among 
themselves but they lie with in the uncertainty band of AKS14 model. For example, we have shown in the right panel of Fig.~\ref{figp1},
the uncertainty in $x g_{1p}(x,Q^2)$ arising due to the uncertainty in the PPDFs for BB10~\cite{Blumlein:2010rn}, NNPDFpol1.1~\cite{Hekhorn:2024tqm} and  
AKS14~\cite{Arbabifar:2023hok} parameterizations along with the results
for $x g_{1p}(x,Q^2)$ obtained using the central value of the 
other PPDFs parameterizations~\cite{Gluck:2000dy, Leader:2005ci, Khorramian:2010qa, Borsa:2024mss}.
Hopefully, the model dependence of the PPDFs at NNLO will be small as
envisaged by Chiefa~\cite{Chiefa:2024yyr}, Borsa et al.~\cite{Borsa:2024mss}, Bertone et al.~\cite{Bertone:2024taw}
while comparing the results of PPDFs at NLO and NNLO levels. Among the aforementioned PPDFs, Borsa et al.~\cite{Borsa:2024mss} 
have recently provided the global NNLO grids which are obtained by using the world experimental data for DIS, SIDIS, and proton-proton spin asymmetry.
Therefore, in the present work, to study the impact of higher order perturbative corrections 
on the polarized nucleon structure functions and to investigate the perturbative stability of PPDFs, 
we have used the recent PPDFs provided by Borsa et al.~\cite{Borsa:2024mss}, namely BDSSV24 at NLO and NNLO.

\begin{figure}[h]
\includegraphics[height=6 cm, width=5 cm]{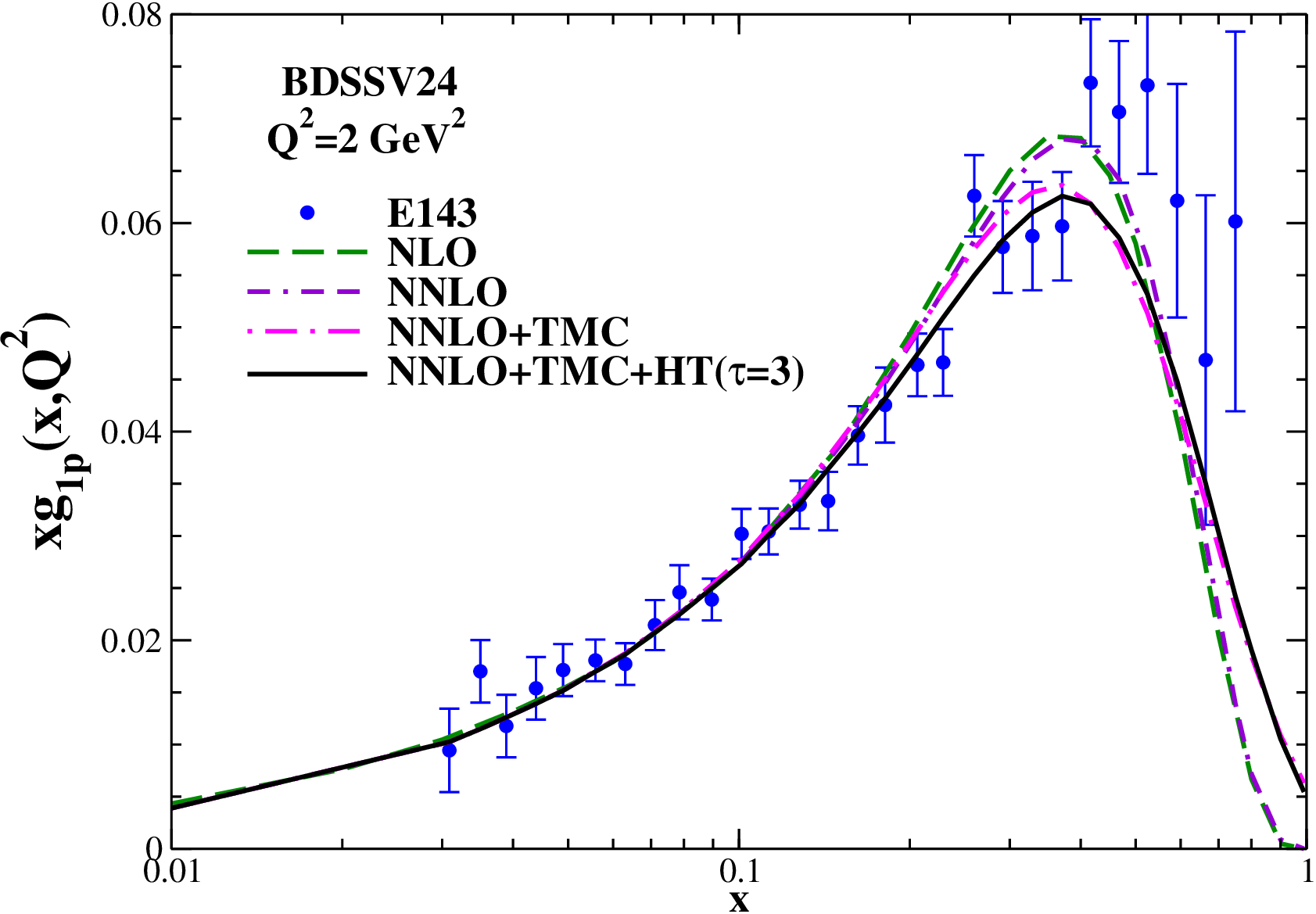}
 \includegraphics[height=6 cm, width=5 cm]{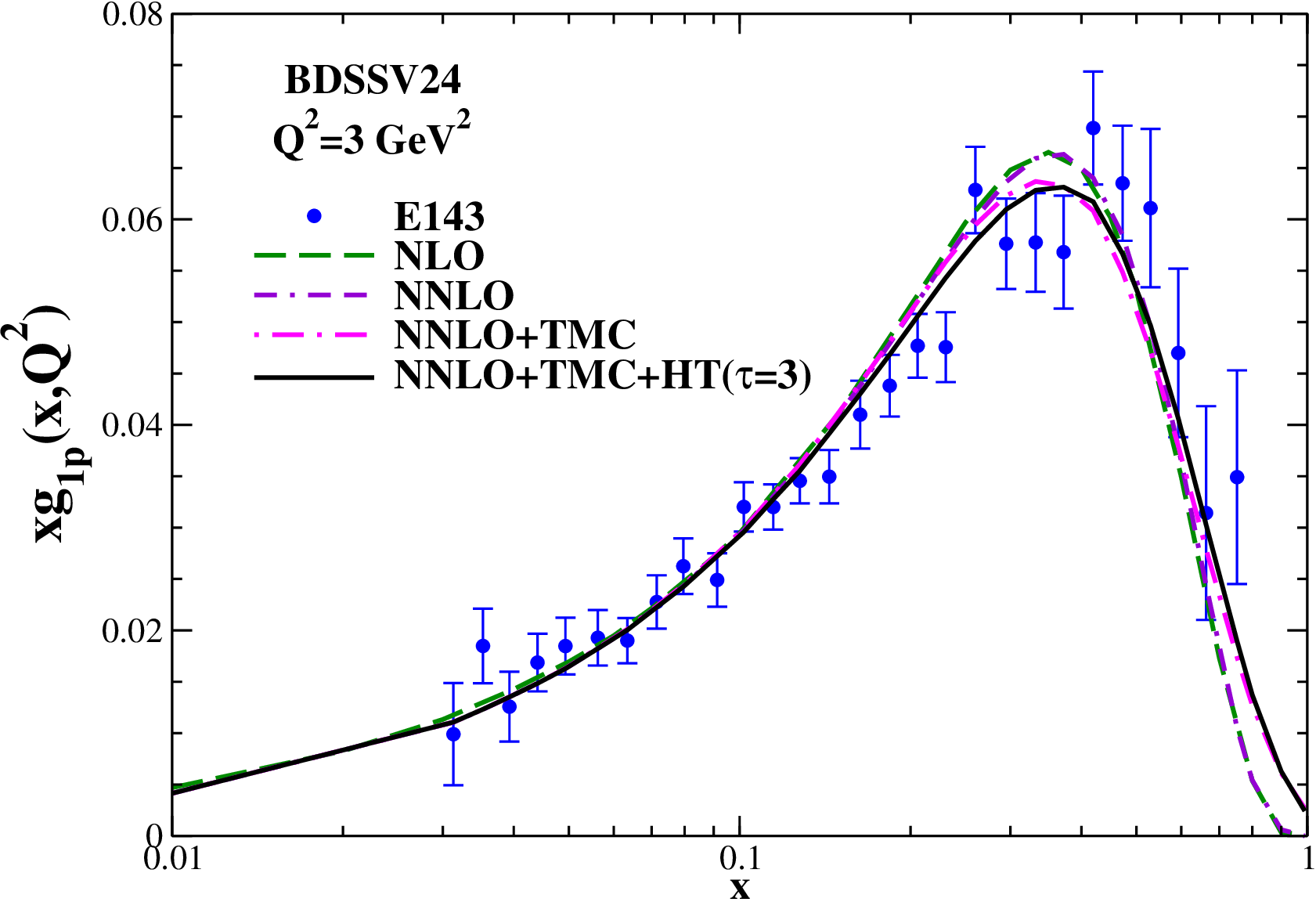}
 \includegraphics[height=6 cm, width=5 cm]{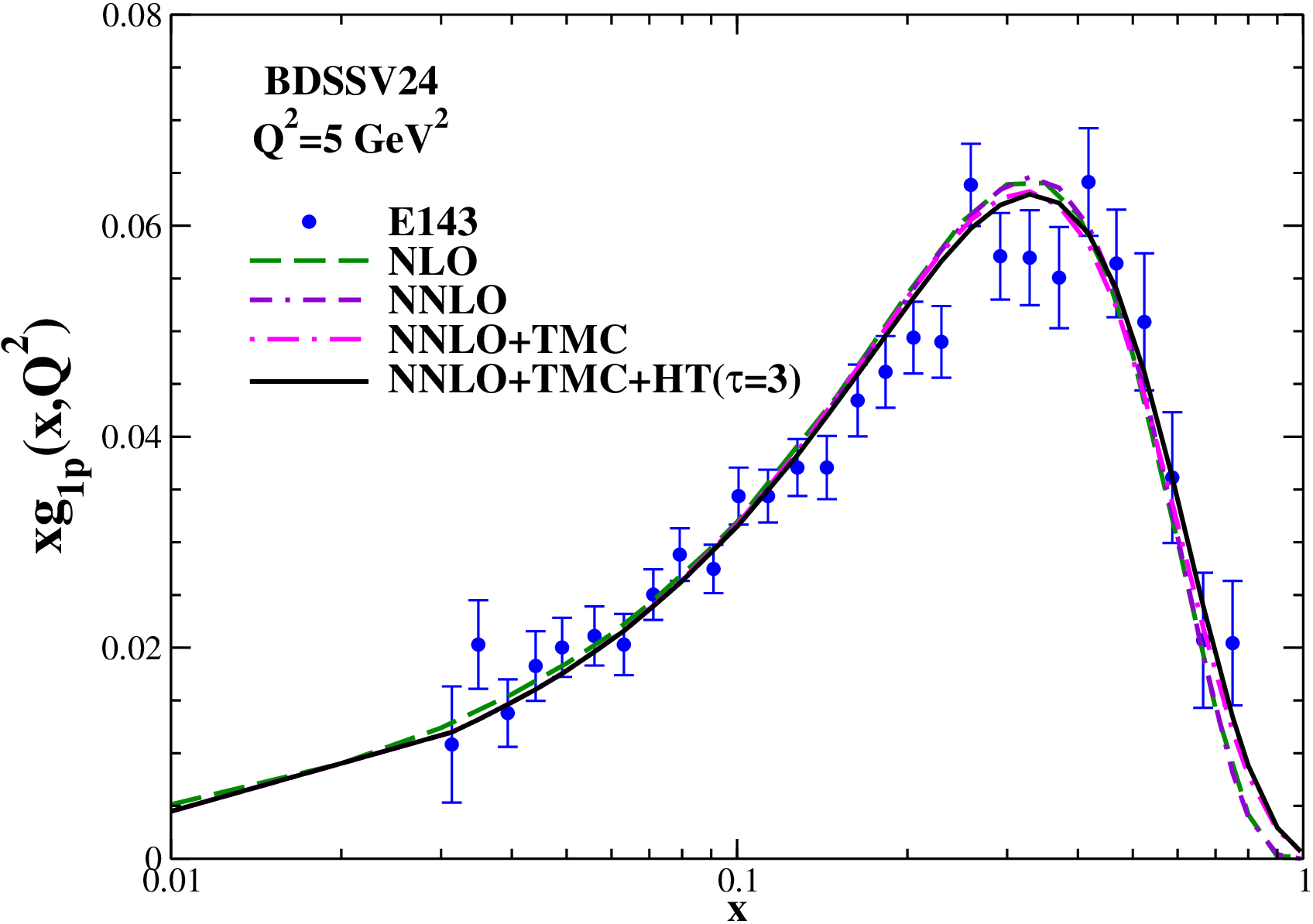}
  \caption{$xg_{1p}(x,Q^2)$ vs $x$ at the different values of $Q^2$ using the PPDFs parameterization of Borsa et al.~\cite{Borsa:2024mss} at {\bf (i)}  
   NLO (dashed line) and NNLO (double dash-dotted line) without TMC and twist-3 correction, and {\bf (ii)} NNLO incorporating the TMC effect 
   without (dash-dotted line) and with (solid line)
   twist-3 correction. Solid lines represent the results with our full model. These numerical results are obtained without applying
  any constraint on the center of mass energy $W$ to compare it with the E143 experimental data~\cite{E143:1998hbs}.}
  \label{fig1}
\end{figure}

To explicitly show the effect of nonperturbative corrections like TMC and HT, as well as the effect of higher order perturbative
corrections in QCD (NLO vs NNLO), in Fig.~\ref{fig1}, we have presented the results for $xg_{1p}(x,Q^2)$ vs $x$ at 
$Q^2=$ 2, 3 and 5 GeV$^2$ without applying any cut on the center of mass energy $W$ using the PPDFs parameterization of Borsa et al.~\cite{Borsa:2024mss}. 
From the figure, it may be noticed that:
\begin{enumerate}[(i)]
 \item  The results for $g_{1p}(x,Q^2)$ evaluated at NNLO have very small difference from the results evaluated at NLO in the entire range of $x$. 
 These results imply the perturbative stability of polarized PPDFs beyond NLO at the values of $Q^2$ taken into considerations.
Furthermore, the effect of the model dependence of $g_{1p}(x,Q^2)$ on the coefficient functions obtained from different 
approaches used in literature~\cite{Zijlstra:1993sh, Vogelsang:1996im, Kretzer:1999nn} has also found to be small.

\item  The TMC effect at NNLO is significant in the intermediate to high $x$ region. As expected it decreases with increase in $Q^2$.
Quantitatively, the inclusion of 
 TMC effect results in a reduction (from no TMC effect) of about
 $3\%$ at $x=0.3$, $9\%$ at $x=0.5$ and an enhancement of $\sim 66\%$ at $x=0.75$ for $Q^2=2$ GeV$^2$, which reduces to $1\%$ at $x=0.3$, $\sim 2\%$ at $x=0.5$,
 and $48\%$ at $x=0.75$ for $Q^2=5$ GeV$^2$.
 
 \item  The effect of the twist-3 correction (black solid lines) is quite small and leads to a mild suppression in the value of $xg_{1p}(x,Q^2)$ in the peak region
 compared to the results obtained using the TMC effect alone ( magenta dash-dotted line). For example, twist-3 correction is found to be 
 4\% at $x=0.3$ for $Q^2=2$ GeV$^2$, reducing to $\sim 1\%$ for $Q^2=5$ GeV$^2$. 
 \item  The numerical results are compared with the experimental data for $xg_{1p}(x,Q^2)$ from E143~\cite{E143:1998hbs}. From the figure, it can be observed that
 our results for $x g_{1p}(x,Q^2)$ show qualitatively good agreement with the experimental data at all the values of $Q^2=2,3$ and 5 GeV$^2$ 
 considered here. 
\end{enumerate}

 In this paper, we are discussing the various observables in the scattering of polarized 
 electrons on polarized protons and their description using the PPDFs at NNLO with the effect of TMC and HT in order to benchmark the PPDFs at NNLO.
 
\begin{figure}[h]
 \includegraphics[height=12 cm, width=0.95\textwidth]{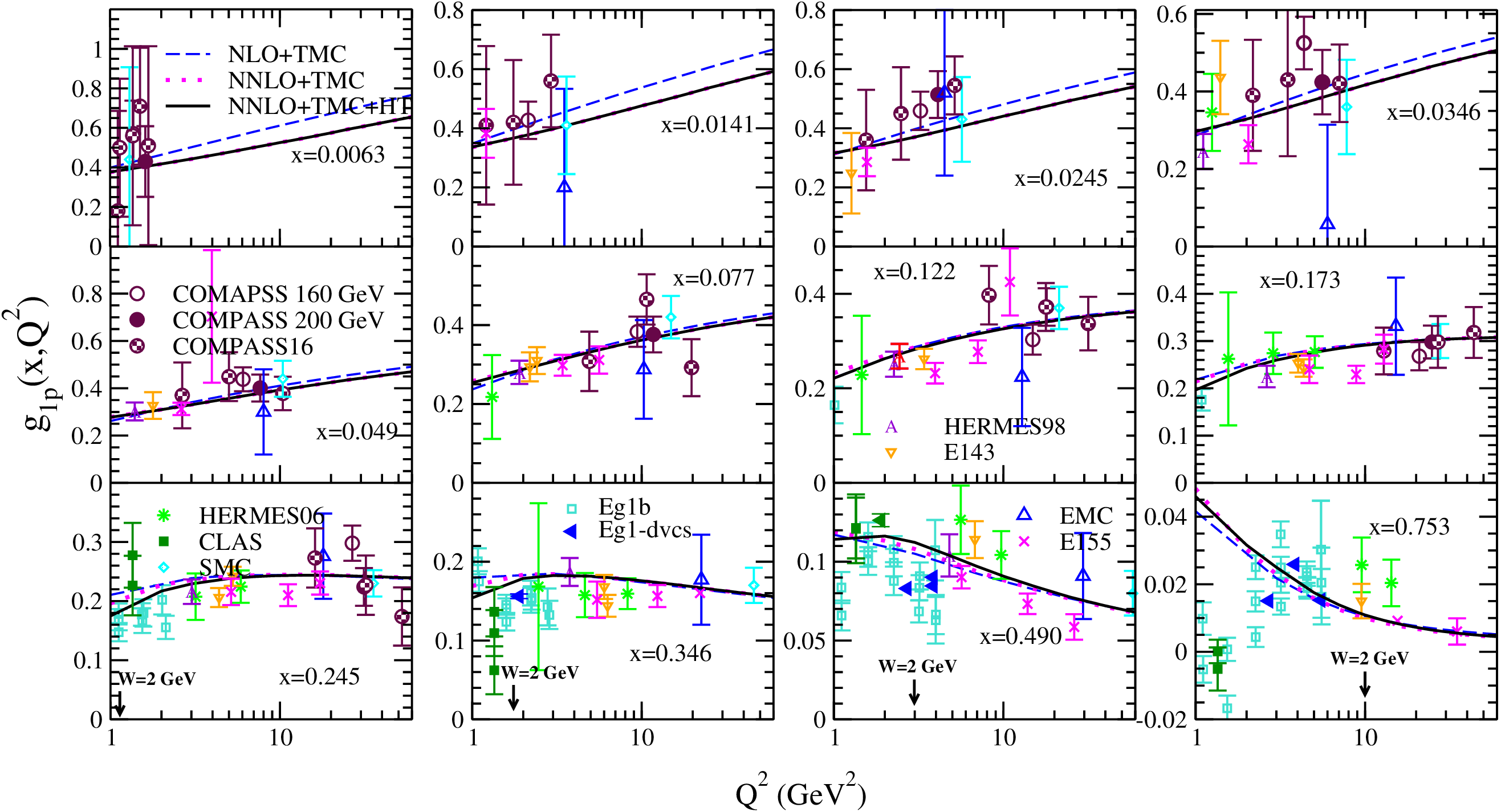}
   \caption{$g_{1p}(x,Q^2)$ vs $Q^2$ at the different values of $x$ without applying any cut on $W$. These results are shown at NLO with TMC effect (dashed line),
   and at NNLO with the TMC (dotted line) and also including the HT (twist-3) effect (solid line)
   for $1 \le Q^2 \le 60$ GeV$^2$ and are compared with 
   the available experimental data from EMC~\cite{EuropeanMuon:1987isl} (triangle up symbol), E143~\cite{E143:1998hbs} (triangle down symbol),
   E155~\cite{E155:2000qdr} (cross symbol),
   SMC~\cite{SpinMuonSMC:1997voo} (open diamond symbol), 
   COMPASS~\cite{COMPASS:2010wkz, COMPASS:2015mhb} (circle symbols), HERMES~\cite{HERMES:1998cbu, HERMES:2006jyl} (star and character symbols), 
   CLAS~\cite{CLAS:2003rjt} (solid square symbol), Eg1-dvcs~\cite{CLAS:2014qtg} (left solid triangle) and EG1b~\cite{CLAS:2017qga} (empty square symbol) experiments.
   In the last panel, the downward arrows on the x-axis are indicating the $Q^2$, where $W$ is $\sim 2$ GeV (notice that $W<2$ GeV while moving to the left from the arrow position).}
   \label{fig3}
\end{figure}
 In Fig.~\ref{fig3}, we have shown the explicit dependence of $g_{1p}(x,Q^2)$ on $Q^2$ by plotting it
 at fixed values of $x$ in the kinematic region of 
 $0.0063\le x< 0.8$ for $Q^2$ varied in the range of $1\le Q^2 \le 60$ GeV$^2$ without applying any cut on $W$. The numerical results are obtained with TMC effect at NLO (dashed lines)
 and NNLO (dotted lines), and at NNLO with TMC and twist-3 corrections (solid lines) using BDSSV24 PPDFs parameterization~\cite{Borsa:2024mss}.
  We observe a difference in the results due to the higher order perturbative evolution (NLO vs NNLO) of parton densities at lower
 values of $x~(x<0.05)$
 which becomes small with the increase in $x$ and $Q^2$. For example, at $x=0.0141$ the reduction in the results evaluated at NNLO from the results
 at NLO is found to be $10-12\%$ for $3\le Q^2\le10$ GeV$^2$, however, at $x=0.0346$, it becomes $6-12\%$. The effect of twist-3 correction is found to be 
 almost negligible for $x<0.173$ at all the values of $Q^2$, while for $x\ge 0.173$ and $Q^2< 5$ GeV$^2$ it is significant.
 We have calculated the $\chi^2/dof$ for the available world 
 data ~\cite{EuropeanMuon:1987isl, E143:1998hbs,E155:2000qdr,COMPASS:2010wkz, COMPASS:2015mhb,HERMES:1998cbu, HERMES:2006jyl,CLAS:2003rjt,CLAS:2014qtg, CLAS:2017qga}
 for $g_{1p}(x,Q^2)$ presented in Fig.~\ref{fig3} in the kinematic region of $Q^2\ge 1$ GeV$^2$ and $W\ge 2$ GeV, usually considered as the safe DIS region in the literature~\cite{SajjadAthar:2020nvy}, 
 and found it to be 1.5 for the theoretical results calculated at NNLO with TMC and twist-3 corrections. This is mainly due to the 
 discrepancy in the theoretical and experimental results of $g_{1p}(x,Q^2)$ in the region of low $Q^2$. In the last panel, we have also shown the value of $W\simeq 2$ GeV corresponding to a given 
 $Q^2$ and $x$ by a downward arrow ($\downarrow$) on the x-axis. While comparing the numerical
 results for $g_{1p}(x,Q^2)$
 with the data available from the EMC~\cite{EuropeanMuon:1987isl}, E143~\cite{E143:1998hbs}, E155~\cite{E155:2000qdr},
SMC~\cite{SpinMuonSMC:1997voo}, COMPASS~\cite{COMPASS:2010wkz, COMPASS:2015mhb}, HERMES~\cite{HERMES:1998cbu, HERMES:2006jyl}, CLAS~\cite{CLAS:2003rjt}
and EG1b~\cite{CLAS:2017qga} experiments, we have found that our full model at NNLO shows reasonable agreement with the experimental observations.

\begin{figure}[h]
\begin{center}
\includegraphics[height=8 cm, width=12 cm]{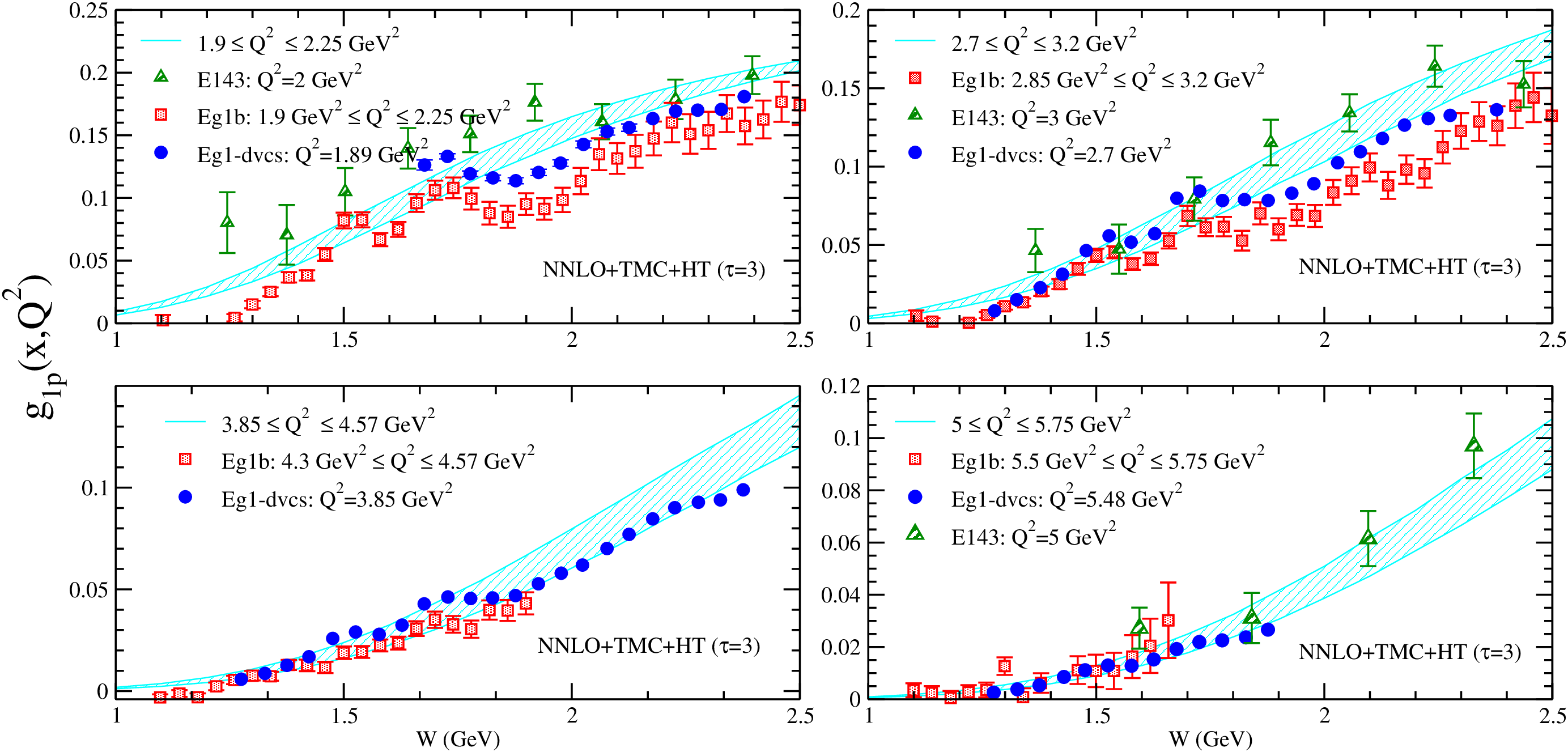}
\end{center}
\caption{ $g_{1p}(x,Q^2)$ vs $W$ using BDSSV24 PPDFs 
parameterization~\cite{Borsa:2024mss} with our full model for the different values of $Q^2$ (in GeV$^2$) (shaded band). 
The lower curve in the band corresponds to the small $Q^2$ value, while the upper curve corresponds to the maximum value of $Q^2$ mentioned in the
individual legends of the figure.
The numerical results are compared with the
Eg1b~\cite{CLAS:2017qga}, Eg1-dvcs~\cite{CLAS:2014qtg} and E143~\cite{E143:1998hbs} experimental data.}
  \label{figp2}
\end{figure}

In Fig.~\ref{figp2}, we have presented the results for $g_{1p}(x,Q^2)$ vs $W$ in the different $Q^2$ bins
varied in the range $\sim 2$ GeV$^2\le Q^2 \simeq 6$ GeV$^2$ and compare them with the experimental data showing different peaks corresponding to 
the nucleon resonances spanning from first to the third resonance regions. This comparison basically highlights, to which extent the DIS curve scales over the 
experimental data showing nucleon resonances and the applicability of DIS formalism in the kinematic region of low $W$ and low $Q^2$.
Present numerical results are obtained using BDSSV24 PPDFs
parameterization~\cite{Borsa:2024mss} with our full model. These numerical results are compared with the experimental data of 
Eg1b~\cite{CLAS:2017qga}, Eg1-dvcs~\cite{CLAS:2014qtg} and E143~\cite{E143:1998hbs}. It may be observed that the numerical results qualitatively explain 
the experimental data from these experiments
in the entire range of $W$ for $Q^2\ge 4$ GeV$^2$, however, for $Q^2<4$ GeV$^2$ numerical results are lower in magnitude than the experimental data
even at the higher values of the center of mass energy $W$. It implies that in the region of $1.08 \le W \le 2$ GeV, where the non-resonant and resonance induced 
inelastic production of mesons may also contribute, the DIS is not adequate as long as $Q^2$ is small. It may also be mentioned 
the data are not consistent among themselves. More study is needed both theoretically and 
experimentally to understand this region.

\begin{figure}[h] 
\includegraphics[height=6.2 cm, width=5.2 cm]{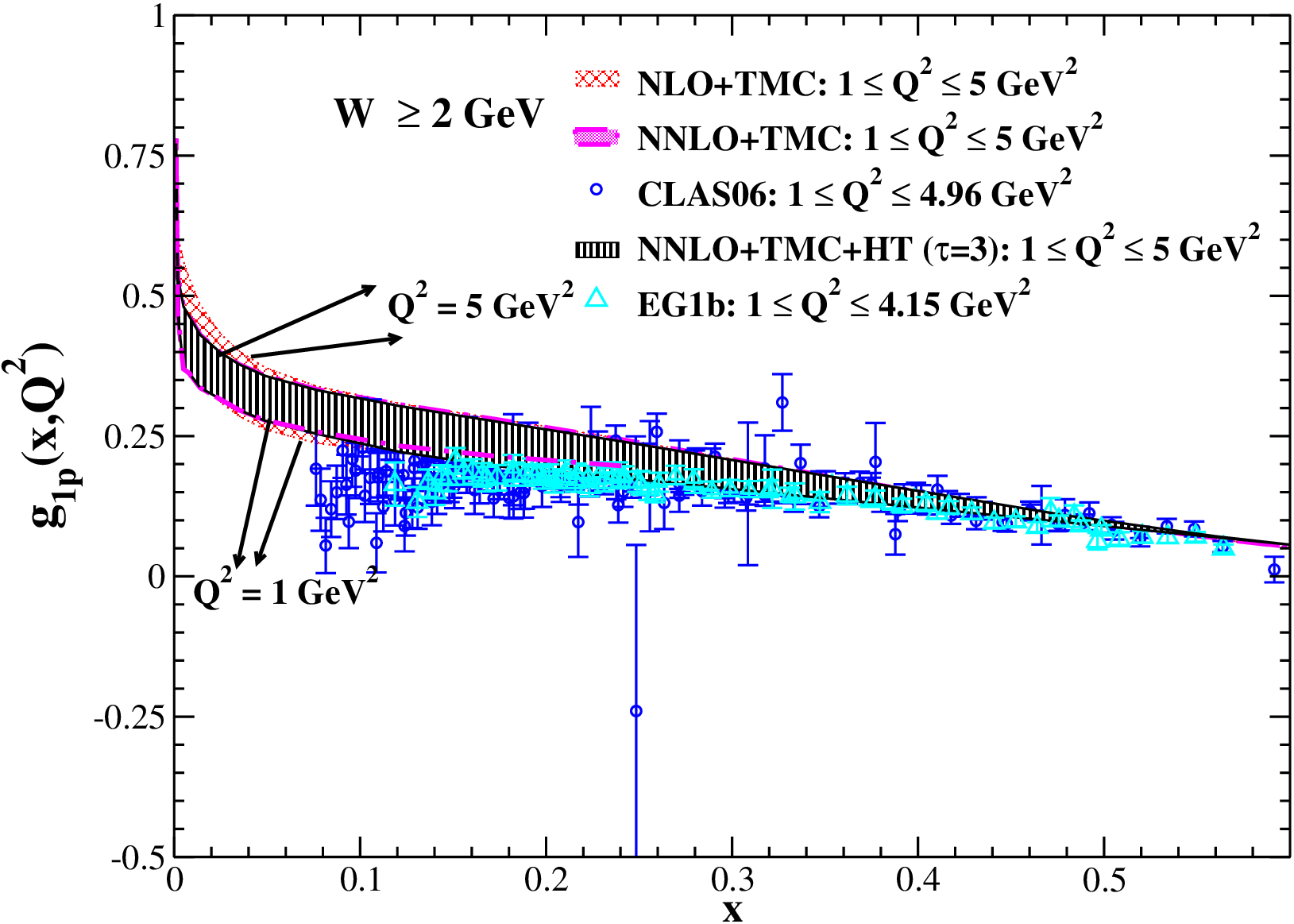}
 \includegraphics[height=6.2 cm, width=5.2 cm]{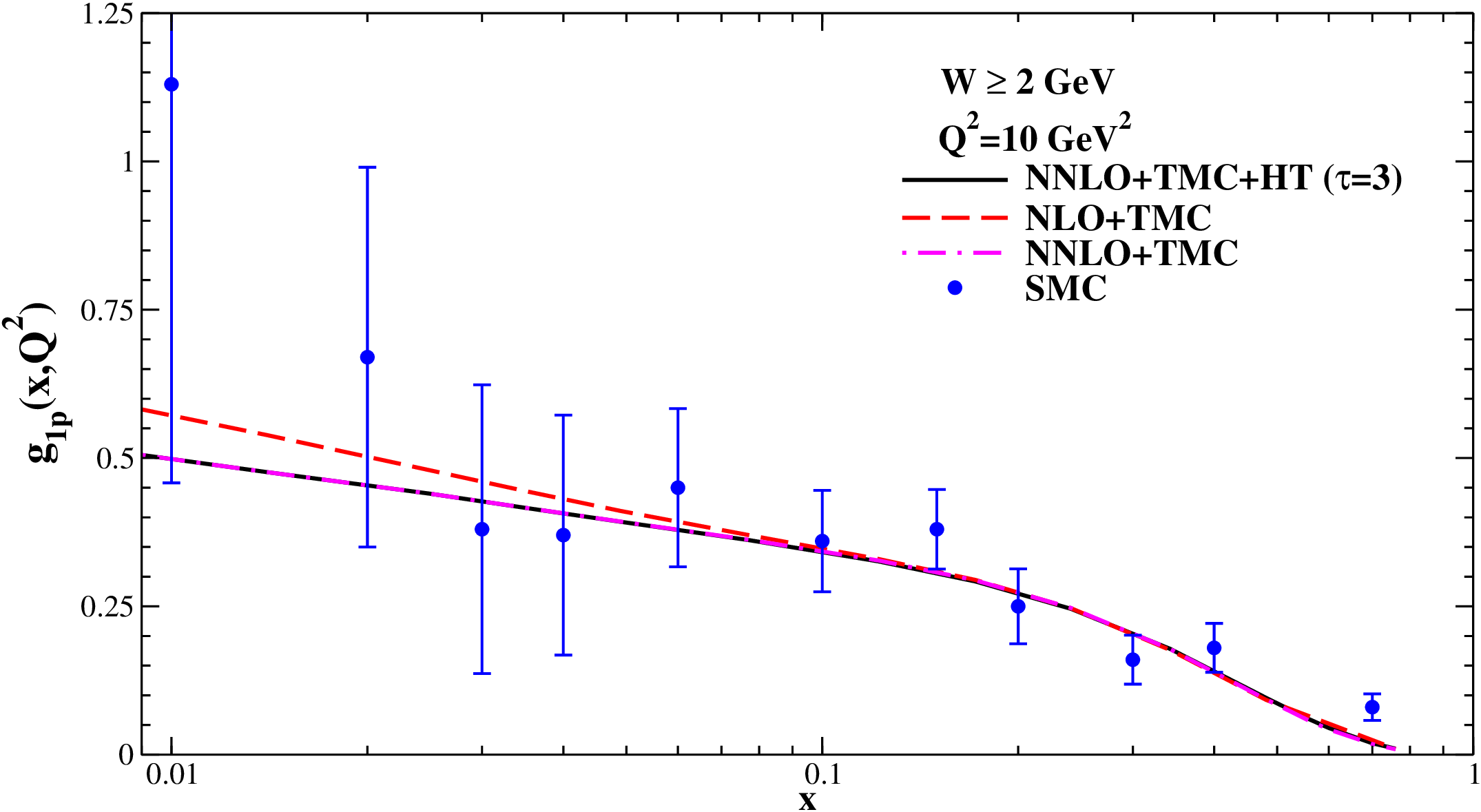}
  \includegraphics[height=6.3 cm, width=5.2 cm]{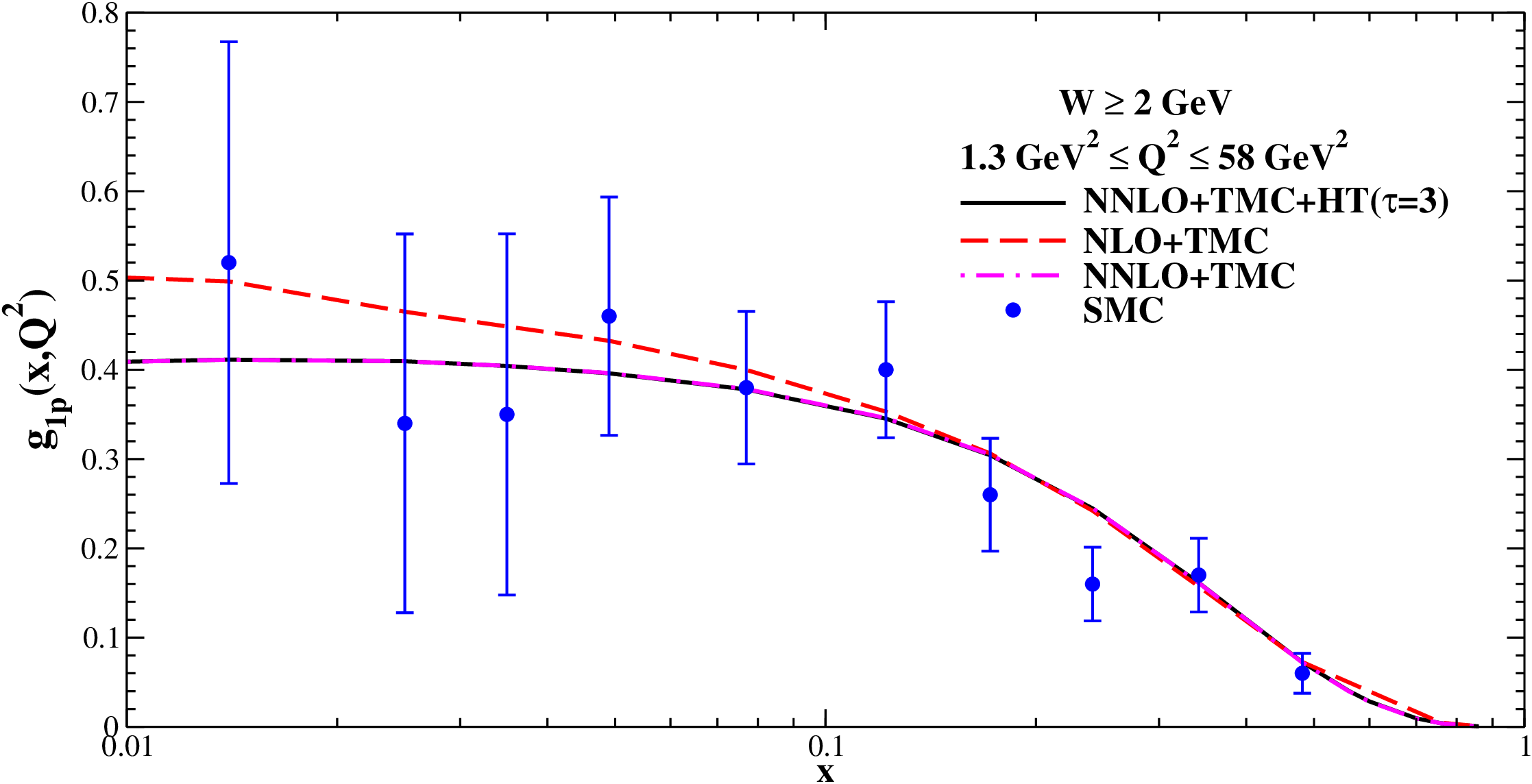}
  \caption{$g_{1p}(x,Q^2)$ vs $x$; (i) Left panel: at NLO with the TMC effect (a band with cross pattern: red color), at 
  NNLO with the TMC effect (a band with dotted pattern: magenta color) and at NNLO with the TMC and twist-3 corrections (a band with vertical lines pattern: black color)
   for $1\le Q^2\le 5$ GeV$^2$, (ii) Middle panel: at NLO (dashed line) and NNLO (dash-dotted line) with 
  the TMC effect only and at NNLO with TMC and twist-3 corrections (solid line) at $Q^2=10$ GeV$^2$, 
  (iii) Right panel: at NLO (dashed line) and NNLO (dash-dotted line) with 
  the TMC effect only and at NNLO with TMC and twist-3 corrections (solid line) for
  $1.3 \le Q^2 \le 58$ GeV$^2$ corresponding to the values of $x$ and $Q^2$ for the experimental data from SMC~\cite{SpinMuonSMC:1997voo}.
  These numerical results are obtained by applying a kinematic cut of 2 GeV on the center of mass energy $W$ (i.e. $W\ge 2$ GeV) and are 
  compared with SMC~\cite{SpinMuonSMC:1997voo} (solid circles), CLAS06~\cite{CLAS:2006ozz} (open circles)
  and EG1b~\cite{CLAS:2017qga} (up triangles) experimental data. }
  \label{fig2}
\end{figure}

 In Fig.~\ref{fig2}, we have shown the results for $g_{1p}(x,Q^2)$ vs $x$ which are obtained for a wide range of $Q^2$ by applying 
 a cut of $W\ge 2$ GeV in order to compare them with the data from CLAS~\cite{CLAS:2006ozz}, EG1b~\cite{CLAS:2017qga} and SMC~\cite{SpinMuonSMC:1997voo}
 experiments. The numerical results are obtained by performing the PPDFs evolution
 up to NNLO in the region of $1\le Q^2\le 5$ GeV$^2$ (left panel: shown by shaded bands), 
 at $Q^2=10$ GeV$^2$ (middle panel) and for a wide range of $Q^2$ ($1.3\le Q^2 \le 58$
 GeV$^2$) in the right panel incorporating the TMC effect and the twist-3 corrections. We find that $g_{1p}(x,Q^2)$ decreases as $x$ increases, 
 while it increases with the increase in $Q^2$. 
 One may notice that at all the values of $Q^2$ considered here, the incorporation of NNLO terms results in a reduction at low $x ~(\lesssim 0.1)$, 
while for $x > 0.1$ these results almost overlap with the results obtained at NLO. Furthermore, at NNLO the numerical 
results without (magenta dash-dotted line) and with (black solid line) the twist-3 corrections
almost overlap at $Q^2=5$ GeV$^2$ while at $Q^2=1$ GeV$^2$ twist-3 correction is observed to be significant and leads to a difference of about
$5-10\%$ for $0.1\le x \le 0.25$.
The numerical results show good agreement with the experimental data 
of CLAS~\cite{CLAS:2006ozz}, EG1b~\cite{CLAS:2017qga} and SMC~\cite{SpinMuonSMC:1997voo} within the statistical errors
 in the entire range of $x$.

\subsubsection{$g_{1n}(x,Q^2)$}
\begin{figure}[h]
 \includegraphics[height=7 cm, width=7.5 cm]{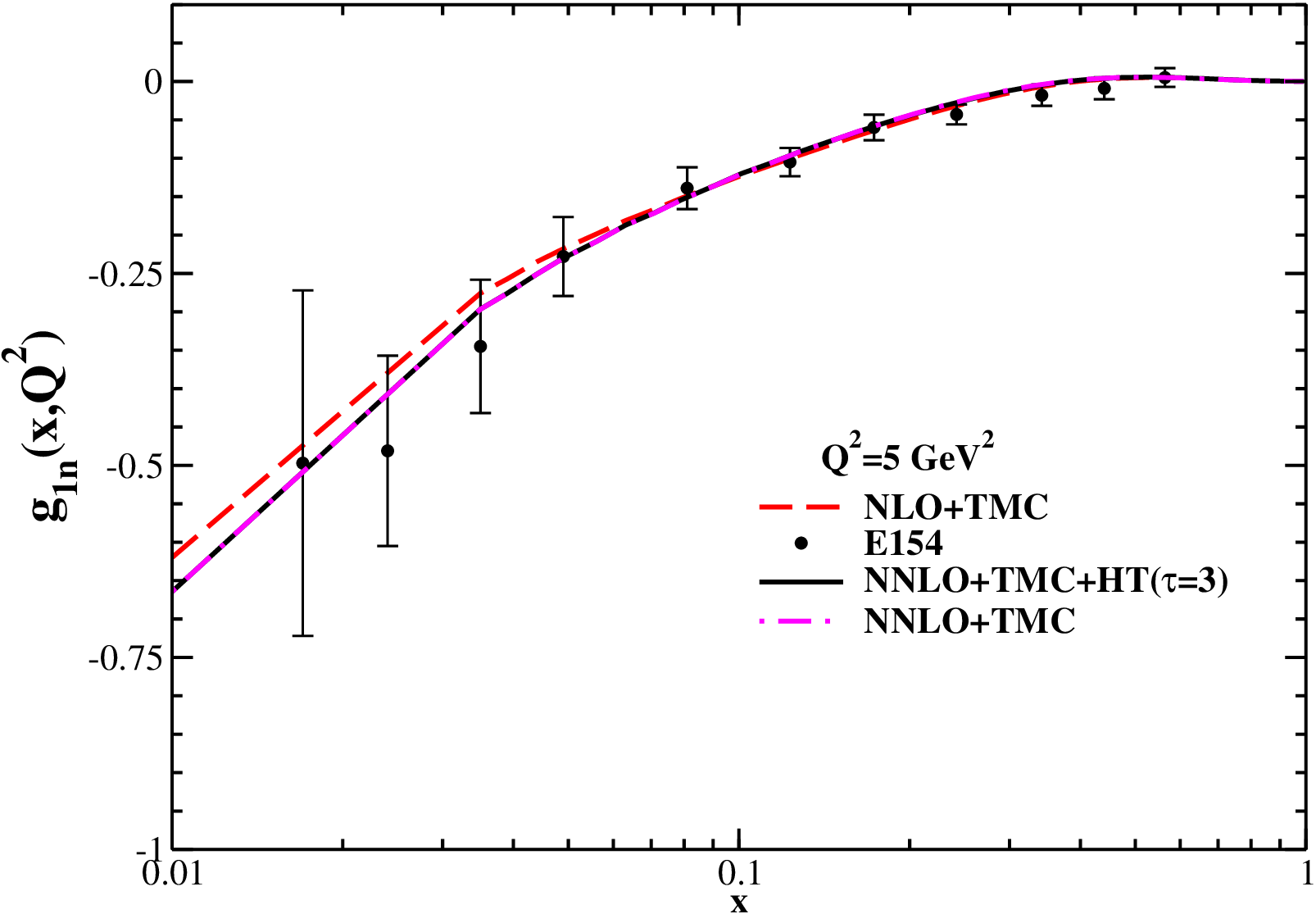}
  \includegraphics[height=7 cm, width=7.5 cm]{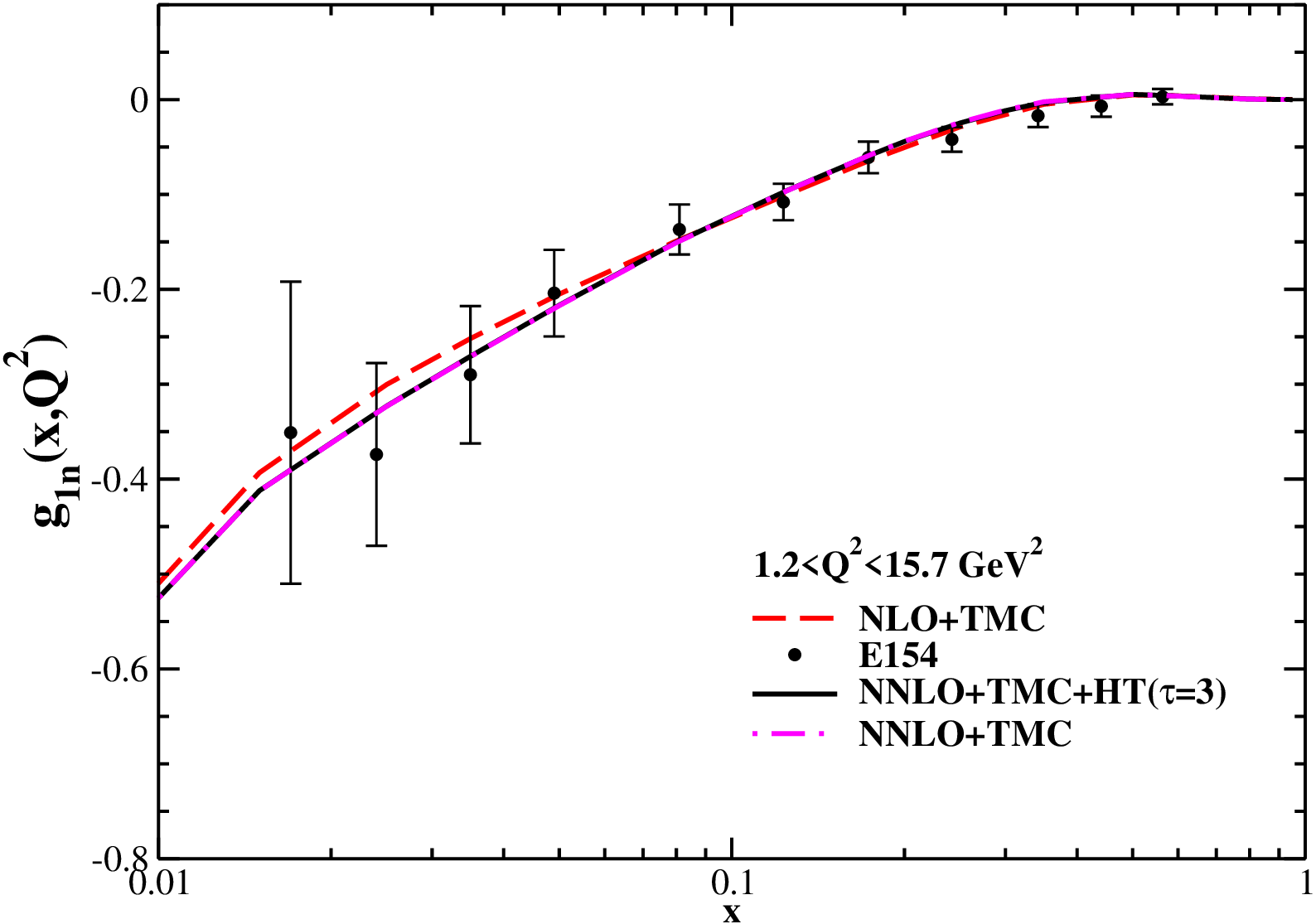}
  \caption{$g_{1n}(x,Q^2)$ vs $x$ at $Q^2=5$ GeV$^2$ (left panel) and for $1.2 \le Q^2 \le 15.7$ GeV$^2$ (right panel) corresponding to the 
 values of $x$ and $Q^2$ for experimental data 
  from E154~\cite{E154:1997xfa} without applying any cut on $W$. The numerical results are obtained at NLO with the TMC effect (dashed line),
  at NNLO with the TMC effect only (dash-dotted line)
  and with the TMC effect and twist-3 correction (solid line), which overlap almost over the NNLO result.}
     \label{fig4}
\end{figure}

In Fig.~\ref{fig4}, the numerical results are presented for the polarized neutron structure function $g_{1n}(x,Q^2)$ vs $x$ at NLO with the TMC effect only
and at NNLO with the TMC and twist-3 effects using BDSSV24 PPDFs parameterization~\cite{Borsa:2024mss}. In the left panel, the results are
shown for a fixed value of $Q^2$ (viz. $Q^2=5$ GeV$^2$), while in the right panel, $Q^2$ lies in the range $1.2 \le Q^2 \le 15.7$ GeV$^2$, 
corresponding to the experimental data from E154~\cite{E154:1997xfa} without applying any cut
on $W$. For a neutron target, $g_{1n}(x,Q^2)$ has a negative sign at low $x$ which is due to the dominance of 
down quarks, and it increases and approaches zero with the increase in $x$ for a fixed $Q^2$. This is in contrast to 
the proton structure function $g_{1p}(x,Q^2)$ which has a positive sign and decreases with the increase in $x$. 
Additionally, we find that the higher order perturbative corrections of the polarized parton densities at NNLO lead to a very small 
decrease in $g_{1n}(x,Q^2)$ at low $x~(x<0.1)$. However, for $x>0.04$, the corrections are small.
We observe that, in the case of neutron as well, the twist-3 correction is almost negligible within the present kinematic range of
$x$ and $Q^2$. Our results at NNLO show good agreement with the E154 experimental data~\cite{E154:1997xfa} across the entire region of $x$ for the values of 
$Q^2$ considered here. 

\subsubsection{ $g_{2p}(x,Q^2)$ and $g_{2n}(x,Q^2)$}

\begin{figure}[h]
 \includegraphics[height=9 cm, width=8.5 cm]{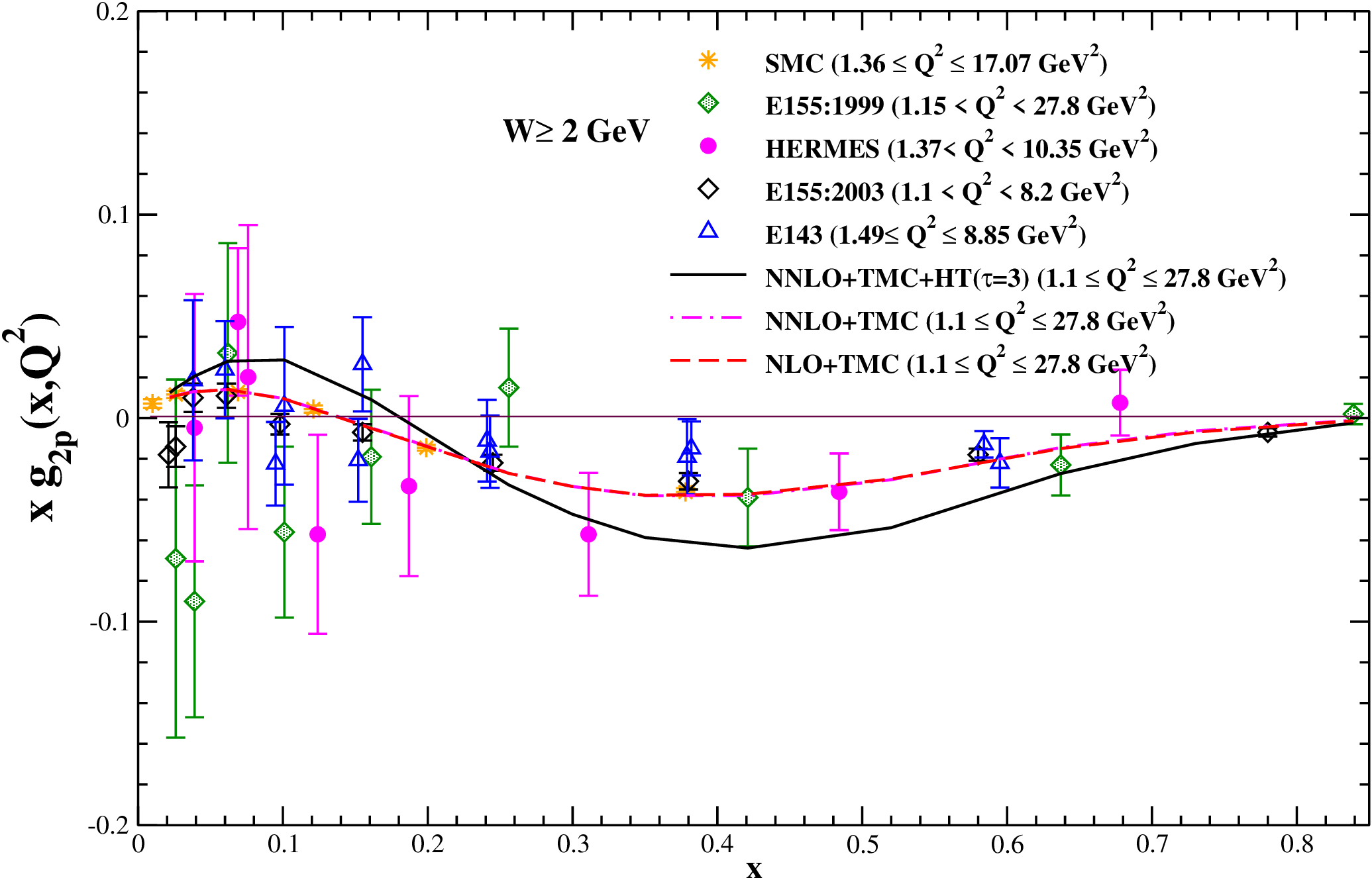}
  \includegraphics[height=9 cm, width=8.5 cm]{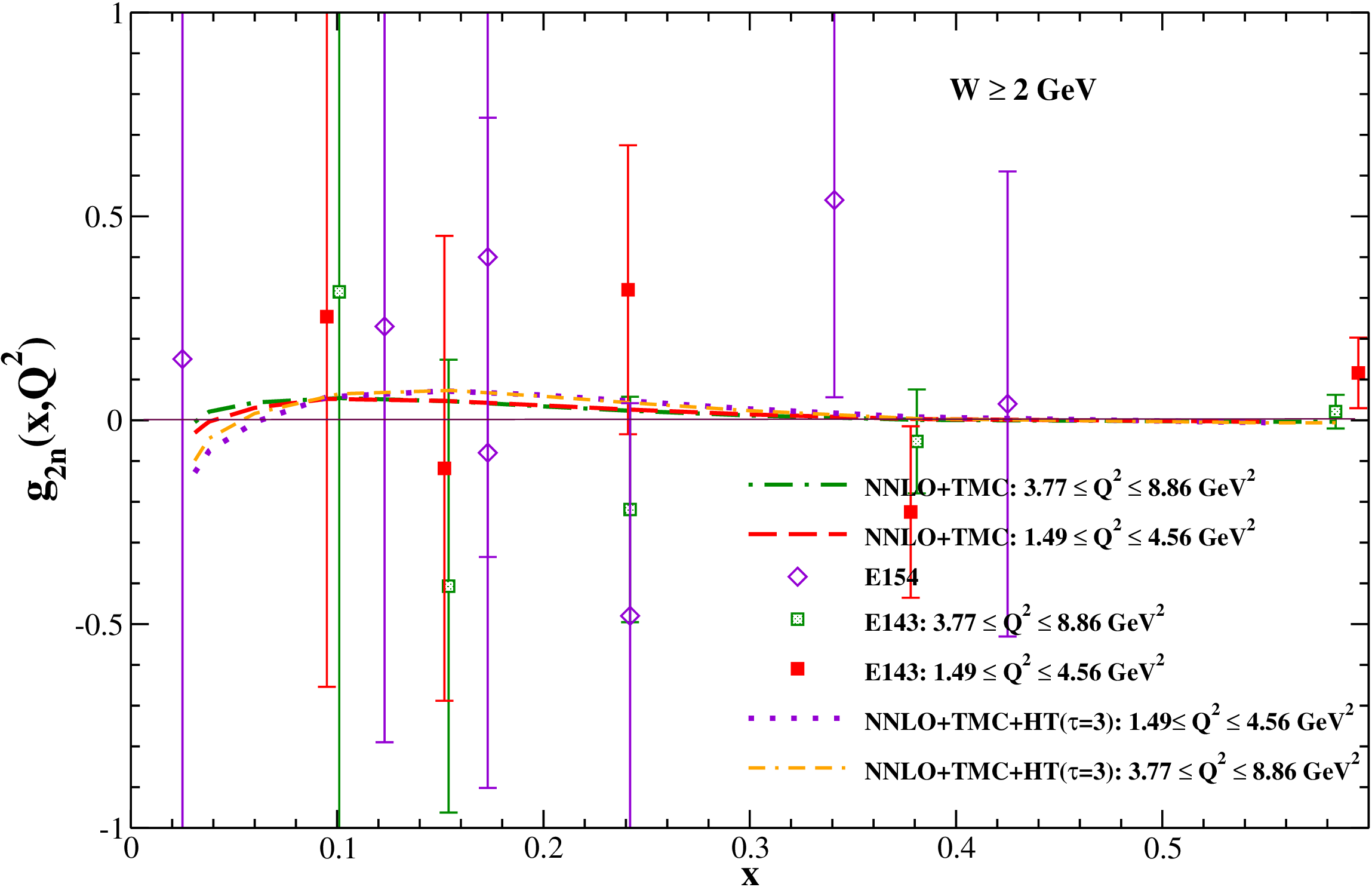}
  \caption{(i) Left panel: $x g_{2p}(x,Q^2)$ vs $x$ at NLO (dashed line) and NNLO (dash-dotted line) with TMC effect only and at NNLO with 
  TMC and HT effects (solid line), and (ii) Right panel: $g_{2n}(x,Q^2)$ vs $x$ at NNLO with TMC effect only (dash dotted and dashed lines)
  and with both the TMC and HT effects (dotted and double dash-dotted lines), in the 
 kinematic region of $W\ge 2$ GeV and $Q^2$ as mentioned in the legends of
 the figures. The numerical results are also compared with the experimental data of E143~\cite{E143:1998hbs}, E154~\cite{E154:1997eyc}, E155~\cite{E155:1999eug, E155:2002iec}, 
 SMC~\cite{SpinMuonSMC:1997mkb}, HERMES~\cite{HERMES:2011xgd} and CEBAF~\cite{Kramer:2005qe} experiments.    }
  \label{fig5}
  \end{figure}

   In Fig.~\ref{fig5} (left panel), we have shown the results for $x g_{2p}(x,Q^2)$ vs $x$ in the $Q^2$ range of $1.1\le Q^2\le 27.8$ GeV$^2$ with a 
   cut on $W\ge2$ GeV corresponding 
   to the kinematic regions of the various experiments, i.e., E143~\cite{E143:1998hbs}, E155~\cite{E155:1999eug, E155:2002iec}, SMC~\cite{SpinMuonSMC:1997mkb} 
   and HERMES~\cite{HERMES:2011xgd}, and in the right panel of this figure, the numerical results are presented for $g_{2n}(x,Q^2)$ vs $x$ at the different values of 
   $Q^2$ corresponding to the available experimental data
  from E143~\cite{E143:1998hbs}, E154~\cite{E154:1997eyc} and CEBAF~\cite{Kramer:2005qe} collaborations. These numerical results are evaluated at 
  NLO and NNLO with the target mass correction effect using the Wandzura-Wilczek relation given in Eq.~\ref{wwrel}. We have also incorporated the twist-3 correction
at NNLO using Eq.~\ref{g2ht}. 
One may observe that the inclusion of higher order perturbative corrections to $x g_{2p}(x,Q^2)$
at NLO results in nonzero positive contribution in the low $x$ region $(x<0.15)$, which becomes negative for $x\ge 0.15$.
We find that the contribution from the NNLO terms is almost negligible.
It is noteworthy that the contribution of the nonperturbative twist-3 corrections is finite in the entire range of $x$, quantitatively about 21\% at $x=0.25$,
54\% at $x=0.35$, and $\sim 70\%$ at $x=0.5$ at NNLO, and decreases with increasing $Q^2$ which is consistent with the 
observation of Monfared et al.~\cite{TaheriMonfared:2014var} and Mirjalili et al.~\cite{Mirjalili:2022cal} that the effect of twist-3 corrections in 
$g_{2p}(x,Q^2)$ is larger than the effect observed in $g_{1p}(x,Q^2)$. 
Our results at NNLO are in agreement with the experimental results 
from experiments at SLAC, CERN and DESY~\cite{SpinMuonSMC:1997mkb, E143:1998hbs, E155:1999eug, E155:2002iec, HERMES:2011xgd}.

In the right panel, the numerical results for $g_{2n}(x,Q^2)$ obtained at NNLO 
are shown for the different values of $Q^2$ lying in the range $1.49\le Q^2 \le 8.86$ GeV$^2$.
We have observed that for $1.49\le Q^2 \le 8.86$ GeV$^2$, there is almost no contribution from the higher order corrections
to $g_{2n}(x,Q^2)$ for $x>0.3$, whereas for $x\le 0.3$, the higher order perturbative corrections 
as well as the nonperturbative twist-3 corrections are not negligible. 
  The effect of the target mass corrections is observed to be small in the region of $x\le 0.3$ (not shown here explicitly). 
  We have also compared the numerical results for $g_{2n}(x,Q^2)$ with 
the data from E143~\cite{E143:1998hbs} and E154~\cite{E154:1997eyc} experiments. Since, the experimental data have quite large statistical errors, 
hence our results either without or with the higher twist corrections show reasonable agreement.

\subsection{Spin asymmetries}\label{sec_asym}


 In general, the polarized nucleon structure functions are experimentally extracted from the cross section measurements and the asymmetries measurements. 
The virtual photon-nucleon asymmetries are given by~\cite{E143:1998hbs,HERMES:2011xgd}:
\begin{eqnarray}
 A_{1N}(x,Q^2) &=& \frac{d\sigma_{1/2}-d\sigma_{3/2}}{d\sigma_{1/2}+d\sigma_{3/2}},\\
  A_{2N}(x,Q^2)&=& \frac{2 d\sigma_{LT}}{d\sigma_{1/2}+d\sigma_{3/2}},
\end{eqnarray}
where $(d\sigma_{3/2})d\sigma_{1/2}$ is the 
 absorption cross section of a transversely polarized photon with spin polarized (anti)parallel
 to the spin of the longitudinally polarized nucleon, and $d\sigma_{LT}$ represents the interference term between the transversely and longitudinally polarized photon-nucleon
amplitudes. These absorption cross sections are related to the nucleon structure functions as~\cite{E143:1998hbs,HERMES:2011xgd}:
\begin{eqnarray}
 d\sigma_{3/2}&=&\frac{4\pi \alpha^2}{M \sqrt{\nu^2+Q^2}}\;\Big(F_{1N}(x,Q^2)- g_{1N}(x,Q^2)+\frac{Q^2}{\nu^2}\;g_{2N}(x,Q^2)\Big),\\
  d\sigma_{1/2}&=&\frac{4\pi \alpha^2}{M \sqrt{\nu^2+Q^2}}\;\Big(F_{1N}(x,Q^2)+ g_{1N}(x,Q^2)-\frac{Q^2}{\nu^2}\;g_{2N}(x,Q^2)\Big),\\
   d\sigma_{LT}&=&\frac{4\pi \alpha^2}{M \sqrt{\nu^2+Q^2}}\;\sqrt{\left(\frac{Q^2}{\nu^2}\right)}\Big( g_{1N}(x,Q^2)+g_{2N}(x,Q^2)\Big),\\
\end{eqnarray}
where $\nu=\frac{Q^2}{2Mx}$. After some simplification, asymmetries are obtained in terms of the polarized and unpolarized nucleon 
structure functions as~\cite{E143:1998hbs,HERMES:2011xgd}: 

\begin{eqnarray}
\label{asymg}
 A_{1N}(x,Q^2) &=& \frac{g_{1N}(x,Q^2)-\gamma^2\;g_{2N}(x,Q^2)}{F_{1N}(x,Q^2)},\\
  A_{2N}(x,Q^2) &=& \frac{\gamma\;\big(g_{1N}(x,Q^2)+g_{2N}(x,Q^2)\big)}{F_{1N}(x,Q^2)},
  \label{asymg2}
\end{eqnarray}
where $\gamma=\sqrt{\frac{4M^2 x^2}{Q^2}}$. 
In the limit of high $Q^2\to \infty$ or low $x\to 0$ the kinematic factor 
$\gamma$ becomes very small ($\gamma<<1$), hence the asymmetries $A_{1N}(x,Q^2)$ and $A_{2N}(x,Q^2)$ 
get simplified and are given by
$$ A_{1N}(x,Q^2) =\frac{g_{1N}(x,Q^2)}{F_{1N}(x,Q^2)}\;,\;\;\;A_{2N}(x,Q^2)  \to 0.$$
In order to calculate the asymmetries $A_{1N, 2N}(x,Q^2)$, the numerical results for the unpolarized structure 
function $F_{1N}(x,Q^2)$ have been taken from our earlier work~\cite{Zaidi:2019mfd}. The numerical results for these asymmetries are discussed below.

\subsubsection{$A_{1p}(x,Q^2)$ and $A_{2p}(x,Q^2)$}
\begin{figure}
 \includegraphics[height=6 cm, width=7 cm]{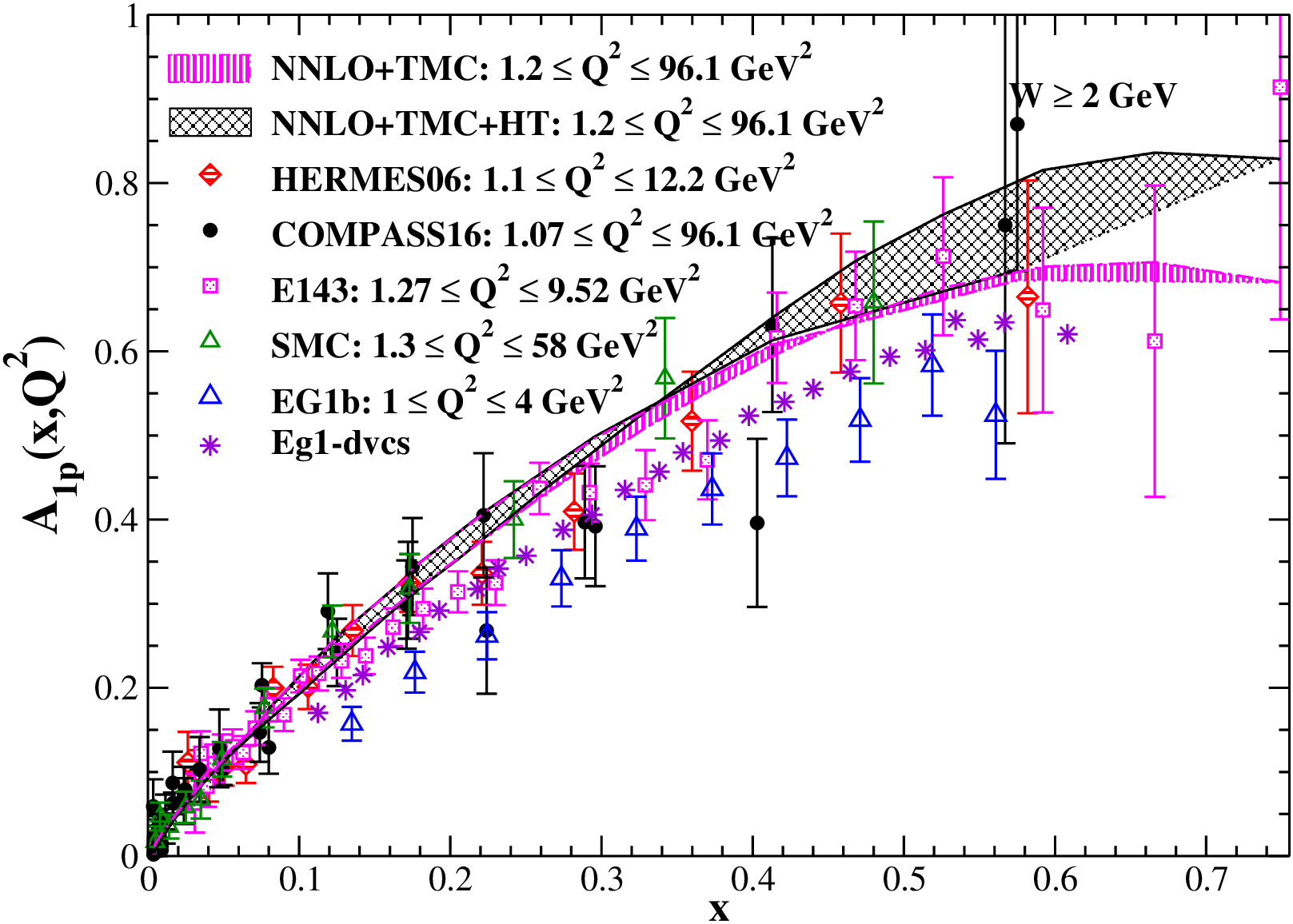} 
  \includegraphics[height=6 cm, width=7 cm]{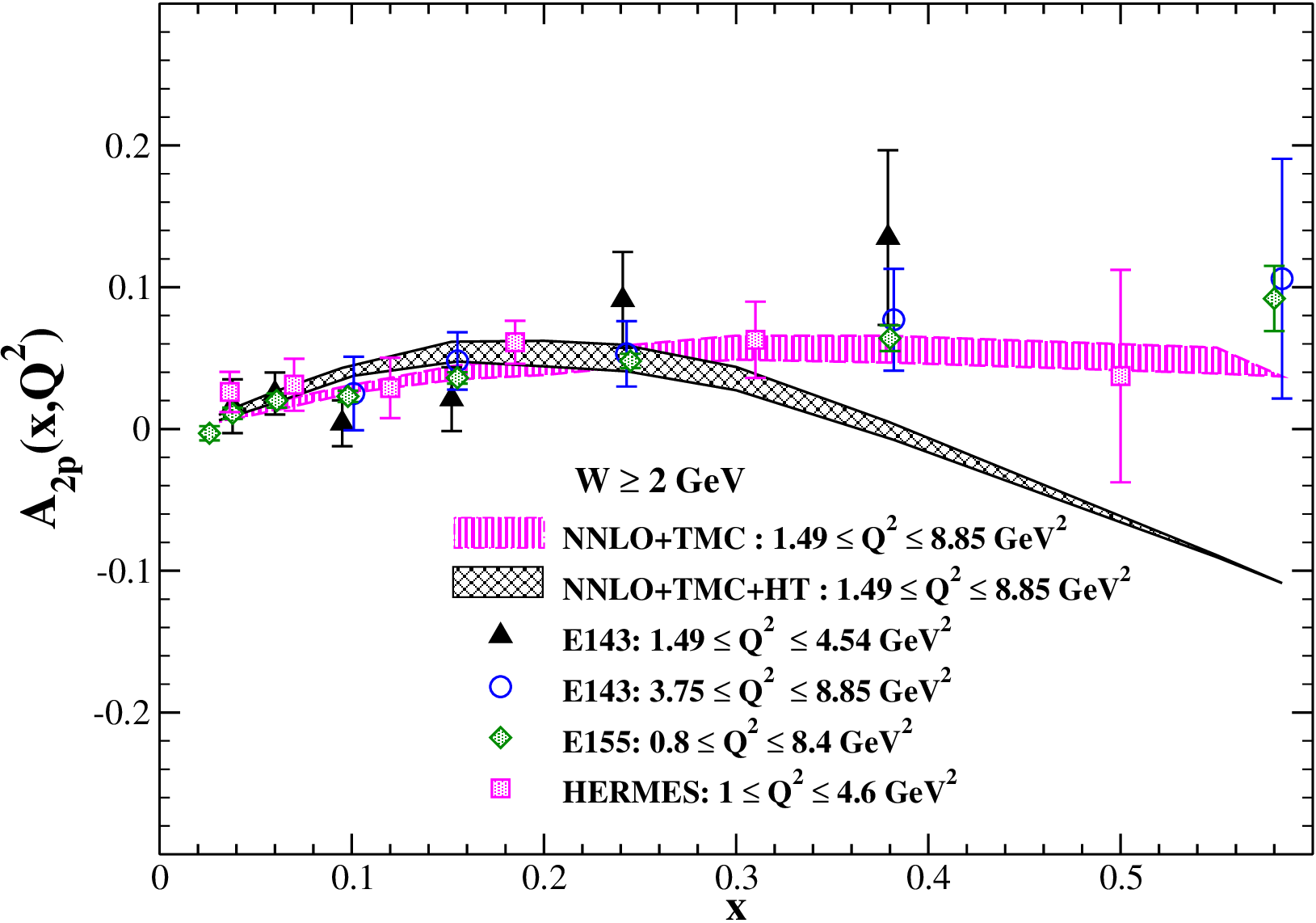}
   \caption{The results for $A_{1p}(x,Q^2)$ (left panel) and $A_{2p}(x,Q^2)$ (right panel) vs $x$ are shown with a kinematic 
   cut of $W\ge 2$ GeV. These results are obtained at NNLO without (vertical lines pattern band) and with (crossed pattern band) the twist-3 corrections 
   corresponding to the available data from E143~\cite{E143:1998hbs}, E155~\cite{E155:2002iec}, SMC~\cite{SpinMuonSMC:1997voo}, 
   HERMES~\cite{HERMES:2006jyl, HERMES:2011xgd}, Eg1-dvcs~\cite{CLAS:2014qtg},
    COMPASS16~\cite{COMPASS:2015mhb} and EG1b~\cite{CLAS:2017qga} experiments
   and mentioned in the legends of the figure. }
   \label{fig7}
\end{figure}

In the literature, various models~\cite{Farrar:1975yb, Close:1988br, Brodsky:1994kg, Leader:1997kw, Isgur:1998yb}
suggest that as $x$ approaches 1, $A_{1p}(x,Q^2)\to1$, implying that the valence quarks carry most of 
the momentum fraction of the target proton, and is polarized in the direction of the proton spin. Therefore, a theoretical understanding of perturbative and nonperturbative QCD corrections in $A_{1p}(x,Q^2)$
becomes important, especially in the high $x$ region, as it paves the way to understand the role of valence quarks and their 
orbital angular momentum contribution to the proton spin. Furthermore, $A_{2p}(x,Q^2)$ is important for obtaining
information about the transverse polarization of quark spins.

In Fig.~\ref{fig7}, the numerical results are presented for the spin asymmetries $A_{1p}(x,Q^2)$ (left panel) and $A_{2p}(x,Q^2)$ (right panel) vs $x$
 using Eqs.~\ref{asymg} and \ref{asymg2} at NNLO, incorporating the TMC and HT effects, and 
applying a kinematical cut of $W\ge 2$ GeV for the different ranges of $Q^2$, as mentioned in the legends of the figures. 
 It can be observed that the asymmetry $A_{1p}(x,Q^2)$ increases with $x$ specially for $x<0.6$
 and then approaches almost a constant value, which implies that in the high $x$ region, quarks carry most of the spin fraction along with most
 of the momentum fraction of the nucleon. We find that higher order 
 QCD corrections (up to NNLO) to the NLO results of $A_{1p}(x,Q^2)$ (not shown here explicitly) 
 are small. Furthermore, we have observed that the effect of twist-3 correction is significant here, for example,
 at $0.1\le x \le 0.3$ twist-3 correction is about 2-3\% for $3\lesssim Q^2\le\lesssim 6$ GeV$^2$ and it increases to 8-16\% for 
 $0.4\le x \le 0.6$ and $6\lesssim Q^2\lesssim 8$ GeV$^2$. However, twist-3 correction becomes small with the increase in $Q^2$. We observe that
 the asymmetry does not reach 1. Whereas, in a naive quark model, valence quarks are expected to be fully polarized and for a fully polarized quark-parton
 model, one expects
 the asymmetry to be close to 1 for $x\simeq 1$. If the asymmetry in the polarized structure function does not reach 1, 
 it is a strong indication that additional contributions, such as sea quarks, gluons, and relativistic effects, are playing a role in 
 modifying the nucleon’s spin structure. This deviation from 1 provides insight into the complexity of nucleon spin dynamics and helps to test the 
 theoretical models.
 We have also compared the numerical results with the experimental data from 
 E143~\cite{E143:1998hbs}, SMC~\cite{SpinMuonSMC:1997voo}, HERMES\cite{HERMES:2006jyl} and COMPASS16~\cite{COMPASS:2015mhb}, 
 finding reasonable agreement with the data in the region of very low $x$.
 For the intermediate range of $x$ ($0.15\le x \le 0.6$), 
 our results overestimate the experimental results from Eg1-dvcs~\cite{CLAS:2014qtg} and Eg1b~\cite{CLAS:2017qga} experiments
 which contains data lying in the region of low $Q^2$, i.e., $1\le Q^2\le 4$ GeV$^2$, and it should be noted that the adequacy of applying DIS 
 in the region of low $Q^2$ has been a subject of considerable discussion in recent literature~\cite{SajjadAthar:2020nvy}.
  More work is needed in this direction, a discussion of
 which is beyond the scope of the present work.
 
  In the right panel, we show the numerical results for $A_{2p}(x,Q^2)$ and its comparison with the experimental 
  data from E143~\cite{E143:1998hbs}, E155~\cite{E155:2002iec} and HERMES~\cite{HERMES:2011xgd}.
  It may be noticed that $A_{2p}(x,Q^2)$ is small in magnitude than $A_{1p}(x,Q^2)$, as it is 
  suppressed by a factor of $\gamma$ (see Eq.~\ref{asymg2}). Consequently, we find that at low $x(<0.1)$, $A_{2p}(x,Q^2)$
  is very small. While comparing the numerical results obtained  
 for the different values of $Q^2$, we observe that $A_{2p}$ decreases with increasing $Q^2$. 
 Here also, the twist-3 correction is found to be significant especially in the high $x$ region for $1.49\le Q^2 \le 8.85$ GeV$^2$.
 The numerical results have been compared with the corresponding 
 experimental data from  E143~\cite{E143:1998hbs}, E155~\cite{E155:2002iec} and HERMES~\cite{HERMES:2011xgd}.
 
 To summarize, we find that the theoretical results for the proton asymmetries $A_{1p}(x,Q^2)$ and $A_{2p}(x,Q^2)$ at NNLO with TMC and HT effects
 are in good agreement with the experimental data.
 
 \subsubsection{$A_{1n}(x,Q^2)$ and $A_{2n}(x,Q^2)$}
 
 \begin{figure}
 \includegraphics[height=9 cm, width=8.5 cm]{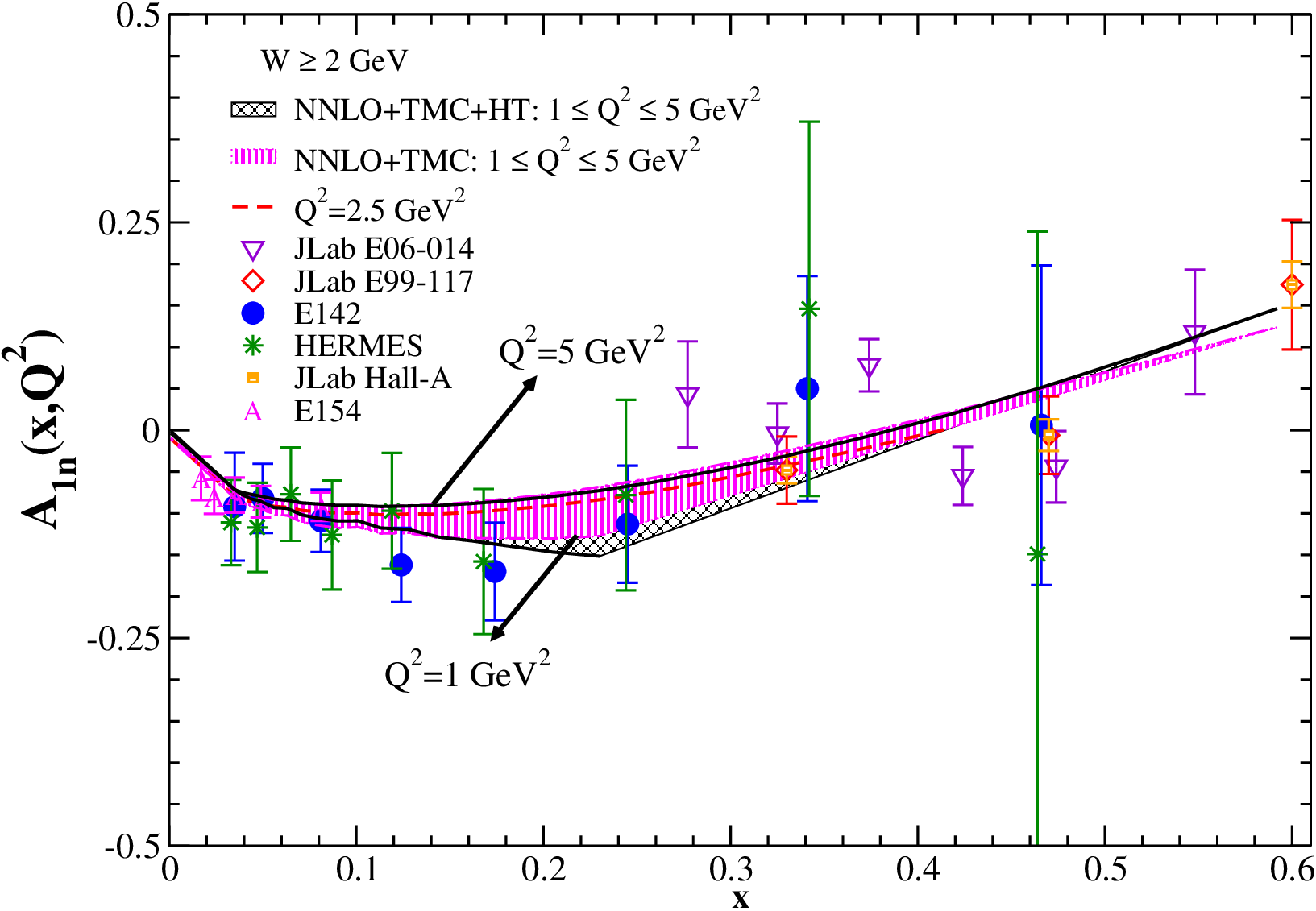}
 \includegraphics[height=9 cm, width=8.5 cm]{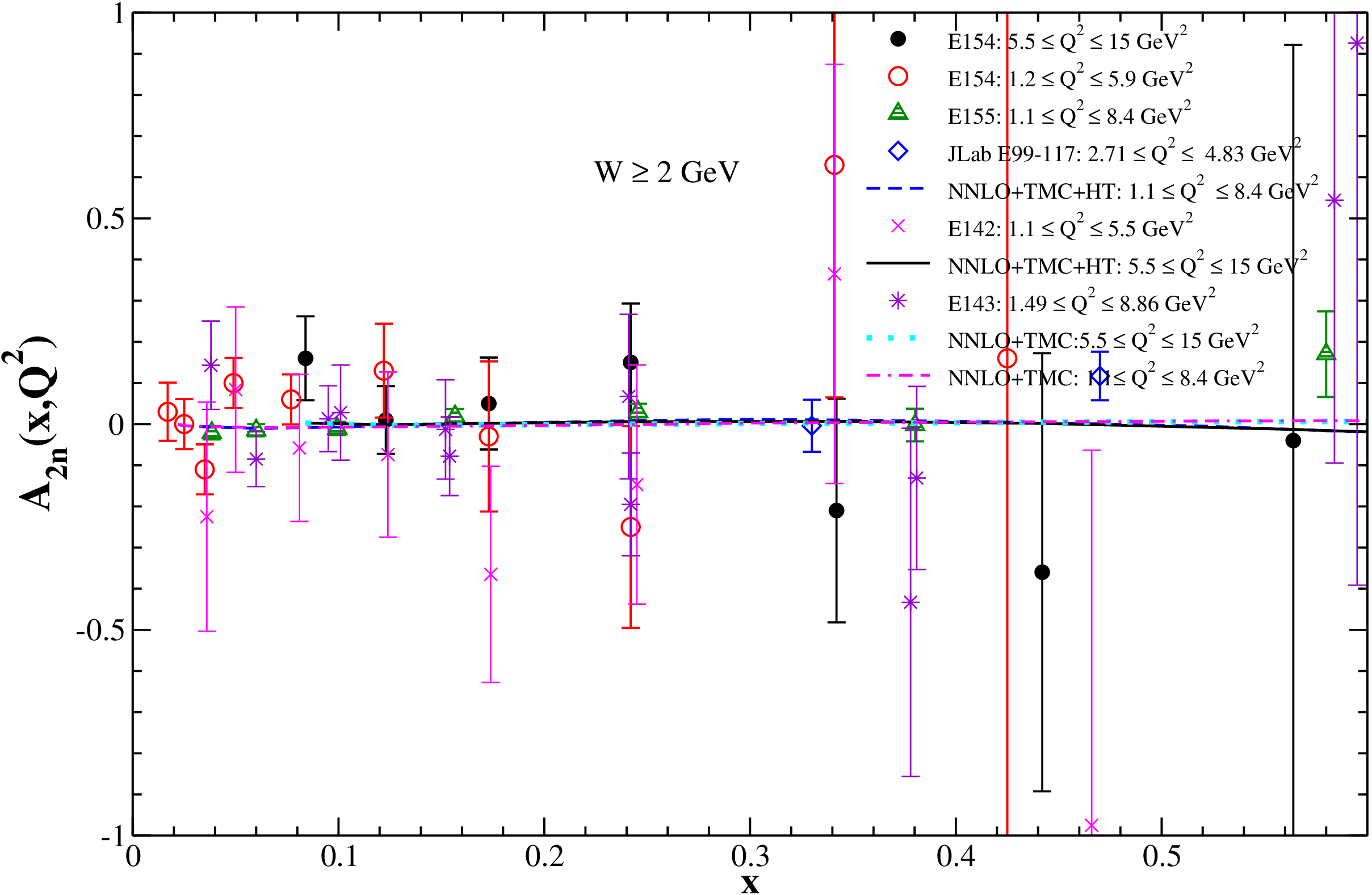}
   \caption{(i) Left panel: The results for $A_{1n}(x,Q^2)$ vs $x$ are shown
  at NNLO with TMC effect only (vertical lines pattern band) and with our full model (cross pattern band).
   The lower curve in the band corresponds to the minimum $Q^2$ value, while the upper curve corresponds to the maximum value of $Q^2$ mentioned in the
legends of the figure. Red dashed line corresponds to the result at NNLO with TMC effect for $Q^2=2.5$ GeV$^2$ (ii) Right panel: $A_{2n}(x,Q^2)$ vs $x$ at NNLO with TMC effect only (double dash-dotted and dotted lines)
  and with both the TMC and HT effects (dash and solid lines).
   These results are obtained by applying a cut of 2 GeV on the center of mass energy $W$ in the different ranges of $Q^2$
   corresponding to the available data from E142~\cite{E142:1996thl}, E143~\cite{E143:1998hbs}, E154~\cite{E154:1997xfa}, E155~\cite{E155:2002iec}, 
   HERMES~\cite{HERMES:1997hjr}, 
   JLab Hall-A~\cite{JeffersonLabHallA:2003joy},
   E99-117~\cite{JeffersonLabHallA:2004tea} and E06-014~\cite{JeffersonLabHallA:2016neg} experiments
   and mentioned in the legends of the figure. }
   \label{fig7b}
\end{figure}

In Fig.~\ref{fig7b}, we present the numerical results for the asymmetries $A_{1n}(x,Q^2)$ and $A_{2n}(x,Q^2)$ for neutrons.
These numerical results are obtained using BDSSV24 PPDFs parameterization~\cite{Borsa:2024mss} at NNLO with the TMC and twist-3 effects by incorporating a 
kinematical cut of $W\ge 2$ GeV. In the left panel of the figure, the results for $A_{1n}(x,Q^2)$ are shown in kinematic region of 
$1\le Q^2\le 5$ GeV$^2$ (shaded bands) and the results corresponding to the central value of $Q^2(=2.5$ GeV$^2$) is shown by dashed line. 
We find that $A_{1n}(x,Q^2)$ is negative and decreases until $x \lesssim 0.4$, then, it gradually increases and becomes positive, 
whereas the spin asymmetry for the proton target, $A_{1p}(x,Q^2)$ has positive contribution in the entire range of Bjorken $x$
due to the dominance of valence up quarks. The effect of twist-3 correction is found to be significant, quantitatively, for $Q^2=1$ GeV$^2$
it is $\sim 7\%$ at $x=0.1$ and $12\%$ at $x=0.2$, while for $Q^2=5$ GeV$^2$ it is $<1\%$ at $x=0.1$, $\sim 5\%$ at $x=0.2$, $14\%$ at $x=0.3$ and 
$18\%$ at $x=0.6$. The numerical results are compared with the data from experiments performed 
at JLab~\cite{JeffersonLabHallA:2003joy, JeffersonLabHallA:2004tea, JeffersonLabHallA:2016neg}, SLAC~\cite{E142:1996thl, E154:1997xfa, E155:2002iec} 
and DESY~\cite{HERMES:1997hjr} and are found to be in reasonable agreement in the entire range of $x$.
 In the right panel
of the figure, the numerical results for $A_{2n}(x,Q^2)$ are shown for the different values of $Q^2$ varied in the ranges $1.1\le Q^2\le 8.4$ GeV$^2$ 
and $5.5\le Q^2 \le 15$ GeV$^2$. We observe that higher order corrections to 
$A_{2n}(x,Q^2)$ are very small. The experimental results from E142~\cite{E142:1996thl}, E143~\cite{E143:1998hbs}, 
E154~\cite{E154:1997xfa}, E155~\cite{E155:2002iec} and E99-117~\cite{JeffersonLabHallA:2004tea} are also shown, but they have very large statistical
 errors and are consistent with zero.

 \begin{figure}
 \includegraphics[height=7 cm, width=7 cm]{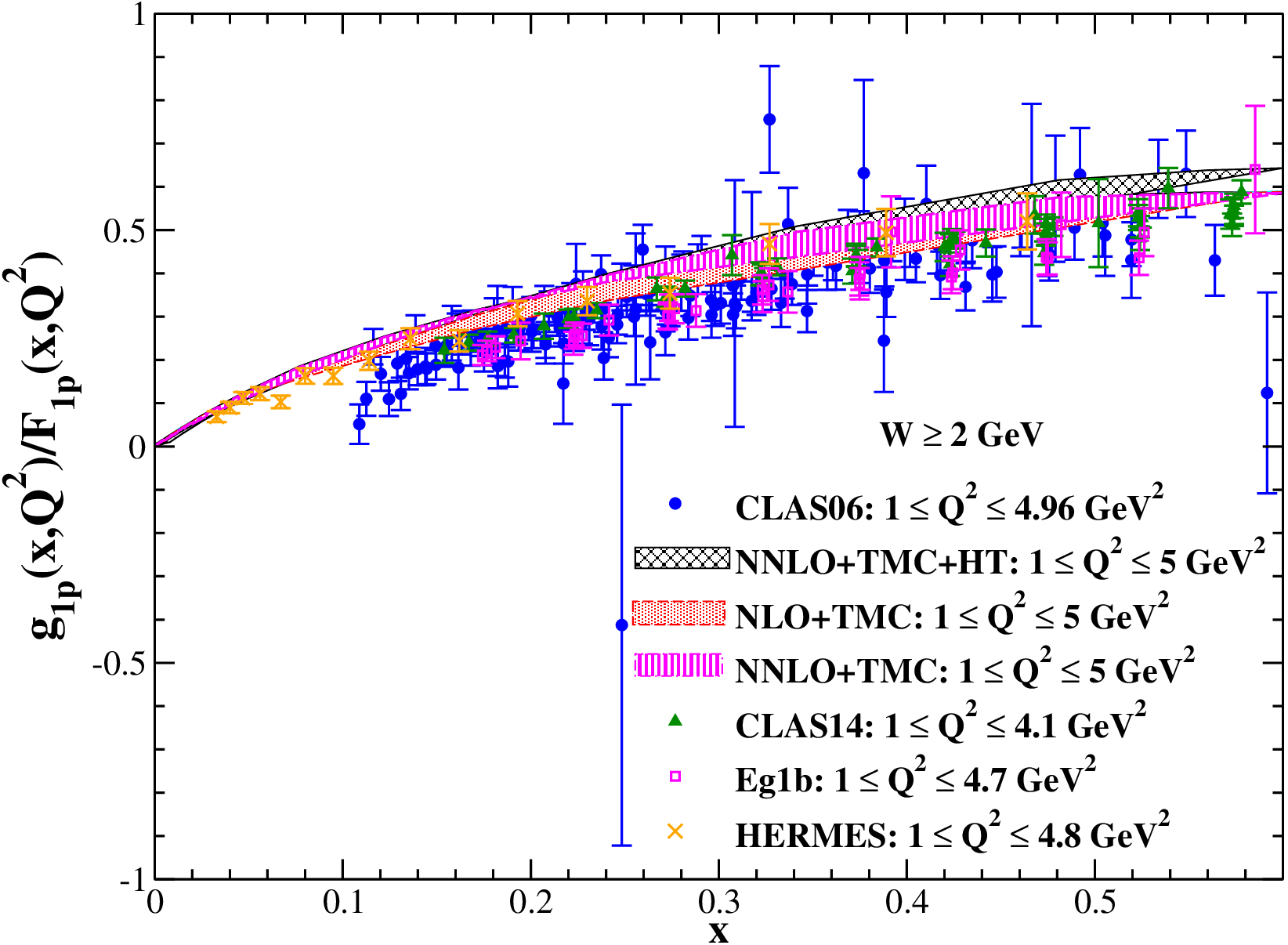}
 \includegraphics[height=7 cm, width=7 cm]{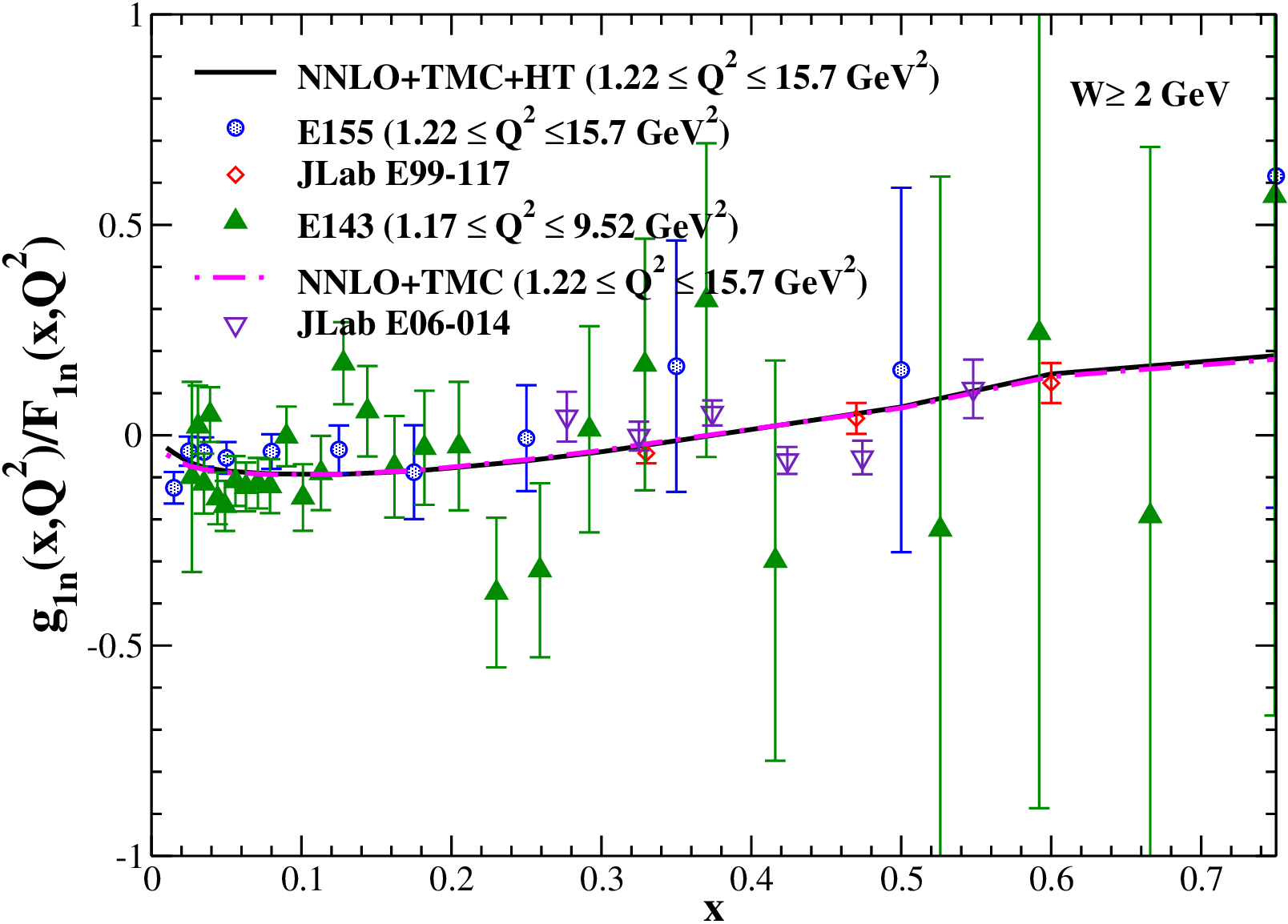}
 \caption{{\bf Left panel:} Results for the ratio $g_{1p}(x,Q^2)/F_{1p}(x,Q^2)$ at NLO (dotted pattern band) and NNLO (vertical lines pattern band) 
 incorporating the TMC effect, and at NNLO with TMC and HT effects (cross pattern band) for $1.0 \le Q^2 \le 5.0$ GeV$^2$. These results are compared with the 
  experimental data of HERMES~\cite{HERMES:1998cbu}, CLAS06~\cite{CLAS:2006ozz}, CLAS14~\cite{CLAS:2014qtg} and EG1b~\cite{CLAS:2017qga} experiments. 
 {\bf Right panel:} Results for the ratio $g_{1n}(x,Q^2)/F_{1n}(x,Q^2)$ vs $x$ at NNLO with the TMC effect only (dash-dotted line) and with the 
 further inclusion of HT effect (solid line) for $1.2 \le Q^2 \le 15.7$ GeV$^2$ corresponding to
 the values of $x$ and $Q^2$ for the experimental data of E143~\cite{E143:1998hbs}, E155~\cite{E155:2000qdr}, E99-117~\cite{JeffersonLabHallA:2004tea}
 and E06-014~\cite{JeffersonLabHallA:2016neg}. The numerical results are evaluated with a cut of $W\ge 2$ GeV.}
  \label{fig8}
\end{figure}

 \subsubsection{The ratios $\frac{g_{1p}(x,Q^2)}{F_{1p}(x,Q^2)}$ and $\frac{g_{1n}(x,Q^2)}{F_{1n}(x,Q^2)}$}
In view of the experimental results available on the ratio $\frac{g_{1N}(x,Q^2)}{F_{1N}(x,Q^2)}\;\;(\textrm{where} ~N=p,n)$ from experiments performed at 
SLAC and JLab~\cite{E143:1998hbs, E155:2000qdr, CLAS:2006ozz, CLAS:2014qtg, JeffersonLabHallA:2004tea, JeffersonLabHallA:2016neg}, we have also
studied the effect of higher order 
corrections to this ratio and compared with experiments. The E143 experiment at SLAC~\cite{E143:1998hbs} indicated that for $Q^2\ge 1$ GeV$^2$ the ratio $\frac{g_{1p}(x,Q^2)}{F_{1p}(x,Q^2)}$
 is almost independent of $Q^2$, and show weak $Q^2$ dependence implying that the higher order perturbative QCD corrections are small. However, 
 the results on the proton asymmetry $A_{1p}(x,Q^2)$ and the ratio $\frac{g_{1p}(x,Q^2)}{F_{1p}(x,Q^2)}$ from 
 the EG1b experiment at JLab~\cite{CLAS:2017qga} show that in the 
 low and moderate regions of $Q^2$ ($0.5\lesssim Q^2 \lesssim 5$ GeV$^2$) these corrections are significant. Therefore, the theoretical understanding of 
 perturbative and nonperturbative QCD corrections in $\frac{g_{1p}(x,Q^2)}{F_{1p}(x,Q^2)}$ become important.

In Fig.~\ref{fig8}, we have presented the results for the ratio of polarized to unpolarized proton and neutron structure functions, i.e.,
$\frac{g_{1p}(x,Q^2)}{F_{1p}(x,Q^2)}$ (left panel) and $\frac{g_{1n}(x,Q^2)}{F_{1n}(x,Q^2)}$ (right panel) respectively, evaluated up to NNLO with the TMC and HT effects. These
results are obtained at different
$Q^2$ corresponding to the kinematic range of E143~\cite{E143:1998hbs}, E155~\cite{E155:2000qdr} and CLAS~\cite{CLAS:2006ozz, CLAS:2014qtg, CLAS:2017qga} experiments.
The numerical results for $\frac{g_{1p}(x,Q^2)}{F_{1p}(x,Q^2)}$
exhibit similar qualitative behavior as that observed in the case of spin asymmetry $A_{1p}(x,Q^2)$ while quantitatively there are some differences.
We find that with the increase in $x$ there is an increase in the value of $\frac{g_{1p}(x,Q^2)}{F_{1p}(x,Q^2)}$ and at high $x~(x>0.5)$, it saturates. 
Moreover, we have observed that the ratio $\frac{g_{1p}(x,Q^2)}{F_{1p}(x,Q^2)}$
has significant difference between the results evaluated at NLO and NNLO for $Q^2=1$ GeV$^2$, for example, the inclusion 
of NNLO terms leads to an enhancement of about 18\% at $x=0.12$, 16\% at $x=0.17$ and $\sim 12\%$ at $x=0.24$.  While this difference becomes small
at $Q^2=5$ GeV$^2$, quantitatively, 1-3\% for $0.1\le x \le 0.6$. Furthermore, twist-3 correction is found to be significant especially in the region of 
high $x$, for example, at $Q^2=1$ GeV$^2$ it is about 2-7\% for $0.1\le x\le 0.25$, however, at $Q^2=5$ GeV$^2$ twist-3 corrections is about
$1\%$ for $0.1\le x\le 0.25$ while it becomes 6\% at $x=0.5$ and 10\% at $x=0.6$.
The band corresponds to the $Q^2$ varied in the range of
$1.0 \le Q^2 \le 5.0$ GeV$^2$ and $W\ge 2$ GeV, are in qualitatively good agreement with the experimental data reported
from CLAS~\cite{CLAS:2006ozz, CLAS:2014qtg}, Eg1b~\cite{CLAS:2017qga} and HERMES~\cite{HERMES:1998cbu}. To conclude, the higher order corrections 
are found to be important in the kinematic region of low to moderate $Q^2$ which is in agreement with the experimental observation 
by Eg1b at JLab~\cite{CLAS:2017qga}.

In the right panel, the 
numerical results for $\frac{g_{1n}(x,Q^2)}{F_{1n}(x,Q^2)}$ are shown at NNLO with TMC and HT effects and compared with the experimental results~\cite{E143:1998hbs, E155:2000qdr, JeffersonLabHallA:2004tea, JeffersonLabHallA:2016neg} in 
the kinematic region of $1.2 \le Q^2 \le 15.7$ GeV$^2$. This ratio increases with the increase in $x$. Here also, the higher order perturbative
corrections are found to be significant (not shown here explicitly). Quantitatively the inclusion of NNLO terms results in a reduction of
about $12\%$ at $x=0.5$ and $40\%$ at $x=0.75$ from the results evaluated at NLO, while for $x<0.5$ the ratio $\frac{g_{1n}(x,Q^2)}{F_{1n}(x,Q^2)}$ at NLO 
almost overlap with the results at NNLO. However, the twist-3 correction is almost negligible for the present kinematic range of $x$ and $Q^2$.
The experimental data of E143~\cite{E143:1998hbs} and E155~\cite{E155:2000qdr} have larger uncertainties in the region of
intermediate and high $x (>0.4)$. Consequently, our numerical results 
are consistent with the experimental data across the entire range of $x$ which have large statistical errors for $x>0.2$.

\subsection{Sum Rule Integrals (SRI)}\label{res_sr}
There are various sum rules viz. Ellis-Jaffe, Bjorken, and Burkhardt-Cottingham, involving the polarized nucleon structure functions $g_{1p,1n}(x,Q^2)$
 and $g_{2p,2n}(x,Q^2)$ integrated over the entire kinematic region of $x$, i.e., $0\le x\le 1$ for a given $Q^2$. 
In experimental situations the lower and upper limits of $x$ in the integrals are not reachable and some extrapolation procedure
is applied to evaluate the integrals between the limits of 0 and 1 in some cases. 
However, in most of the cases, the experimental results for the integral are quoted at the average value of 
the realistic limits of $x_{min}$ and $x_{max}$, i.e., $x_{mid}\Big(=\frac{x_{min}+x_{max}}{2}\Big)$. In Figs.~\ref{fig10}-\ref{fig13}, the numerical results 
for some of the sum rule integrals are presented and compared with the available experimental data.

\subsubsection{Ellis-Jaffe and Bjorken sum rules}
The Ellis-Jaffe sum rule predicts the first moment of the proton and neutron spin structure functions assuming that the gluons do not contribute to 
the nucleon spin. Using the SU(3) symmetry along with the assumption of neglecting the polarizations of sea quarks ($\Delta s=0=\Delta \bar q$),
following sum rule was obtained by Ellis and Jaffe~\cite{Ellis:1973kp} for protons and neutrons in the QPM:
  \begin{equation}
   \left.
 \begin{array}{c}
   \label{eq1}
 S_{EJ}^p=\int_0^1\;dx\;g_{1p} (x)=\frac{1}{12}\;\mid \frac{g_A}{g_V}\mid\;\Big[1+\frac{5}{3}\;\Big(\frac{3F-D}{F+D} \Big) \Big],\\
  S_{EJ}^n=\int_0^1\;dx\;g_{1n} (x)=\frac{1}{12}\;\mid \frac{g_A}{g_V}\mid\;\Big[-1+\frac{5}{3}\;\Big(\frac{3F-D}{F+D} \Big) \Big],
 \end{array}\right\}
 \end{equation}
 where $F$ and $D$ are the $\beta-$decay parameters of the baryon octet decays, and $g_A$ and $g_V$ are the axial and vector coupling constants
of neutron $\beta-$decay. Experiments have shown that this sum rule is violated, indicating that gluon spin contributions are significant in the nucleon.

 Bjorken sum rule relates the difference in the proton and neutron spin-dependent structure functions to the nucleon axial charge($g_A$), measurable
 in neutron beta decay. It takes the difference
of proton and neutron first moments ($\Gamma_{1p}$ and $\Gamma_{1n}$) evaluated
in the limit of very high $Q^2$ by using the techniques of Gell-Mann's current algebra~\cite{Gell-Mann:1962yej, Gell-Mann:1964hhf, Feynman:1964fk} 
and isospin invariance. It relates 
The Bjorken sum rule is given in terms of the integrals $\Gamma_{1p}$ and $\Gamma_{1n}$~\cite{Bjorken:1966jh, Bjorken:1969mm} as:
\begin{equation}\label{eq2}
S_{Bj}=\Gamma_{1p}-\Gamma_{1n}=\int_0^1\;dx\;g_{1p} (x)-\int_0^1\;dx\;g_{1n} (x).
\end{equation}

\begin{figure}
 \includegraphics[height=8 cm, width=14 cm]{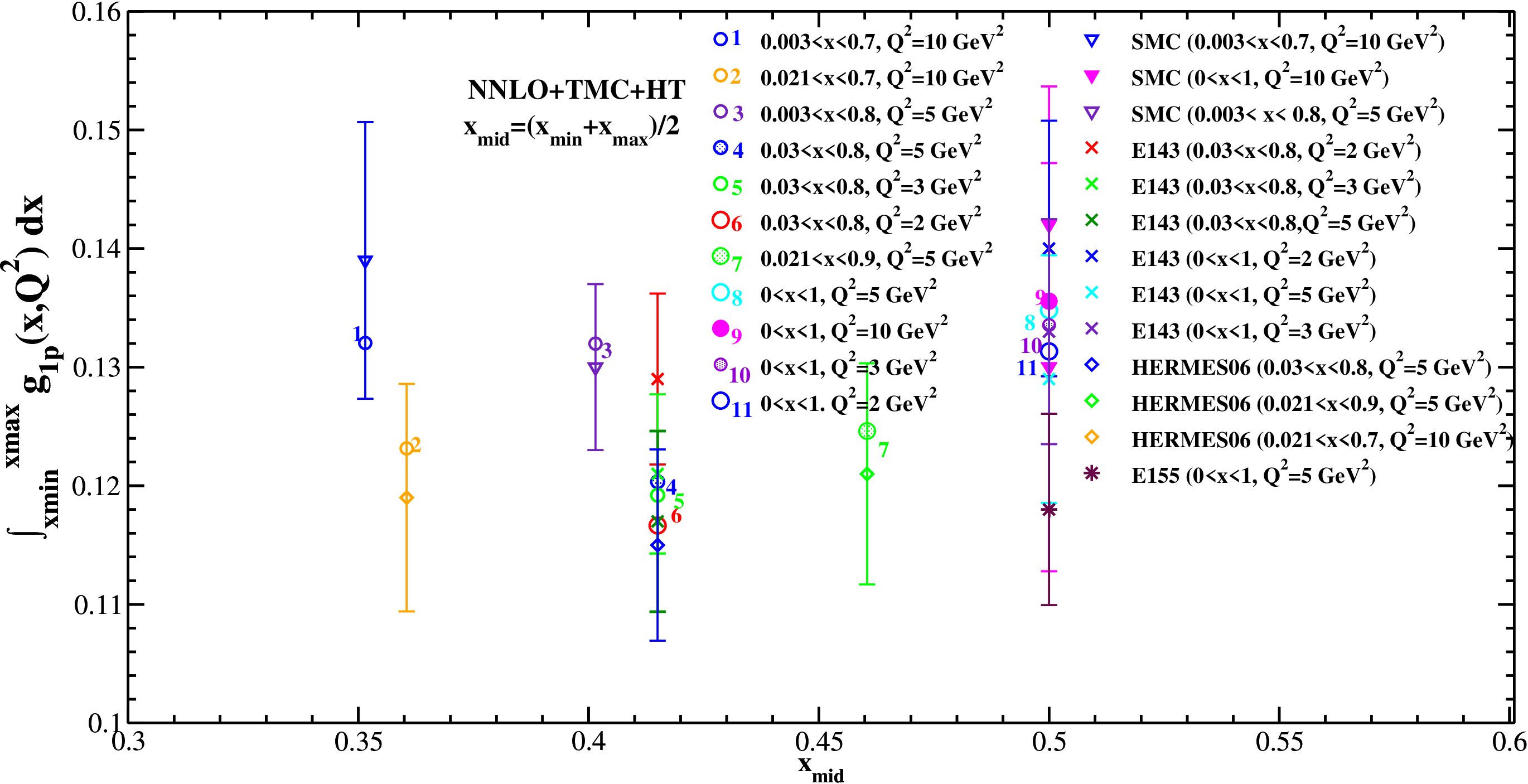}
 \caption{$S_{EJ}^{p}=\int_{x_{min}}^{x_{max}}  g_{1p}(x,Q^2) dx$ 
 vs $x_{mid}\Big(=\frac{x_{min}+x_{max}}{2}\Big)$ with our full model for the different values of $Q^2$. Our results have been shown by circle 
 symbol identified by a number. 
The numerical results are obtained in the integration limits of $0 \le x\le 1$ at $Q^2=2,3,5$ and 10 GeV$^2$, $0.021 \le x\le 0.7$ at $Q^2=10$ GeV$^2$,
$0.021 \le x\le 0.9$ and $0.003\le x \le 0.8$ at $Q^2=5$ GeV$^2$, $0.003\le x \le 0.7$ at $Q^2=10$ GeV$^2$ and $0.03\le x \le 0.8$ at $Q^2=2,3$ and 5 GeV$^2$
  corresponding to the experimental data from E143~\cite{E143:1998hbs} (cross symbol), E155~\cite{E155:2000qdr} (star symbol), SMC~\cite{SpinMuonSMC:1997voo, SpinMuonSMC:1997mkb} (triangle down symbol)
  and HERMES06~\cite{HERMES:2006jyl} (diamond symbol) experiments. The numerical results are obtained 
without applying any cut on $W$.}
    \label{fig10}
\end{figure}

\begin{figure}
  \includegraphics[height=8 cm, width=14 cm]{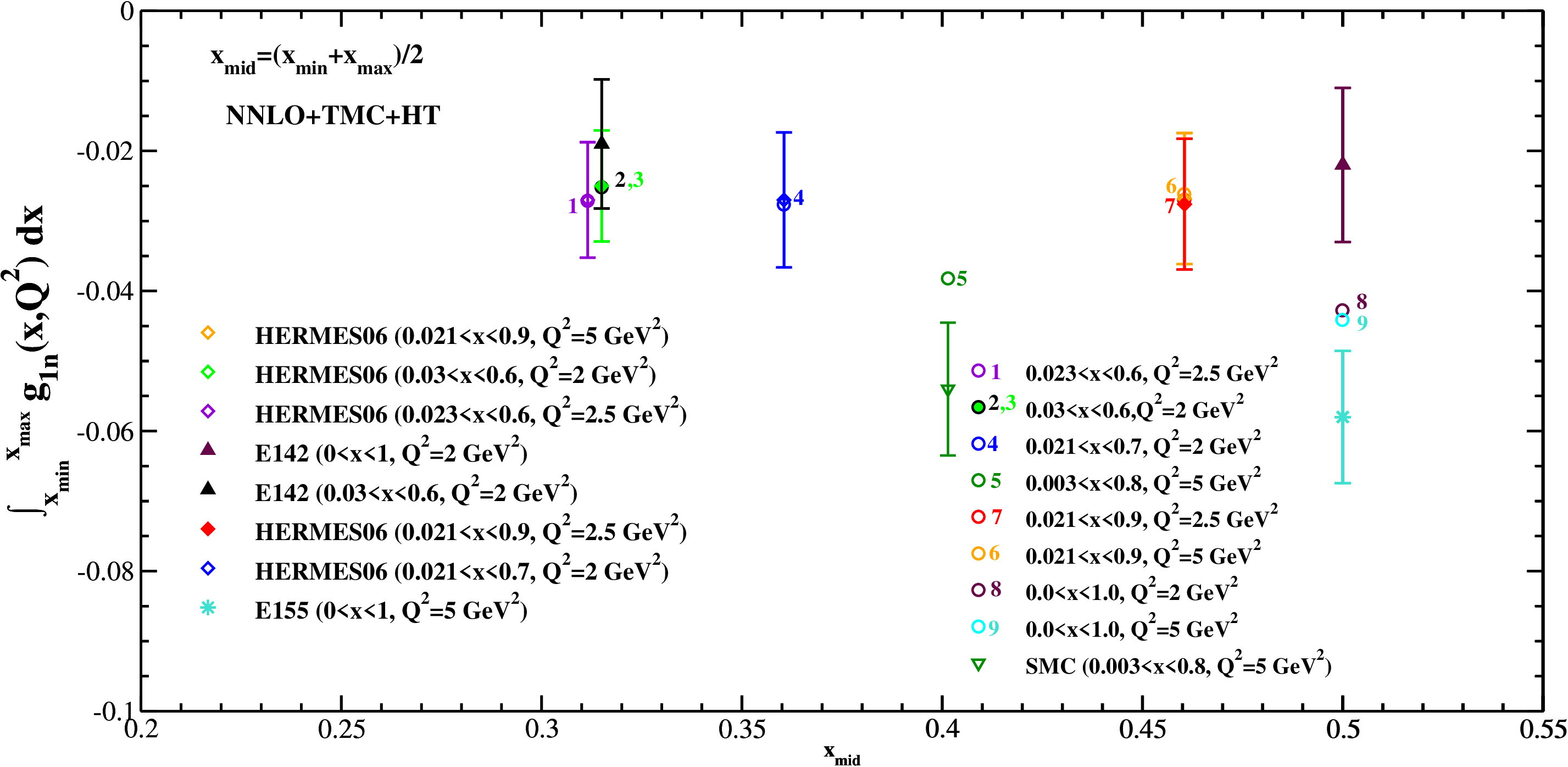}
 \caption{$S_{EJ}^{n}=\int_{x_{min}}^{x_{max}} g_{1n}(x,Q^2) dx$  
 vs $x_{mid}\Big(=\frac{x_{min}+x_{max}}{2}\Big)$ with our full model for the different values of $Q^2$. Our results have been shown by the 
 circle symbol identified by a number. 
The numerical results are obtained in the integration limits of $0.03 \le x\le 0.6$ and $0.021 \le x\le 0.7$ at $Q^2=2$ GeV$^2$, $0.023 \le x\le 0.6$ at $Q^2=2.5$ GeV$^2$ and $0.021 \le x\le 0.9$ at $Q^2=2.5$ GeV$^2$
 as well as at $Q^2=5$ GeV$^2$ corresponding to the experimental data from E142~\cite{E142:1993hql} (triangle up symbol), E155~\cite{E155:2000qdr} (star symbol),
  SMC~\cite{SpinMuon:1998eqa} (triangle down symbol) and HERMES06~\cite{HERMES:2006jyl} (diamond symbol) experiments. The numerical results are obtained 
without applying any cut on $W$.}
    \label{fig10a}
\end{figure}

In Figs. \ref{fig10} and \ref{fig10a}, the results are presented for the integral of the proton and 
neutron polarized structure functions $g_{1p}(x,Q^2)$ and $g_{1n}(x,Q^2)$ 
with our full model, i.e., at NNLO with the TMC effect and twist-3 corrections, in 
the different limits of integration over $x$ as described in the legends of the figures.
 These numerical results are obtained without applying any cut on $W$, and  
are shown by the circle symbol (which has been assigned a number to correspond 
to the kinematical constrain applied by the respective experiment) and have been compared with the experimental 
data from E142~\cite{E142:1993hql}, E143~\cite{E143:1998hbs}, E155~\cite{E155:2000qdr}, 
SMC~\cite{SpinMuonSMC:1997voo, SpinMuon:1998eqa, SpinMuonSMC:1997mkb}, HERMES~\cite{HERMES:2006jyl}, etc.
We find that the contribution of NNLO terms in the sum rule integral 
$S_{EJ}^p$ is small (shown in Fig.~\ref{fig10}), quantitatively, it is about 1\% from 
the NLO results for the kinematic range of $1\le Q^2 \le 5$ GeV$^2$. 
 The twist-3 corrections at NNLO are found to be about 6\% for $Q^2=1$ GeV$^2$ and reduces to 1\% for $2\le Q^2 \le 10$ GeV$^2$.
However, the numerical results for $S_{EJ}^n$ (shown in Fig.~\ref{fig10a}) at NNLO 
show a difference of about 
5\% at $Q^2=1$ GeV$^2$ which reduces to $\sim 1\%$ for $Q^2>1$ GeV$^2$ from the results at NLO. In the case of polarized neutron target, twist-3 corrections 
is almost negligible at NNLO.
It may be noticed that the present results are in good agreement within the uncertainty of experimental data for the polarized proton 
target. For the polarized neutron target, the theoretical results are also consistent with the 
experimental data.

\begin{figure}
  \includegraphics[height=8 cm, width=13 cm]{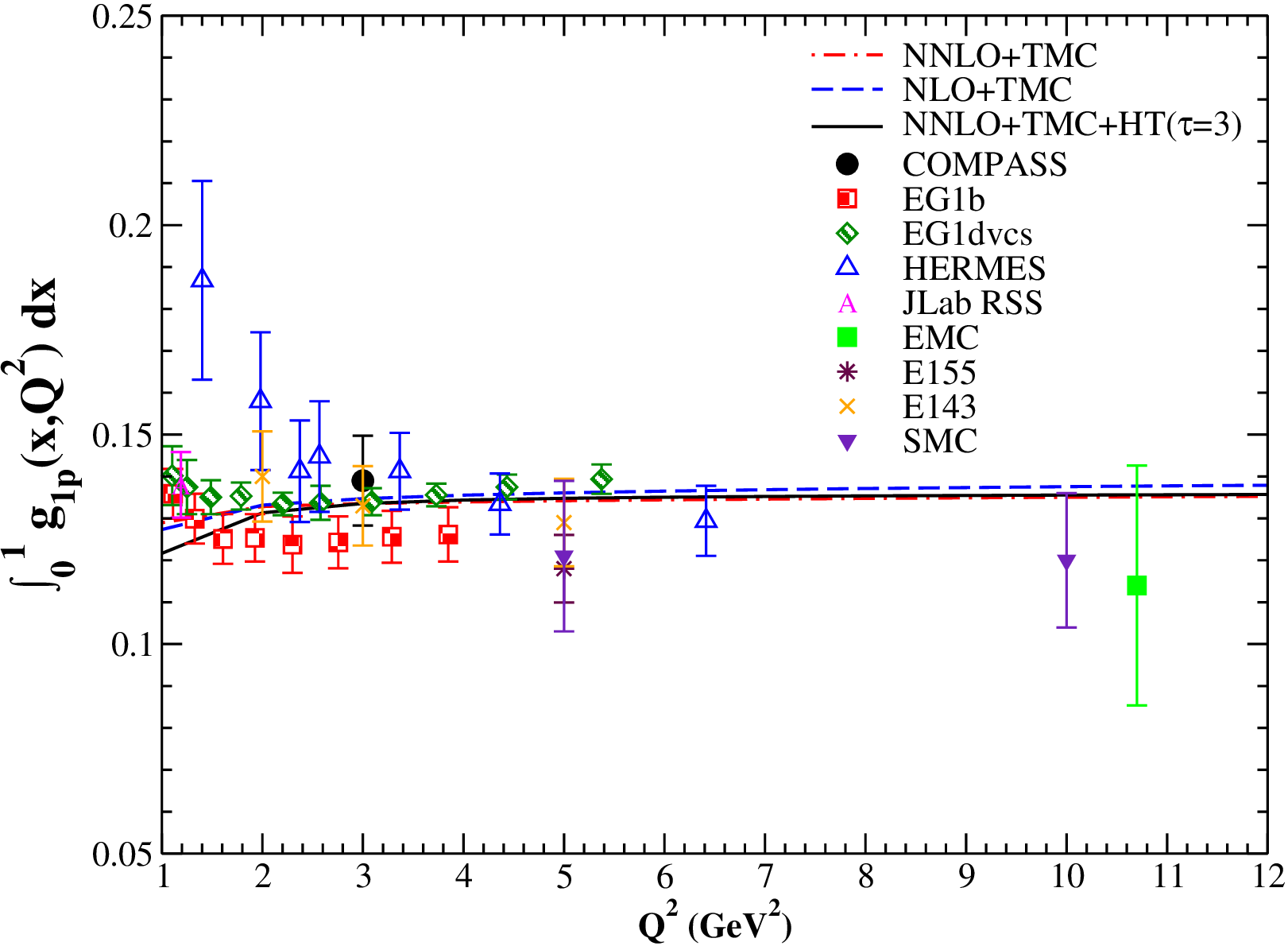}
 \caption{$S_{EJ}^{p}=\int_{0}^{1}  g_{1p}(x,Q^2) dx$ 
 vs $Q^2$ at NLO (dashed line), NNLO with the TMC effect only (dash-dotted line) and NNLO with TMC and twist-3 corrections (solid line). 
The numerical results are compared with the experimental data from EMC~\cite{EuropeanMuon:1987isl} (solid square), 
E143~\cite{E143:1998hbs} (cross symbol), E155~\cite{E155:2000qdr} (star symbol), 
COMPASS~\cite{COMPASS:2015mhb} (solid circle), HERMES~\cite{HERMES:2006jyl} (triangle up), EG1dvcs~\cite{CLAS:2014qtg} (diamond symbol), EG1b~\cite{CLAS:2017qga} (square symbol),
SMC~\cite{SpinMuon:1998zdf} (solid triangle down) and JLab RSS~\cite{RSS:2006tbm} (character symbol) experiments. The numerical results are obtained 
without applying any cut on $W$.}
    \label{fig9a}
\end{figure}

In Fig.~\ref{fig9a}, the numerical results are presented for $\int_{0}^{1}  g_{1p}(x,Q^2) dx$ vs $Q^2$ at NLO and at NNLO with TMC and twist-3 corrections.
These results are obtained 
without applying any constraint on $W$. We find that the higher order perturbative corrections are 
small ($\approx 1\%$) for $Q^2>1$ GeV$^2$, while at $Q^2=1$ GeV$^2$ it is found to be about 6\% at NNLO. We have compared the present
results with the experimental data from SLAC~\cite{E143:1998hbs, E155:2000qdr}, CERN~\cite{EuropeanMuon:1987isl, SpinMuon:1998zdf, COMPASS:2015mhb},
DESY~\cite{HERMES:2006jyl}
and JLab~\cite{RSS:2006tbm, CLAS:2014qtg, CLAS:2017qga} collaborations. The theoretical results underestimate the experimental
data from HERMES~\cite{HERMES:2006jyl} in the region of $1\le Q^2 \le 3$ GeV$^2$.
One may notice that the numerical results with
higher order NNLO corrections are in reasonable agreement with the data from other experiments, especially, from Eg1-dvcs experiment at JLab~\cite{CLAS:2014qtg}
in the region of $1\le Q^2 \le 6$ GeV$^2$.

\begin{figure}
 \includegraphics[height=7 cm, width=14 cm]{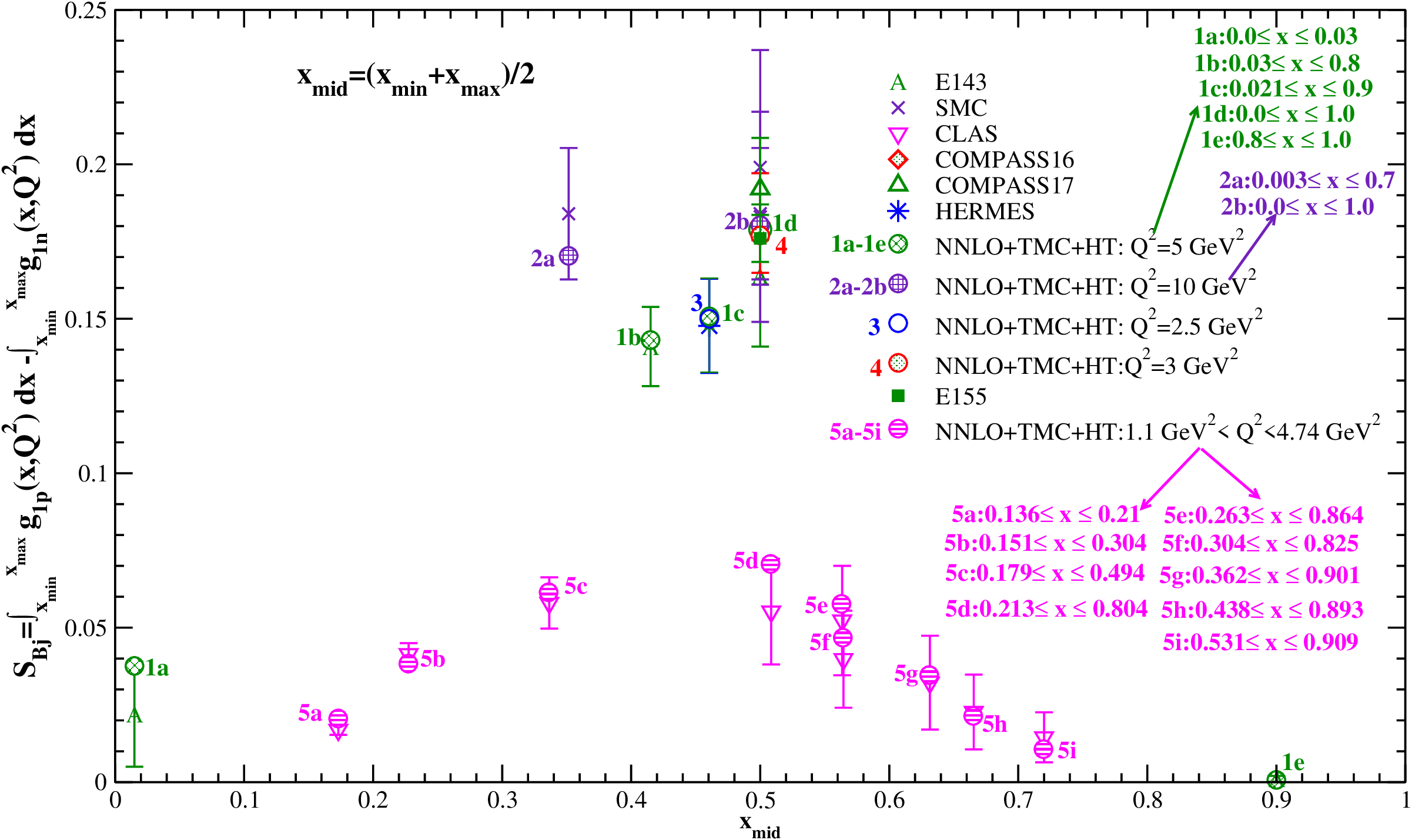}
 \caption{$S_{Bj}=\int_{x_{min}}^{x_{max}} g_{1p}(x,Q^2) dx-\int_{x_{min}}^{x_{max}} g_{1n}(x,Q^2) dx$ vs $x_{mid}\Big(=\frac{x_{min}+x_{max}}{2}\Big)$ 
 with our full model at the different 
 values of $Q^2$ without applying any cut on $W$. Numerical results are shown by the circle symbol identified by a number. The integration limits (as mentioned in the legends of the figure) are chosen corresponding to the 
 experimental data available in the literature. Numerical results are compared with the data reported from 
 E143~\cite{E143:1998hbs}, E155~\cite{E155:2000qdr}, SMC~\cite{SpinMuonSMC:1997mkb, SpinMuon:1998eqa}, 
 COMPASS16~\cite{COMPASS:2015mhb}, COMPASS17~\cite{COMPASS:2016jwv}, HERMES~\cite{HERMES:2006jyl} 
 and CLAS~\cite{Deur:2014vea} experiments.}
  \label{fig11}
\end{figure}

In Fig.~\ref{fig11}, the numerical results are presented for $\int_{x_{min}}^{x_{max}} g_{1p}(x,Q^2) dx-\int_{x_{min}}^{x_{max}} g_{1n}(x,Q^2) dx$ vs $x_{mid}$ (Eq.~\ref{eq2}) 
which have been evaluated using full model. The inclusion of NNLO terms leads to a small difference in the results
obtained at NLO (not shown here explicitly) which is about 1\% for $1\le Q^2\le 5$ GeV$^2$
in the integration limits of $x$ between $0$ to $1$. However, the effect of twist-3 corrections at NNLO is found to be
about 4\% at $Q^2=1$ GeV$^2$ while it is $\lesssim 1\%$ for $2\le Q^2\le 10$ GeV$^2$
in the integration limits of $x$ between 0 and 1. Furthermore, it may also be observed
that these numerical results at NNLO are in reasonable agreement with the corresponding experimental data 
of E143~\cite{E143:1998hbs}, E155~\cite{E155:2000qdr}, SMC~\cite{SpinMuonSMC:1997mkb, SpinMuon:1998eqa}, HERMES~\cite{HERMES:2006jyl},
COMPASS16~\cite{COMPASS:2015mhb}, COMPASS17~\cite{COMPASS:2016jwv} and CLAS~\cite{Deur:2014vea}.
Quantitatively, the numerical value of $S_{Bj}$ at NNLO is $0.179$ in the integration limit of $x$ between 0 and 1, 
and it is consistent with the value of Bjorken sum 
rule obtained in Ref.~\cite{Kuhn:2008sy}, which is $0.182\pm0.002$ at $\langle Q^2 \rangle=5$ GeV$^2$.

\begin{figure}
 \includegraphics[height=7 cm, width=10 cm]{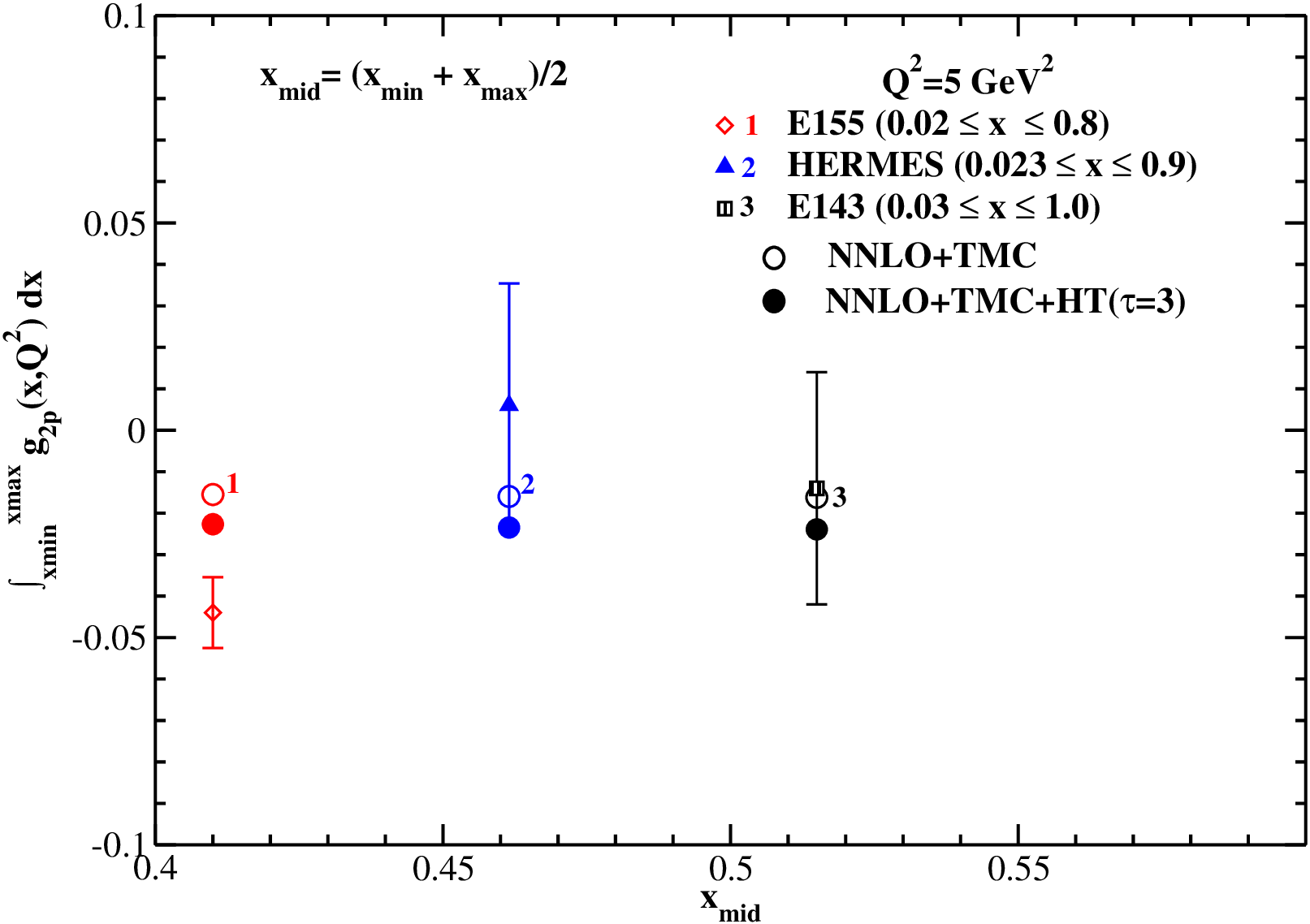}
 \caption{$\int_{x_{min}}^{x_{max}} g_{2p}(x,Q^2) dx$ vs $x_{mid}\Big(=\frac{x_{min}+x_{max}}{2}\Big)$ at NNLO with TMC effect (empty circles)
 as well as with the twist-3 correction (solid circles) at $Q^2=5$ GeV$^2$ without applying any cut on $W$. Numerical results are shown 
 by the circle symbol identified by a number and compared with 
 E143~\cite{E143:1998hbs}, E155~\cite{E155:2002iec} and HERMES~\cite{HERMES:2011xgd} experimental data.}
  \label{fig13}
\end{figure}

\subsubsection{Burkhardt-Cottingham sum rule} Burkhardt-Cottingham (BC) sum rule states that the integral of the transverse spin
structure function $g_2(x)$ over all $x$ must
vanish. It is derived for the polarized spin structure function $g_{2N}(x)$
given by~\cite{Burkhardt:1970ti}:
\begin{equation}\label{eq4}
 \int_0^1\;g_{2N}(x)\;dx=0\;;\hspace{4 mm} N=p,n
\end{equation}
from the virtual Compton scattering dispersion relations in the limit of high $Q^2$. 
This sum rule is obtained by considering the Wandzura-Wilczek relation~\cite{Wandzura:1977qf} incorporating the twist-2 contribution (Eq.~\ref{wwrel})
as well as using Eq.~\ref{g2ht} which further incorporates the twist-3 contribution.

In Fig.~\ref{fig13}, the results are presented for the BC sum rule at a fixed $Q^2$, specifically at $Q^2=5$ GeV$^2$,
without (empty circles) and with (solid circles) the twist-3 correction. 
We find that at NNLO this sum rule integral is non-zero though small in magnitude and the inclusion of TMC effect leads to a difference of 
1-2\% while the further inclusion of twist-3 correction results in a suppression of about 46-47\%
from the results evaluated with TMC effects only in the present kinematic range of $x$ and $Q^2$. Our numerical results 
are consistent with the experimental data from E143~\cite{E143:1998hbs} and HERMES~\cite{HERMES:2011xgd}, 
although they overestimate the experimental findings of E155~\cite{E155:2002iec} in their kinematic range of $x_{min}$ and $x_{max}$.

\section{Conclusions}\label{summary}

To conclude, we have studied the effects of perturbative and nonperturbative QCD corrections to the QPM in the evaluation of polarized 
nucleon structure functions. The higher order perturbative corrections are calculated at the NLO and NNLO using BDSSV24 PPDFs in the 3-flavor $\overline{\textrm{MS}}$ scheme. 
The nonperturbative effects of target mass correction and the twist-3 correction are also incorporated.
The results are presented for the polarized proton and neutron structure functions, $g_{1p,1n}(x,Q^2)$, $g_{2p,2n}(x,Q^2)$, the nucleon asymmetries 
$A_{1p,1n}(x,Q^2)$, $A_{2p,2n}(x,Q^2)$, along with the Ellis-Jaffe, Bjorken and Burkhardt-Cottingham sum rule integrals and
compared these theoretical results with the experimental data available in the literature. We conclude that:
\begin{itemize}
\item In the evaluation of polarized nucleon structure functions at NLO significant model dependence arises
due to the choice of polarized PDFs in literature, for example, GRSV01~\cite{Gluck:2000dy}, LSS05~\cite{Leader:2005ci}, BB10~\cite{Blumlein:2010rn}, 
KATAO10~\cite{Khorramian:2010qa} and BDSSV24~\cite{Borsa:2024mss}.
 This, in turn, leads to 
uncertainties in the evaluation of nucleon asymmetries $A_{1N,2N}(x,Q^2)$ and the sum rule integrals. 
However, these PPDFs parameterizations have uncertainty bands and there is some overlap among the 
different PPDFs parameterizations in certain kinematic regions of $x$ and $Q^2$ 
when errors are taken into account along with the central values.

 \item The nucleon structure functions and other observables evaluated at NLO and NNLO using BDSSV24 parameterization 
 for the polarized PDFs show very small difference with each other suggesting 
 perturbative stability except at very small $x< .04$ and low $Q^2 < 2$ GeV$^2$. 
 The evidence of the perturbation stability demonstrated with the use of BDSSV24 parameterizations of the polarized PDFs 
exists even in the presence of the nonperturbative effects which are quite significant except in the region of very low $x$, i.e., $x\lesssim 0.04$.
 The theoretical values for these observables evaluated at NLO and NNLO both show reasonable agreement with the experimental values. 

 
 \item The inclusion of nonperturbative twist-3 correction in the evaluation of the polarized nucleon structure
 functions $g_{1p,1n}(x,Q^2)$ does not result in any significant changes for $Q^2\ge 5$ GeV$^2$, but are found to be $\sim 5\%$ in the lower $Q^2$ region
 for mid and high $x (\ge 0.3)$. However, in $g_{2p,2n}(x,Q^2)$ it 
 provides significant contribution in the considered range of $x$ and $Q^2$, i.e., $0\le x \le 0.8$ and $Q^2\ge 1$ GeV$^2$.

 \item We find that the nucleon asymmetries $A_{1p}(x,Q^2)$, $A_{2p}(x,Q^2)$ and the ratios 
$\frac{g_{1p,1n}(x,Q^2)}{F_{1p,1n}(x,Q^2)}$ increase with the increase in $x$. The inclusion of higher order perturbative and nonperturbative corrections 
bring the numerical results into better agreement with the experimental data. However, we find that the asymmetry $A_{1p}(x,Q^2)$ does not 
reach 1, i.e., $A_{1p}(x,Q^2) \ne 1$ which may be due to the additional contribution from the sea quarks, gluons and the relativistic effects.
 
\item The theoretical results for the Ellis-Jaffe, Bjorken, and Ellis-Jaffe, Bjorken, and Burkhardt-Cottingham sum rules integrals
are qualitatively in agreement with the experimental data, though the experimental
error bars are large 
and more precise data are needed especially in the region of very low $x$ as well as at high $x$ in order to avoid the
uncertainties arising due to the extrapolation
of polarized parton densities in the evaluation of the sum rule integrals. The numerical results for the different sum rule integrals may provide important
information about the helicities of quarks, flavor asymmetry of sea quarks, axial and strong coupling constants in the presence of higher order corrections.
The future experiments like 
Electron-Ion Collider at BNL~\cite{Nadel-Turonski:2025sfv}, EIC in China~\cite{Anderle:2021wcy}, JLab 22GeV upgrade~\cite{Cotton:2024rpn}, etc.,
may shed more light on the validity of these rules.
 
 
 \item The perturbative evolution of the polarized parton densities at higher orders in the perturbative expansion plays important role 
in the kinematic region of Bjorken $x$ and $Q^2$ considered in the present work while moving from LO to NLO. However, the inclusion of NNLO terms
 does not make any significant changes in the evaluation of the polarized nucleon structure functions except
for low $x\lesssim 0.04$ and $1\le Q^2\le 60$ GeV$^2$, and for high $x~(\gtrsim 0.7)$ and $1\le Q^2\lesssim 5$ GeV$^2$. Moreover, while evaluating
 the ratios of polarized to unpolarized nucleon structure functions, these 
higher order corrections are found to be important at low and moderate values of $Q^2$.
We find that the inclusion of higher order perturbative corrections at NLO leads to better agreement with the experimental data
 in the case of proton target, while in the case of neutron target, the statistical errors on various observables
 are too large to make any definite conclusions.

 \item The PPDFs of BDSSV24 at NNLO~\cite{Borsa:2024mss} used in the determination of polarized nucleon structure functions and other observables
has been successful in explaining the experimental results on these observables on nucleon targets. 
Furthermore, the benchmarking of the BDSSV24 parameterization of polarized PDFs demonstrated in this work may be 
useful in future studies of various observables in the DIS of polarized charged leptons and (anti)neutrinos from polarized nucleons and nuclei.
\end{itemize}

The present work may be helpful in understanding the role of higher order corrections in explaining the 
upcoming experimental results from the JLab, EIC, EIcC and CERN collaborations 
in the relevant kinematic range of $x$ and $Q^2$.
Since future experiments at JLab and EIC at BNL are planned to use the light to heavy nuclear targets such as $^{2}D$, $^{3}He$, $^{12}C$, $^{14}N$, $^{16}O$,
$^{56}Fe$, etc., therefore, the understanding of nuclear medium effects on the polarized structure functions is important.
We plan to study these nuclear medium effects in different nuclear targets by taking into
account the nuclear effects due to the Fermi motion and binding energy of the nucleons, nucleon-nucleon correlations, shadowing/antishadowing and mesonic 
cloud contributions in a wide kinematic range of $x$ and $Q^2$ relevant for the future experiments. 
 Furthermore, the present work can be useful in theoretical understanding of the various observables
 in the scattering of (anti)neutrinos from the polarized nucleon and nuclear targets planned to be studied
 in future experiments at the proposed neutrino factories.

\section*{Acknowledgment}
We are thankful to I. Borsa for providing us the PPDFs grids of BDSSV24 parameterization on request.
F. Zaidi is thankful to Council of Scientific \& Industrial Research, Govt. of India for providing Senior Research Associateship (SRA) under the Scientist's Pool Scheme, file no. 13(9240-A)2023-POOL and to 
the Department of Physics, Aligarh Muslim University, Aligarh for providing the necessary facilities to pursue this research work.
M. S. A. is thankful to the Department of Science and Technology (DST), Government of India for providing 
financial assistance under Grant No. SR/MF/PS-01/2016-AMU/G.


\begin{thebibliography}{100}
\bibitem{EuropeanMuon:1987isl}
J.~Ashman \textit{et al.} [European Muon],
Phys. Lett. B \textbf{206}, 364 (1988).

\bibitem{Feynman:1969ej}
R.~P.~Feynman,
Phys. Rev. Lett. \textbf{23}, 1415-1417 (1969)

\bibitem{Feynman:1973xc}
R.~P.~Feynman,
``Photon-hadron interactions.''

\bibitem{E142:1993hql}
P.~L.~Anthony \textit{et al.} [E142],
Phys. Rev. Lett. \textbf{71}, 959-962 (1993).

\bibitem{E142:1996thl}
P.~L.~Anthony \textit{et al.} [E142],
Phys. Rev. D \textbf{54}, 6620-6650 (1996).

\bibitem{E143:1998hbs}
K.~Abe \textit{et al.} [E143],
Phys. Rev. D \textbf{58}, 112003 (1998).

\bibitem{E154:1997xfa}
K.~Abe \textit{et al.} [E154],
Phys. Rev. Lett. \textbf{79}, 26-30 (1997).

\bibitem{E154:1997eyc}
K.~Abe \textit{et al.} [E154],
Phys. Lett. B \textbf{404}, 377-382 (1997).

\bibitem{E155:1999eug}
P.~L.~Anthony \textit{et al.} [E155],
Phys. Lett. B \textbf{458}, 529-535 (1999).

\bibitem{E155:2000qdr}
P.~L.~Anthony \textit{et al.} [E155],
Phys. Lett. B \textbf{493}, 19-28 (2000).


\bibitem{E155:2002iec}
P.~L.~Anthony \textit{et al.} [E155],
Phys. Lett. B \textbf{553}, 18-24 (2003).
 
 

\bibitem{SpinMuonSMC:1997voo}
B.~Adeva \textit{et al.} [Spin Muon (SMC)],
Phys. Lett. B \textbf{412}, 414-424 (1997).

\bibitem{SpinMuonSMC:1997mkb}
D.~Adams \textit{et al.} [Spin Muon (SMC)],
Phys. Rev. D \textbf{56}, 5330-5358 (1997).
   
    

\bibitem{SpinMuon:1998zdf}
B.~Adeva \textit{et al.} [Spin Muon],
Phys. Rev. D \textbf{58}, 112002 (1998).

\bibitem{SpinMuon:1998eqa}
B.~Adeva \textit{et al.} [Spin Muon],
Phys. Rev. D \textbf{58}, 112001 (1998).

\bibitem{COMPASS:2010wkz}
M.~G.~Alekseev \textit{et al.} [COMPASS],
Phys. Lett. B \textbf{690}, 466-472 (2010).

\bibitem{COMPASS:2015mhb}
C.~Adolph \textit{et al.} [COMPASS],
Phys. Lett. B \textbf{753}, 18-28 (2016).

\bibitem{COMPASS:2016jwv}
C.~Adolph \textit{et al.} [COMPASS],
Phys. Lett. B \textbf{769}, 34-41 (2017).

\bibitem{HERMES:1997hjr}
K.~Ackerstaff \textit{et al.} [HERMES],
Phys. Lett. B \textbf{404}, 383-389 (1997).

\bibitem{HERMES:1998cbu}
A.~Airapetian \textit{et al.} [HERMES],
Phys. Lett. B \textbf{442}, 484-492 (1998).

\bibitem{HERMES:2006jyl}
A.~Airapetian \textit{et al.} [HERMES],
Phys. Rev. D \textbf{75}, 012007 (2007).

\bibitem{HERMES:2011xgd}
A.~Airapetian \textit{et al.} [HERMES],
Eur. Phys. J. C \textbf{72}, 1921 (2012).
 
 

\bibitem{CLAS:2003rjt}
R.~Fatemi \textit{et al.} [CLAS],
Phys. Rev. Lett. \textbf{91}, 222002 (2003).

\bibitem{JeffersonLabHallA:2003joy}
X.~Zheng \textit{et al.} [Jefferson Lab Hall A],
Phys. Rev. Lett. \textbf{92}, 012004 (2004).

\bibitem{JeffersonLabHallA:2004tea}
X.~Zheng \textit{et al.} [Jefferson Lab Hall A],
Phys. Rev. C \textbf{70}, 065207 (2004).

\bibitem{Kramer:2005qe}
K.~Kramer, D.~S.~Armstrong, T.~D.~Averett, W.~Bertozzi, S.~Binet, C.~Butuceanu, A.~Camsonne, G.~D.~Cates, J.~P.~Chen and S.~Choi, \textit{et al.}
Phys. Rev. Lett. \textbf{95}, 142002 (2005).

\bibitem{CLAS:2006ozz}
K.~V.~Dharmawardane \textit{et al.} [CLAS],
Phys. Lett. B \textbf{641}, 11-17 (2006).

\bibitem{RSS:2006tbm}
F.~R.~Wesselmann \textit{et al.} [RSS],
Phys. Rev. Lett. \textbf{98}, 132003 (2007).

\bibitem{CLAS:2014qtg}
Y.~Prok \textit{et al.} [CLAS],
Phys. Rev. C \textbf{90}, no.2, 025212 (2014).

\bibitem{Deur:2014vea}
A.~Deur, Y.~Prok, V.~Burkert, D.~Crabb, F.~X.~Girod, K.~A.~Griffioen, N.~Guler, S.~E.~Kuhn and N.~Kvaltine,
Phys. Rev. D \textbf{90}, no.1, 012009 (2014).

\bibitem{JeffersonLabHallA:2016neg}
D.~Flay \textit{et al.} [Jefferson Lab Hall A],
Phys. Rev. D \textbf{94}, no.5, 052003 (2016).

\bibitem{CLAS:2017qga}
R.~Fersch \textit{et al.} [CLAS],
Phys. Rev. C \textbf{96}, no.6, 065208 (2017).

\bibitem{Aschenauer:2013iia}
E.~C.~Aschenauer, T.~Burton, T.~Martini, H.~Spiesberger and M.~Stratmann,
Phys. Rev. D \textbf{88}, 114025 (2013).

\bibitem{Delahaye:2018yfq}
J.~P.~Delahaye, C.~M.~Ankenbrandt, S.~A.~Bogacz, P.~Huber, H.~G.~Kirk, D.~Neuffer, M.~A.~Palmer, R.~Ryne and P.~V.~Snopok,
JINST \textbf{13}, T06003 (2018).

\bibitem{Bogacz:2022xsj}
A.~Bogacz, V.~Brdar, A.~Bross, A.~de Gouv\^ea, J.~P.~Delahaye, P.~Huber, M.~Hostert, K.~J.~Kelly, K.~Long and M.~Palmer, \textit{et al.}
[arXiv:2203.08094 [hep-ph]].

\bibitem{Borsa:2022irn}
I.~Borsa, D.~de Florian and I.~Pedron,
Eur. Phys. J. C \textbf{82}, no.12, 1167 (2022).

\bibitem{Riedl:2022pad}
C.~Riedl,
Acta Phys. Polon. B \textbf{53}, 5-A2 (2022).

\bibitem{Altarelli:1988nr}
G.~Altarelli and G.~G.~Ross,
Phys. Lett. B \textbf{212}, 391-396 (1988).

\bibitem{Anselmino:1994gn}
M.~Anselmino, A.~Efremov and E.~Leader,
Phys. Rept. \textbf{261}, 1-124 (1995)
[erratum: Phys. Rept. \textbf{281}, 399-400 (1997)].

\bibitem{Ball:1995td}
R.~D.~Ball, S.~Forte and G.~Ridolfi,
Phys. Lett. B \textbf{378}, 255-266 (1996).

\bibitem{Altarelli:1998nb}
G.~Altarelli, R.~D.~Ball, S.~Forte and G.~Ridolfi,
Acta Phys. Polon. B \textbf{29}, 1145-1173 (1998)

\bibitem{Lampe:1998eu}
B.~Lampe and E.~Reya,
Phys. Rept. \textbf{332}, 1-163 (2000).

\bibitem{Hughes:1999wr}
E.~W.~Hughes and R.~Voss,
Ann. Rev. Nucl. Part. Sci. \textbf{49}, 303-339 (1999).

\bibitem{Bass:2004xa}
S.~D.~Bass,
Rev. Mod. Phys. \textbf{77}, 1257-1302 (2005).

\bibitem{Bissey:2005kd}
F.~Bissey, F.~G.~Cao and A.~I.~Signal,
Phys. Rev. D \textbf{73}, 094008 (2006).

\bibitem{Kuhn:2008sy}
S.~E.~Kuhn, J.~P.~Chen and E.~Leader,
Prog. Part. Nucl. Phys. \textbf{63}, 1-50 (2009).


\bibitem{Blumlein:2012bf}
J.~Blumlein,
Prog. Part. Nucl. Phys. \textbf{69}, 28-84 (2013).

\bibitem{Bertone:2024taw}
V.~Bertone \textit{et al.} [MAP],
[arXiv:2404.04712 [hep-ph]].


\bibitem{Borsa:2024mss}
I.~Borsa, M.~Stratmann, W.~Vogelsang, D.~de Florian and R.~Sassot,
Phys. Rev. Lett. \textbf{133}, no.15, 151901 (2024).

\bibitem{Anselmino:1996cd}
M.~Anselmino, P.~Gambino and J.~Kalinowski,
Phys. Rev. D \textbf{55}, 5841-5844 (1997).

\bibitem{Forte:2001ph}
S.~Forte, M.~L.~Mangano and G.~Ridolfi,
Nucl. Phys. B \textbf{602}, 585-621 (2001).

\bibitem{Timoshin:2014ina}
E.~S.~Timoshin and S.~I.~Timoshin,
Phys. Part. Nucl. Lett. \textbf{11}, 86-90 (2014).

\bibitem{Bjorken:1968dy}
J.~D.~Bjorken,
Phys. Rev. \textbf{179}, 1547-1553 (1969).

\bibitem{Friedman:1990ur}
J.~I.~Friedman, H.~W.~Kendall and R.~E.~Taylor,
SLAC-REPRINT-1991-019.

\bibitem{Whitlow:1990gk}
L.~W.~Whitlow, S.~Rock, A.~Bodek, E.~M.~Riordan and S.~Dasu,
Phys. Lett. B \textbf{250}, 193-198 (1990).
  

\bibitem{NewMuon:1996fwh}
M.~Arneodo \textit{et al.} [New Muon],
Nucl. Phys. B \textbf{483}, 3-43 (1997).

\bibitem{HERMES:2011yno}
A.~Airapetian \textit{et al.} [HERMES],
JHEP \textbf{05}, 126 (2011).

\bibitem{Altarelli:1977zs} 
  G.~Altarelli and G.~Parisi,
    Nucl.\ Phys.\ B {\bf 126}, 298 (1977);\\
   V.~N.~Gribov and L.~N.~Lipatov,
    Sov.\ J.\ Nucl.\ Phys.\  {\bf 15}, 438 (1972),
  [Yad.\ Fiz.\  {\bf 15}, 781 (1972)];\\
    L.~N.~Lipatov,
    Sov.\ J.\ Nucl.\ Phys.\  {\bf 20}, 94 (1975),
  [Yad.\ Fiz.\  {\bf 20}, 181 (1974)];\\
   Y.~L.~Dokshitzer,
    Sov.\ Phys.\ JETP {\bf 46}, 641 (1977)
  [Zh.\ Eksp.\ Teor.\ Fiz.\  {\bf 73}, 1216 (1977)].
  

\bibitem{Vogt:2008yw}
A.~Vogt, S.~Moch, M.~Rogal and J.~A.~M.~Vermaseren,
Nucl. Phys. B Proc. Suppl. \textbf{183}, 155-161 (2008).

\bibitem{Moch:2014sna}
S.~Moch, J.~A.~M.~Vermaseren and A.~Vogt,
Nucl. Phys. B \textbf{889}, 351-400 (2014).

\bibitem{Moch:2015usa}
S.~Moch, J.~A.~M.~Vermaseren and A.~Vogt,
Phys. Lett. B \textbf{748}, 432-438 (2015).

\bibitem{Blumlein:2021enk}
J.~Bl\"umlein, P.~Marquard, C.~Schneider and K.~Sch\"onwald,
Nucl. Phys. B \textbf{971}, 115542 (2021).

\bibitem{Blumlein:2021ryt}
J.~Bl\"umlein, P.~Marquard, C.~Schneider and K.~Sch\"onwald,
JHEP \textbf{01}, 193 (2022).

\bibitem{Vogt:2004ns}
A.~Vogt,
Comput. Phys. Commun. \textbf{170}, 65-92 (2005).

\bibitem{Wilson:1969zs}
K.~G.~Wilson,
Phys. Rev. \textbf{179}, 1499-1512 (1969).

\bibitem{Brandt:1970kg}
R.~A.~Brandt and G.~Preparata,
Nucl. Phys. B \textbf{27}, 541-567 (1971).

\bibitem{Christ:1972ms}
N.~H.~Christ, B.~Hasslacher and A.~H.~Mueller,
Phys. Rev. D \textbf{6}, 3543 (1972).

\bibitem{SajjadAthar:2020nvy}
M.~Sajjad Athar and J.~G.~Morf\'\i{}n,
J. Phys. G \textbf{48}, no.3, 034001 (2021).

\bibitem{Salimi-Amiri:2018had}
M.~Salimi-Amiri, A.~Khorramian, H.~Abdolmaleki and F.~I.~Olness,
Phys. Rev. D \textbf{98}, no.5, 056020 (2018).

\bibitem{Khanpour:2017cha}
H.~Khanpour, S.~T.~Monfared and S.~Atashbar Tehrani,
Phys. Rev. D \textbf{95}, no.7, 074006 (2017).

\bibitem{Blumlein:2010rn}
J.~Blumlein and H.~Bottcher,
Nucl. Phys. B \textbf{841}, 205-230 (2010).

\bibitem{Leader:2005ci}
E.~Leader, A.~V.~Sidorov and D.~B.~Stamenov,
Phys. Rev. D \textbf{73}, 034023 (2006).

\bibitem{Nocera:2014gqa}
E.~R.~Nocera \textit{et al.} [NNPDF],
Nucl. Phys. B \textbf{887}, 276-308 (2014).

\bibitem{Ethier:2017zbq}
J.~J.~Ethier, N.~Sato and W.~Melnitchouk,
Phys. Rev. Lett. \textbf{119}, no.13, 132001 (2017).

\bibitem{DeFlorian:2019xxt}
D.~De Florian, G.~A.~Lucero, R.~Sassot, M.~Stratmann and W.~Vogelsang,
Phys. Rev. D \textbf{100}, no.11, 114027 (2019).

\bibitem{Mirjalili:2022cal}
A.~Mirjalili and S.~Tehrani Atashbar,
Phys. Rev. D \textbf{105}, no.7, 074023 (2022).

\bibitem{Blumlein:1998nv}
J.~Blumlein and A.~Tkabladze,
Nucl. Phys. B \textbf{553}, 427-464 (1999).

\bibitem{Braun:2011aw}
V.~M.~Braun, T.~Lautenschlager, A.~N.~Manashov and B.~Pirnay,
Phys. Rev. D \textbf{83}, 094023 (2011).

\bibitem{Chiefa:2024yyr}
A.~Chiefa,
[arXiv:2408.09263 [hep-ph]].

\bibitem{Ellis:1973kp}
J.~R.~Ellis and R.~L.~Jaffe,
Phys. Rev. D \textbf{9}, 1444 (1974)
[erratum: Phys. Rev. D \textbf{10}, 1669 (1974)].

\bibitem{Burkhardt:1970ti}
H.~Burkhardt and W.~N.~Cottingham,
Annals Phys. \textbf{56}, 453-463 (1970).


\bibitem{Zijlstra:1993sh}
E.~B.~Zijlstra and W.~L.~van Neerven,
Nucl. Phys. B \textbf{417}, 61-100 (1994)
[erratum: Nucl. Phys. B \textbf{426}, 245 (1994); erratum: Nucl. Phys. B \textbf{773}, 105-106 (2007); erratum: Nucl. Phys. B \textbf{501}, 599-599 (1997)].

\bibitem{Vogelsang:1996im}
W.~Vogelsang,
Nucl. Phys. B \textbf{475}, 47-72 (1996).

\bibitem{Blumlein:1996vs}
J.~Blumlein and N.~Kochelev,
Nucl. Phys. B \textbf{498}, 285-309 (1997).

\bibitem{Zaidi:2019asc}
F.~Zaidi, H.~Haider, M.~Sajjad Athar, S.~K.~Singh and I.~Ruiz Simo,
Phys. Rev. D \textbf{101}, no.3, 033001 (2020).

\bibitem{Zaidi:2019mfd}
F.~Zaidi, H.~Haider, M.~Sajjad Athar, S.~K.~Singh and I.~Ruiz Simo,
Phys. Rev. D \textbf{99}, no.9, 093011 (2019).

\bibitem{Kodaira:1979ib}
J.~Kodaira, S.~Matsuda, K.~Sasaki and T.~Uematsu,
Nucl. Phys. B \textbf{159}, 99-124 (1979).

\bibitem{Kodaira:1978sh}
J.~Kodaira, S.~Matsuda, T.~Muta, K.~Sasaki and T.~Uematsu,
Phys. Rev. D \textbf{20}, 627 (1979)
doi:10.1103/PhysRevD.20.627


\bibitem{Bodwin:1989nz}
G.~T.~Bodwin and J.~W.~Qiu,
Phys. Rev. D \textbf{41}, 2755 (1990)
doi:10.1103/PhysRevD.41.2755

\bibitem{Vogelsang:1990ug}
W.~Vogelsang,
Z. Phys. C \textbf{50}, 275-284 (1991).


\bibitem{Gorishnii:1985xm}
S.~G.~Gorishnii and S.~A.~Larin,
Phys. Lett. B \textbf{172}, 109-112 (1986).


\bibitem{Wandzura:1977qf}
S.~Wandzura and F.~Wilczek,
Phys. Lett. B \textbf{72}, 195-198 (1977).

\bibitem{Athar:2020kqn}
M.~S.~Athar and S.~K.~Singh,
Cambridge University Press, 2020,
ISBN 978-1-108-77383-6, 978-1-108-48906-5
doi:10.1017/9781108489065

\bibitem{Hekhorn:2024tqm}
F.~Hekhorn, G.~Magni, E.~R.~Nocera, T.~R.~Rabemananjara, J.~Rojo, A.~Schaus and R.~Stegeman,
Eur. Phys. J. C \textbf{84}, no.2, 189 (2024).

\bibitem{Harland-Lang:2014zoa}
L.~A.~Harland-Lang, A.~D.~Martin, P.~Motylinski and R.~S.~Thorne,
Eur. Phys. J. C \textbf{75}, no.5, 204 (2015).

\bibitem{Gluck:2000dy}
M.~Gluck, E.~Reya, M.~Stratmann and W.~Vogelsang,
Phys. Rev. D \textbf{63}, 094005 (2001).

\bibitem{Khorramian:2010qa}
A.~N.~Khorramian, S.~Atashbar Tehrani, S.~Taheri Monfared, F.~Arbabifar and F.~I.~Olness,
Phys. Rev. D \textbf{83}, 054017 (2011).

\bibitem{Arbabifar:2023hok}
F.~Arbabifar, S.~Atashbar Tehrani and H.~Khanpour,
Phys. Rev. C \textbf{108}, no.3, 035203 (2023).

\bibitem{Gluck:1995yr}
M.~Gluck, E.~Reya, M.~Stratmann and W.~Vogelsang,
Phys. Rev. D \textbf{53}, 4775-4786 (1996).

\bibitem{Leader:2001kh}
E.~Leader, A.~V.~Sidorov and D.~B.~Stamenov,
Eur. Phys. J. C \textbf{23}, 479-485 (2002).


\bibitem{Leader:2006xc}
E.~Leader, A.~V.~Sidorov and D.~B.~Stamenov,
Phys. Rev. D \textbf{75}, 074027 (2007).

\bibitem{Leader:2014uua}
E.~Leader, A.~V.~Sidorov and D.~B.~Stamenov,
Phys. Rev. D \textbf{91}, no.5, 054017 (2015).

\bibitem{Blumlein:2002qeu}
J.~Blumlein and H.~Bottcher,
Nucl. Phys. B \textbf{636}, 225-263 (2002)

\bibitem{deFlorian:2000bm}
D.~de Florian and R.~Sassot,
Phys. Rev. D \textbf{62}, 094025 (2000).

\bibitem{deFlorian:2005mw}
D.~de Florian, G.~A.~Navarro and R.~Sassot,
Phys. Rev. D \textbf{71}, 094018 (2005).

\bibitem{deFlorian:2008mr}
D.~de Florian, R.~Sassot, M.~Stratmann and W.~Vogelsang,
Phys. Rev. Lett. \textbf{101}, 072001 (2008).

\bibitem{deFlorian:2014yva}
D.~de Florian, R.~Sassot, M.~Stratmann and W.~Vogelsang,
Phys. Rev. Lett. \textbf{113}, no.1, 012001 (2014).

\bibitem{Hirai:2003pm}
M.~Hirai \textit{et al.} [Asymmetry Analysis],
Phys. Rev. D \textbf{69}, 054021 (2004).

\bibitem{Hirai:2008aj}
M.~Hirai \textit{et al.} [Asymmetry Analysis],
Nucl. Phys. B \textbf{813}, 106-122 (2009).

\bibitem{Cruz-Martinez:2025ahf}
J.~Cruz-Martinez, T.~Hasenack, F.~Hekhorn, G.~Magni, E.~R.~Nocera, T.~R.~Rabemananjara, J.~Rojo, T.~Sharma and G.~van Seeventer,
[arXiv:2503.11814 [hep-ph]].

\bibitem{deFlorian:2009vb}
D.~de Florian, R.~Sassot, M.~Stratmann and W.~Vogelsang,
Phys. Rev. D \textbf{80}, 034030 (2009).

\bibitem{Kretzer:1999nn}
S.~Kretzer and M.~Stratmann,
Eur. Phys. J. C \textbf{10}, 107-119 (1999).







































































































\bibitem{TaheriMonfared:2014var}
S.~Taheri Monfared, Z.~Haddadi and A.~N.~Khorramian,
Phys. Rev. D \textbf{89}, no.7, 074052 (2014)
[erratum: Phys. Rev. D \textbf{89}, no.11, 119901 (2014)].

\bibitem{Farrar:1975yb}
G.~R.~Farrar and D.~R.~Jackson,
Phys. Rev. Lett. \textbf{35}, 1416 (1975).

\bibitem{Close:1988br}
F.~E.~Close and A.~W.~Thomas,
Phys. Lett. B \textbf{212}, 227-230 (1988).

\bibitem{Brodsky:1994kg}
S.~J.~Brodsky, M.~Burkardt and I.~Schmidt,
Nucl. Phys. B \textbf{441}, 197-214 (1995).


\bibitem{Leader:1997kw}
E.~Leader, A.~V.~Sidorov and D.~B.~Stamenov,
Int. J. Mod. Phys. A \textbf{13}, 5573-5592 (1998).

\bibitem{Isgur:1998yb}
N.~Isgur,
Phys. Rev. D \textbf{59}, 034013 (1999).

\bibitem{Gell-Mann:1962yej}
M.~Gell-Mann,
Phys. Rev. \textbf{125}, 1067-1084 (1962).

\bibitem{Gell-Mann:1964hhf}
M.~Gell-Mann,
Physics Physique Fizika \textbf{1}, 63-75 (1964).

\bibitem{Feynman:1964fk}
R.~P.~Feynman, M.~Gell-Mann and G.~Zweig,
Phys. Rev. Lett. \textbf{13}, 678-680 (1964).

\bibitem{Bjorken:1966jh}
J.~D.~Bjorken,
Phys. Rev. \textbf{148}, 1467-1478 (1966).

\bibitem{Bjorken:1969mm}
J.~D.~Bjorken,
Phys. Rev. D \textbf{1}, 1376-1379 (1970).


\bibitem{Nadel-Turonski:2025sfv}
P.~Nadel-Turonski,
Acta Phys. Polon. Supp. \textbf{18}, no.1, 1-A43 (2025).

\bibitem{Anderle:2021wcy}
D.~P.~Anderle, V.~Bertone, X.~Cao, L.~Chang, N.~Chang, G.~Chen, X.~Chen, Z.~Chen, Z.~Cui and L.~Dai, \textit{et al.}
Front. Phys. (Beijing) \textbf{16}, no.6, 64701 (2021).

\bibitem{Cotton:2024rpn}
C.~Cotton, J.~Smith and X.~Zheng,
PoS \textbf{SPIN2023}, 159 (2024).




















































































































































































































































































































































\end{thebibliography}
\end{document}